    \documentclass[
        fleqn,
        usenatbib,
    ]{mnras}
    \usepackage{
        amsmath,
        amssymb,
        newtxtext,
        newtxmath,
        ae, aecompl,
        graphicx,
        booktabs,
        xcolor,
    }

    \newcommand*{\rd}[2]{\frac{\mathrm{d}#1}{\mathrm{d}#2}}
    
    \newcommand*{\pd}[2]{\frac{\partial#1}{\partial#2}}
    
    \newcommand*{\rdil}[2]{\mathrm{d}#1/\mathrm{d}#2}
    
    \newcommand*{\abs}[1]{\left|#1\right|}
    \newcommand*{\ev}[1]{\left\langle#1\right\rangle}
    \newcommand*{\p}[1]{\left(#1\right)}
    \newcommand*{\s}[1]{\left[#1\right]}
    
    \newcommand*{\bm}[1]{\mathbf{#1}}
    \newcommand*{\uv}[1]{\hat{\boldsymbol{\mathbf{#1}}}}

    \newlength{\colummwidth}
\title[Tidal Dissipation and Resonance Capture]{Dynamics of Colombo's Top: Tidal
Dissipation and Resonance Capture, With Applications to Oblique Super-Earths,
Ultra-Short-Period Planets and Inspiraling Hot Jupiters }
\author[Y. Su and D. Lai.]{
Yubo Su,$^1$\thanks{E-mail: yubosu@astro.cornell.edu},
Dong Lai$^1$
\\
$^1$ Cornell Center for Astrophysics and Planetary Science, Department of
Astronomy, Cornell University, Ithaca, NY 14853, USA\\
}

\date{Accepted 2021 October 27. Received 2021 October 25; in original form 2021
July 29.}

\pubyear{2021}

\begin{document}\label{firstpage}
\pagerange{\pageref{firstpage}--\pageref{lastpage}}
\maketitle

\begin{abstract}
We present a comprehensive theoretical study on the spin evolution of a planet
under the combined effects of tidal dissipation and gravitational perturbation
from an external companion.  Such a ``spin + companion'' system (called
Colombo's top) appears in many [exo]planetary contexts. The competition between
the tidal torque (which drives spin-orbit alignment and synchronization) and the
gravitational torque from the companion (which drives orbital precession of the
planet) gives rise to two possible spin equilibria (``Tidal Cassini
Equilibria'', tCE) that are stable and attracting: the ``simple'' tCE1, which
typically has a low spin obliquity, and the ``resonant'' tCE2, which can have a
significant obliquity. The latter arises from a spin-orbit resonance and can be
broken when the tidal alignment torque is stronger than the precessional torque
from the companion.  We characterize the long-term evolution of the planetary
spin (both magnitude and obliquity) for an arbitrary initial spin orientation,
and develop a new theoretical method to analytically obtain the probability of
resonance capture driven by tidal dissipation. Applying our general theoretical
results to exoplanetary systems, we find that a super-Earth (SE) with an
exterior companion can have a substantial probability of being trapped in the
high-obliquity tCE2, assuming that SEs have a wide range of primordial
obliquities. We also evaluate the recently proposed ``obliquity tide'' scenarios
for the formation of ultra-short-period Earth-mass planets and for the orbital
decay of hot Jupiter WASP-12b. We find in both cases that the probability of
resonant capture into tCE2 is generally low and that such a high-obliquity state
can be easily broken by the required orbital decay.

\end{abstract}


\section{Introduction}\label{s:intro}

It is well recognized that the obliquity of a planet, the angle between the spin
and orbital axes, likely reflects its dynamical history. In our Solar
System, planetary obliquities (hereafter just ``obliquities'') range from
$3.1^\circ$ for Jupiter to $26.7^\circ$ for Saturn to $98^\circ$ for Uranus. The
obliquities of exoplanets are challenging to measure, and so far only loose
constraints have been obtained for the obliquity of a faraway ($\gtrsim
50\;\mathrm{AU}$) planetary-mass companion \citep{bryan2020obliquity}.
Nevertheless, there are prospects for better constraints on exoplanetary
obliquities in the coming years, such as using high-resolution spectroscopy to
measure $v\sin i$ for planetary rotation \citep{snellen2014fast,
bryan2018constraints} and using high-precision photometry to measure the
asphericity of a planet \citep{seager2002constraining}. Substantial obliquities
are of increasing theoretical interest for their proposed role in explaining
peculiar thermal phase curves \citep[see e.g.][]{millholland_signatures,
ohno_infer_obl}, in enhancing tidal dissipation in hot Jupiters
\citep{millholland_wasp12b} and super-Earths \citep{millholland2019obliquity},
and in the formation of ultra-short-period planets
\citep[USPs;][]{millholland2020formation}.

While nonzero obliquities are sometimes attributed to one or many giant
impacts/collisions \citep[e.g.][]{original_gi, benz1989tilting,
korycansky1990one, dones1993does, morbidelli_gi, li2020planetary, li2021giant},
some studies suggest that large planetary obliquities may be produced by
spin-orbit resonances. In this scenario, a rotating planet is subjected to a
gravitational torque from its host star, making its spin axis precess around its
orbital (angular momentum) axis. At the same time, the orbital axis precesses
around another fixed axis under the gravitational influence of other masses in
the system, e.g.\ additional planets or a protoplanetary disk. When the two
precession frequencies become comparable, a resonance can occur that excites
the obliquity to large values. This model is known as ``Colombo's Top'' after
the seminal work of \citet{colombo1966}, and subsequent works have investigated
the rich dynamics of this system \citep{peale1969, peale1974possible,
ward1975tidal, henrard1987}. Such resonances have been invoked to explain the
obliquities of both the Solar System gas giants \citep{ward2004I, ward2004II,
ward_jupiter, vokrouhlicky2015tilting, saillenfest2020future,
saillenfest2021large} and the ice giants \citep{hamilton_tilting_ice}.

In a previous paper \citep[hereafter Paper I]{su2020}, we presented a systematic
and general investigation of the dynamics of Colombo's Top when the two
precession frequencies of the system evolve through a commensurability. We
obtained a semi-analytic mapping between the (arbitrary) planetary spin
orientation and the final obliquity after a resonance encounter. We applied our
results to investigate the generation of exoplanetary obliquities via a
dissipating protoplanetary disk. However, our model did not consider the effect
of additional torques in the system. In particular, tidal dissipation in the
planet can cause the planet's spin frequency to approach its orbital frequency
and drive the planet's spin axis towards its orbital axis, complicating the
evolution of Colombo's Top \citep{fabrycky_otides, levrard2007, peale2008obliquity}. In this
paper, we extend these previous works to present a comprehensive study on how
tidal dissipation influences the equilibria (called ``Cassini States'') of the
system and drive its long-term evolution. Our new results (summarized in
Section~\ref{s:summary}) include a stability analysis of tide-modified Cassini
States and a novel, analytic description/calculation of the resonance encounter
process. We apply our general theoretical results to assess how obliquity tides
may affect different types of exoplanetary systems.

Our paper is organized as follows. In Section~\ref{s:theory}, we briefly review
the basic setup and non-dissipative dynamics of Colombo's Top. In
Section~\ref{s:toy_model}, we investigate the effect of adding a simple
alignment torque to Colombo's Top. The resulting dynamics captures the essential
behavior that emerges due to tidal dissipation. In
Section~\ref{s:full_tide_prob}, we solve for the dynamics of the system
including the full effect of tidal dissipation. In Section~\ref{s:applications},
we apply our results to three exoplanetary systems/scenarios of interest: (i) a
super Earth with an exterior companion, (ii) the formation of USPs via obliquity
tides, and (iii) the rapid orbital decay of the hot Jupiter WASP-12b. We
summarize and discuss in Section~\ref{s:summary}.

\section{Spin Evolution Equations and Cassini States: Review}\label{s:theory}

In this section, we briefly review the spin dynamics of a planet in the presence
of a distant perturber and introduce our notations; see Paper I for more
details. We consider a star of mass $M_\star$ hosting an inner oblate planet of
mass $m$ and radius $R$ on a circular orbit with semi-major axis $a$ and an
outer perturber of mass $m_{\rm p}$ on a circular orbit with semi-major axis
$a_{\rm p}$. The two orbits are mutually inclined by the angle $I$. Denote
$\bm{S}$ the spin angular momentum and $\bm{L}$ the orbital angular momentum of
the planet, and $\bm{L}_{\rm p}$ the angular momentum of the perturber. The
corresponding unit vectors are $\uv{s} \equiv \bm{S} / S$, $\uv{l} \equiv \bm{L}
/ L$, and $\uv{l}_{\rm p} \equiv \bm{L}_{\rm p} / L_{\rm p}$. The spin axis
$\uv{s}$ of the planet tends to precess around its orbital (angular momentum)
axis $\uv{l}$, driven by the gravitational torque from the host star acting on
the planet's rotational bulge. On the other hand, $\uv{l}$ and $\uv{l}_{\rm p}$
precess around each other due to gravitational interactions. Assuming $S \ll L$,
the equations of motion for $\uv{s}$ and $\uv{l}$ are
\begin{align}
    \rd{\uv{s}}{t}
        &= \omega_{\rm sl}\p{\uv{s} \cdot \uv{l}}\p{\uv{s} \times \uv{l}}
        \equiv \alpha\p{\uv{s} \cdot \uv{l}}\p{\uv{s} \times
        \uv{l}},\label{eq:dsdt1}\\
    \rd{\uv{l}}{t} &= \omega_{\rm lp}\p{\uv{l} \cdot \uv{l}_{\rm p}}\p{\uv{l}
        \times \uv{l}_{\rm p}} \equiv -g\p{\uv{l} \times \uv{l}_{\rm p}},
        \label{eq:dldt1}
\end{align}
where
\begin{align}
    \omega_{\rm sl} &\equiv \alpha =
        \frac{3GJ_2 mR^2 M_\star}{2a^3 \mathcal{I}\Omega_{\rm s}}
        = \frac{3k_q}{2k}\frac{M_\star}{m}\p{\frac{R}{a}}^3 \Omega_{\rm s},
            \label{eq:wsl}\\
    \omega_{\rm lp} &\equiv -\frac{g}{\cos I}
        = \frac{3m_{\rm p}}{4M_\star}\p{\frac{a}{a_{\rm p}}}^3 n.\label{eq:wlp}
\end{align}
In Eq.~\eqref{eq:wsl}, $\Omega_{\rm s}$ is the spin frequency of the inner
planet, $\mathcal{I} = k mR^2$ (with $k$ the normalized moment of inertia, often
notated as $C_{\rm N}$) is its moment of inertia and $J_2 = k_{\rm q}\Omega_{\rm
s}^2 (R^3/Gm)$ (with $k_{\rm q}$ a constant, related to the hydrostatic Love
number $k_2$ by $k_{\rm q} = k_2/3$) is its rotation-induced (dimensionless)
quadrupole moment [for a fluid body with uniform density, $k=0.4, k_{\rm q} =
0.5$; for the Earth, $k \simeq 0.331$ and $k_{\rm q} \simeq 0.31$; for Jupiter,
$k \simeq 0.27$ and $k_{\rm q} \simeq 0.18$
\citep[e.g.][]{groten2004fundamental, lainey2016quantification}]. In other
studies, $3k_{\rm q} / 2 k$ is often notated as $k_2 / 2C_{\rm N}$
\citep[e.g.][]{millholland_disk}. In Eq.~\eqref{eq:wlp}, $n \equiv
\sqrt{GM_\star/a^3}$ is the inner planet's orbital mean motion,  and we have
assumed $a_{\rm p}\gg a$ and included only the leading-order (quadrupole)
interaction between the inner planet and perturber (Section~\ref{ss:disc_usp}
discusses modifications to Eq.~\ref{eq:wlp} when $a_{\rm p} \gtrsim a$).
Eq.~\eqref{eq:dldt1} neglects the back-reaction torque on $\uv{l}$ from
$\uv{s}$; this is justified since $L \gg S$ (see \citealp{anderson2018teeter}
for the case when $L \sim S$). In Eq.~\eqref{eq:wsl} (and throughout
Sections~\ref{s:theory}--\ref{s:full_tide_prob}), we assume $L_{\rm p} \gg L$ so
that $\uv{l}_{\rm p}$ is a constant (Section~\ref{ss:disc_wasp12b} discusses the
case of $L \simeq L_{\rm p}$). Following the standard notations, we have defined
$\alpha = \omega_{\rm sl}$ and $g \equiv -\omega_{\rm 1p} \cos I$
\citep[e.g.][]{colombo1966}.

As in Paper I, we combine Eqs.~\eqref{eq:dsdt1}--\eqref{eq:dldt1} into a
single equation by transforming into a frame rotating about $\uv{l}_{\rm p}$
with frequency $g$. In this frame, $\uv{l}_{\rm p}$ and $\uv{l}$ are both fixed,
and $\uv{s}$ evolves as
\begin{equation}
    \p{\rd{\uv{s}}{t}}_{\rm rot}
        = \alpha\p{\uv{s} \cdot \uv{l}}\p{\uv{s} \times \uv{l}}
            + g\p{\uv{s} \times \uv{l}_{\rm p}}. \label{eq:dsdt_rot}
\end{equation}
We choose the coordinate system such that $\uv{z} = \uv{l}$ and $\uv{l}_{\rm p}$
lies in the $\uv{x}$-$\uv{z}$ plane. We describe $\uv{s}$ in spherical
coordinates using the polar angle $\theta$, the planet's obliquity, and $\phi$,
the precessional phase of $\uv{s}$ about $\uv{l}$, defined so that when $\phi =
0^\circ$, $\uv{l}_{\rm p}$ and $\uv{s}$ are on opposite sides of $\uv{l}$.

\begin{figure}
    \centering
    \includegraphics[width=\colummwidth]{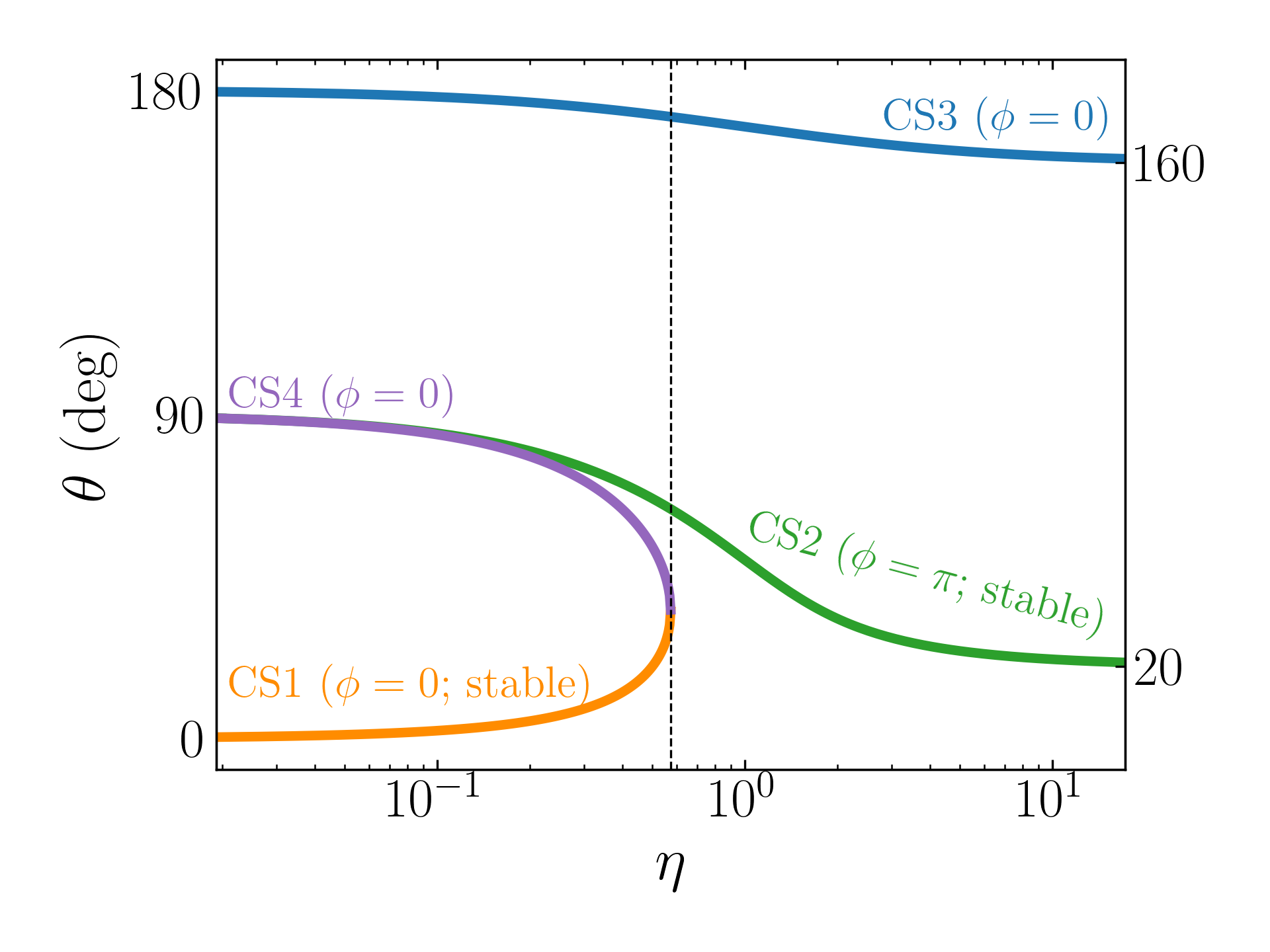}
    \caption{Cassini State obliquities $\theta$ as a function of $\eta \equiv
    -g/\alpha$ (Eq.~\ref{eq:def_eta}) for $I = 20^\circ$. The vertical dashed
    line denotes $\eta_{\rm c}$, where the number of Cassini States changes from
    four to two (Eq.~\ref{eq:def_etac}). The y-axis labels on the right of
    the plot show the asymptotic obliquities for CS2 and CS3, $I$ and $180^\circ
    - I$ respectively. Note that $\theta$ does not follow the standard
    convention (e.g. \citealp{colombo1966}, Paper I) and is simply the angle
    between $\uv{s}$ and $\uv{l}$, while $\phi = 0$ corresponds to $\uv{s}$ and
    $\uv{l}_{\rm p}$ being on opposite sides of $\uv{l}$. While CSs 1--3 are
    ``dynamically'' stable, only CS1 and CS2 are stable and attracting in the
    presence of the spin-orbit alignment torque (see
    Section~\ref{ss:linear_stab}).}\label{fig:cs_locs}
\end{figure}

The equilibria of Eq.~\eqref{eq:dsdt_rot} are referred to as \emph{Cassini
States} \citep[CSs;][]{colombo1966, peale1969}. We follow the notation of
Paper I and introduce the parameter
\begin{equation}
    \eta \equiv -\frac{g}{\alpha} =
        \frac{1}{2}\frac{k}{k_{\rm q}}
            \frac{m_{\rm p}m}{M_\star^2}
            \p{\frac{a}{a_{\rm p}}}^3
            \p{\frac{a}{R}}^3
            \frac{n}{\Omega_{\rm s}}
            \cos I.\label{eq:def_eta}
\end{equation}
For a given value of $\eta$, there can be either two or four CSs, all of which
require $\uv{s}$ lie in the plane of $\uv{l}$ and $\uv{l}_{\rm p}$. In
the standard nomenclature, CSs 1, 3, and 4 have $\theta < 0$, implying that $\uv{s}$ and $\uv{l}_{\rm p}$ are on opposite sides
of $\uv{l}$, while CS2 has $\theta > 0$, implying that $\uv{s}$ and $\uv{l}$ are
on the same side of $\uv{l}$. We depart from the standard convention and simply
label the CSs using the polar angles $\theta$ and $\phi$ (with $\theta \in \s{0,
\pi}$): Figure~\ref{fig:cs_locs} shows the CS obliquities as a function of
$\eta$. CS1 and CS4 do not exist when $\eta > \eta_{\rm c}$, where
\begin{equation}
    \eta_{\rm c} \equiv \p{\sin^{2/3}\!I + \cos^{2/3}\!I}^{-3/2}.
        \label{eq:def_etac}
\end{equation}

\begin{figure}
    \centering
    \includegraphics[width=\colummwidth]{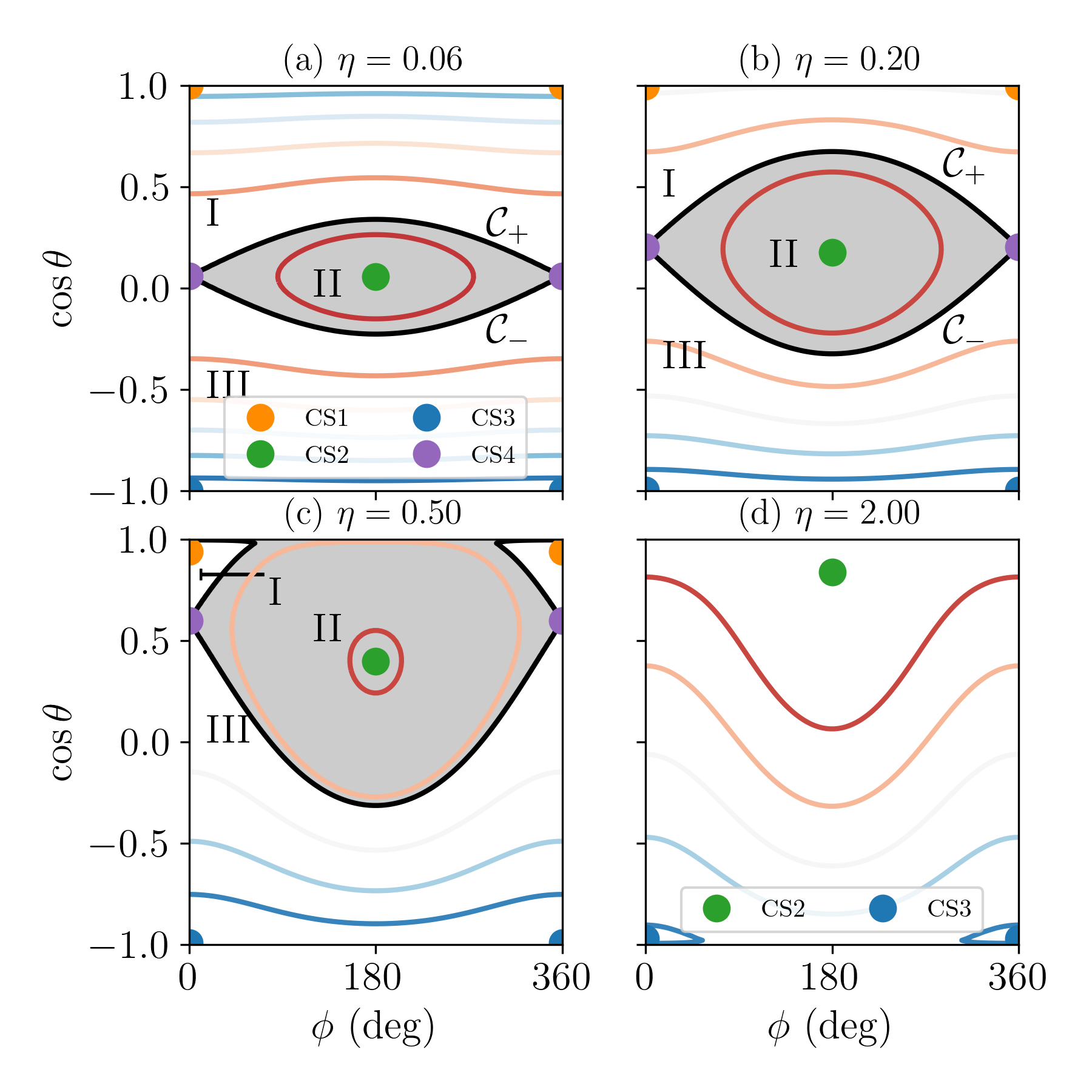}
    \caption{Level curves of the Hamiltonian (Eq.~\ref{eq:H}) for $I =
    20^\circ$, for which $\eta_{\rm c} \approx 0.57$ (Eq.~\ref{eq:def_etac}).
    For $\eta < \eta_{\rm c}$, there are four Cassini States (labeled), while
    for $\eta > \eta_{\rm c}$ there are only two. In the former case, the
    existence of a \emph{separatrix} (solid black lines) separates phase space
    into three numbered zones (I/II/III, labeled). We denote the upper and lower
    legs of the separatrix by $\mathcal{C}_{\pm}$ respectively, as shown in the
    upper two panels. }\label{fig:1contours}
\end{figure}

The Hamiltonian corresponding to Eq.~\eqref{eq:dsdt_rot} is
\begin{align}
    H &= -\frac{\alpha}{2}\p{\uv{s} \cdot \uv{l}}^2
            - g\p{\uv{s} \cdot \uv{l}_{\rm p}}\nonumber\\
        &= -\frac{\alpha}{2} \cos^2\theta
            - g\p{\cos\theta \cos I - \sin I \sin\theta \cos \phi}.\label{eq:H}
\end{align}
Here, $\cos \theta$ and $\phi$ form a canonically conjugate pair of variables.
Figure~\ref{fig:1contours} shows the level curves of this Hamiltonian for $I =
20^\circ$, for which $\eta_{\rm c} \approx 0.574$ (Eq.~\ref{eq:def_etac}). When
$\eta < \eta_{\rm c}$, CS4 exists and is a saddle point. The infinite-period
orbits originating and ending at CS4 form the \emph{separatrix} and divide phase
space into three zones. The angle $\phi$ librates for trajectories in zone II
and circulates for trajectories in zones I and III\@. On the other hand, when
$\eta > \eta_{\rm c}$, the separatrix is absent and all trajectories circulate.
When the separatrix exists, we divide it into two curves: $\mathcal{C}_+$, the
boundary between zones I and II, and $\mathcal{C}_-$, the boundary between zones
II and III\@.

\section{Spin Evolution with Alignment Torque}\label{s:toy_model}

In this section, we consider a simplified dissipative torque that isolates the
important new phenomenon presented in this paper. We assume that the spin
magnitude of the planet is constant, so $\alpha$ and $g$ are both fixed, while
the spin orientation $\uv{s}$ experiences an alignment torque towards $\uv{l}$
on the alignment timescale $t_{\rm al}$:
\begin{equation}
    \p{\rd{\uv{s}}{t}}_{\rm tide}
        = \frac{1}{t_{\rm al}} \uv{s} \times \p{\uv{l} \times \uv{s}}.
        \label{eq:dsdt_tide_toy}
\end{equation}
The full equations of motion for $\uv{s}$ in the coordinates $\theta$ and $\phi$
can be written as
\begin{align}
    \rd{\theta}{t} &= -g\sin I \sin \phi - \frac{1}{t_{\rm al}} \sin \theta,
        \label{eq:dqdt_toy}\\
    \rd{\phi}{t} &= -\alpha \cos\theta
        - g\p{\cos I + \sin I \cot \theta \cos \phi}.\label{eq:dfdt_toy}
\end{align}

\subsection{Modified Cassini States}\label{ss:mcs}

\begin{figure}
    \centering
    \includegraphics[width=\colummwidth]{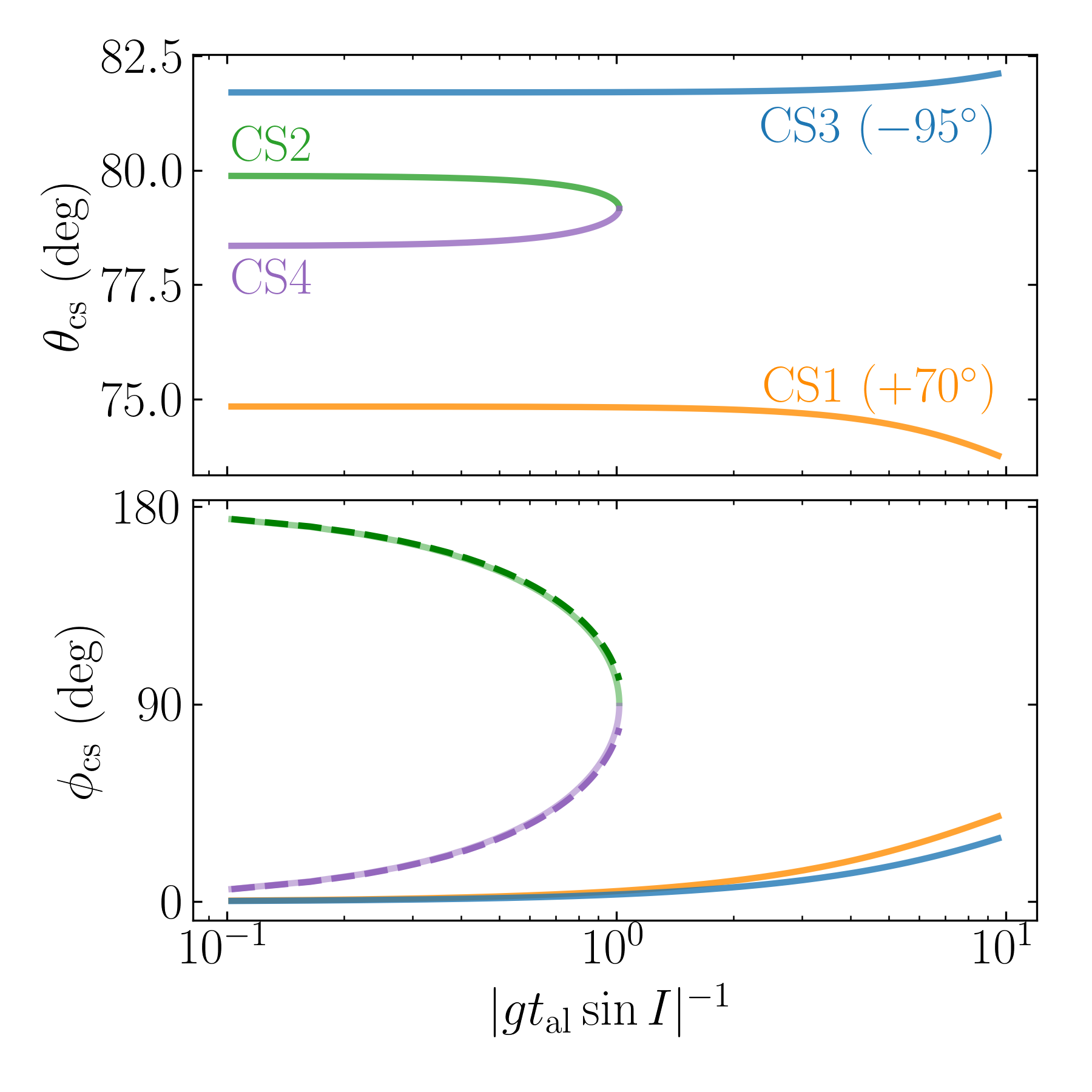}
    \caption{Modified CS obliquities (top) and azimuthal angles (bottom) for $I
    = 20^\circ$ and $\eta = 0.2$, where the CS1 and CS3 obliquities have been
    offset (as labeled) to improve clarity of the plot. In both panels, the
    solid lines give the result when applying a numerical root finding algorithm
    to the full equations of motion,
    Eqs.~(\ref{eq:dqdt_toy}--\ref{eq:dfdt_toy}), while the dotted lines in the
    bottom panel give the CS2 and CS4 azimuthal angles according to
    Eq.~\eqref{eq:mcs_shift}. At $\abs{gt_{\rm al} \sin I} = 1$, CS2 and CS4
    collide and annihilate (see Eq.~\ref{eq:mcs_shift_crit}).}\label{fig:mcs}
\end{figure}

If the alignment torque is weak ($\abs{g}t_{\rm al} \gg 1$), then the fixed
points of Eqs.~\eqref{eq:dqdt_toy}--\eqref{eq:dfdt_toy} are slightly modified
CSs. To leading order, all of the CS obliquities $\theta_{\rm cs}$ are unchanged
while the azimuthal angle $\phi_{\rm cs}$ for each CS now satisfies
\begin{equation}
    \sin \phi_{\rm cs} = \frac{\sin\theta_{\rm cs}}{\sin I \abs{g}t_{\rm al}}.
        \label{eq:mcs_shift}
\end{equation}
We can see that if $t_{\rm al}$ is longer than the critical alignment
timescale $t_{\rm al, c}$, given for a particular $\theta_{\rm cs}$ by
\begin{align}
    t_{\rm al, c} &\equiv \frac{\sin \theta_{\rm cs}}{\abs{g}\sin
    I},\label{eq:mcs_shift_crit}
\end{align}
then Eq.~\eqref{eq:mcs_shift} will always have solutions for $\phi_{\rm cs}$,
and the alignment torque does not change the number of fixed points of the
system. If $t_{\rm al}$ is decreased below $t_{\rm al, c} \sim \abs{g \sin
I}^{-1}$, CS2 and CS4 cease to be fixed points when $\eta \lesssim 1$ \citep[as
noted in][]{levrard2007, fabrycky_otides}, as $\theta_{\rm cs} \sim 90^\circ$ for
these (see Fig.~\ref{fig:cs_locs}). On the other hand, the other CSs have small
$\sin \theta_{\rm cs}$ and are only slightly modified. Figure~\ref{fig:mcs}
shows the obliquity and azimuthal angle for each of the CSs when $\eta = 0.2$,
obtained via numerical root finding of
Eqs.~(\ref{eq:dqdt_toy}--\ref{eq:dfdt_toy}), where it can be seen that CS2 and
CS4 collide and annihilate when $t_{\rm al}$ reaches $t_{\rm al, c}$. The phase
shifts $\phi_{\rm cs}$ for CS2 and CS4 for $t_{\rm al} > t_{\rm al, c}$ can be
predicted to good accuracy using Eq.~\eqref{eq:mcs_shift} and $\theta_{\rm cs}
\approx \pi/2 - \eta \cos I \approx 79^\circ$ \citep{su2020}; these are shown as
the dashed lines in the bottom panel of Fig.~\ref{fig:mcs}. For the remainder of
this section, we will consider the case where $t_{\rm al} \gg t_{\rm al, c}$ and
the CSs only differ slightly from their unmodified locations.

\subsection{Linear Stability Analysis}\label{ss:linear_stab}

We next seek to characterize the stability of small perturbations about each of
the CSs in the presence of the weak alignment torque. We can linearize
Eqs.~(\ref{eq:dqdt_toy}--\ref{eq:dfdt_toy}) about a shifted CS, yelding
\begin{align}
    \rd{}{t}\begin{bmatrix}
        \Delta \theta\\ \Delta \phi
    \end{bmatrix} &= \begin{bmatrix}
        -\frac{\cos \theta}{t_{\rm al}} &
        -g\sin I \cos \phi \\
        \alpha \sin \theta + g\frac{\sin I \cos \phi}{\sin^2\theta} &
        0
    \end{bmatrix}_{\rm cs}\begin{bmatrix}
        \Delta \theta \\ \Delta \phi
    \end{bmatrix},\label{eq:dsdt_hessian}
\end{align}
where the ``cs'' subscript indicates evaluating at a CS, $\Delta \theta = \theta
- \theta_{\rm cs}$, and $\Delta \phi = \phi - \phi_{\rm cs}$. The eigenvalues
$\lambda$ of Eq.~\eqref{eq:dsdt_hessian} satisfy the equation
\begin{equation}
    0 = \p{\lambda + \frac{\cos \theta_{\rm cs}}{t_{\rm al}}}\lambda
        - \lambda_0^2,\label{eq:lambda_orig}
\end{equation}
where
\begin{equation}
    \lambda_0^2 \equiv \p{\alpha
        \sin \theta_{\rm cs} + g\sin I \csc^2\theta_{\rm cs}\cos \phi_{\rm cs}}
            \p{- g \sin I \cos \phi_{\rm cs}}\label{eq:def_l0_sq}.
\end{equation}
When $t_{\rm al}$ is large, we can simplify Eq.~\eqref{eq:lambda_orig} to
\begin{equation}
    \lambda \approx -\frac{\cos \theta_{\rm cs}}{t_{\rm al}}
        \pm \sqrt{\lambda_0^2}. \label{eq:lambda_approx}
\end{equation}

\begin{figure}
    \centering
    \includegraphics[width=\colummwidth]{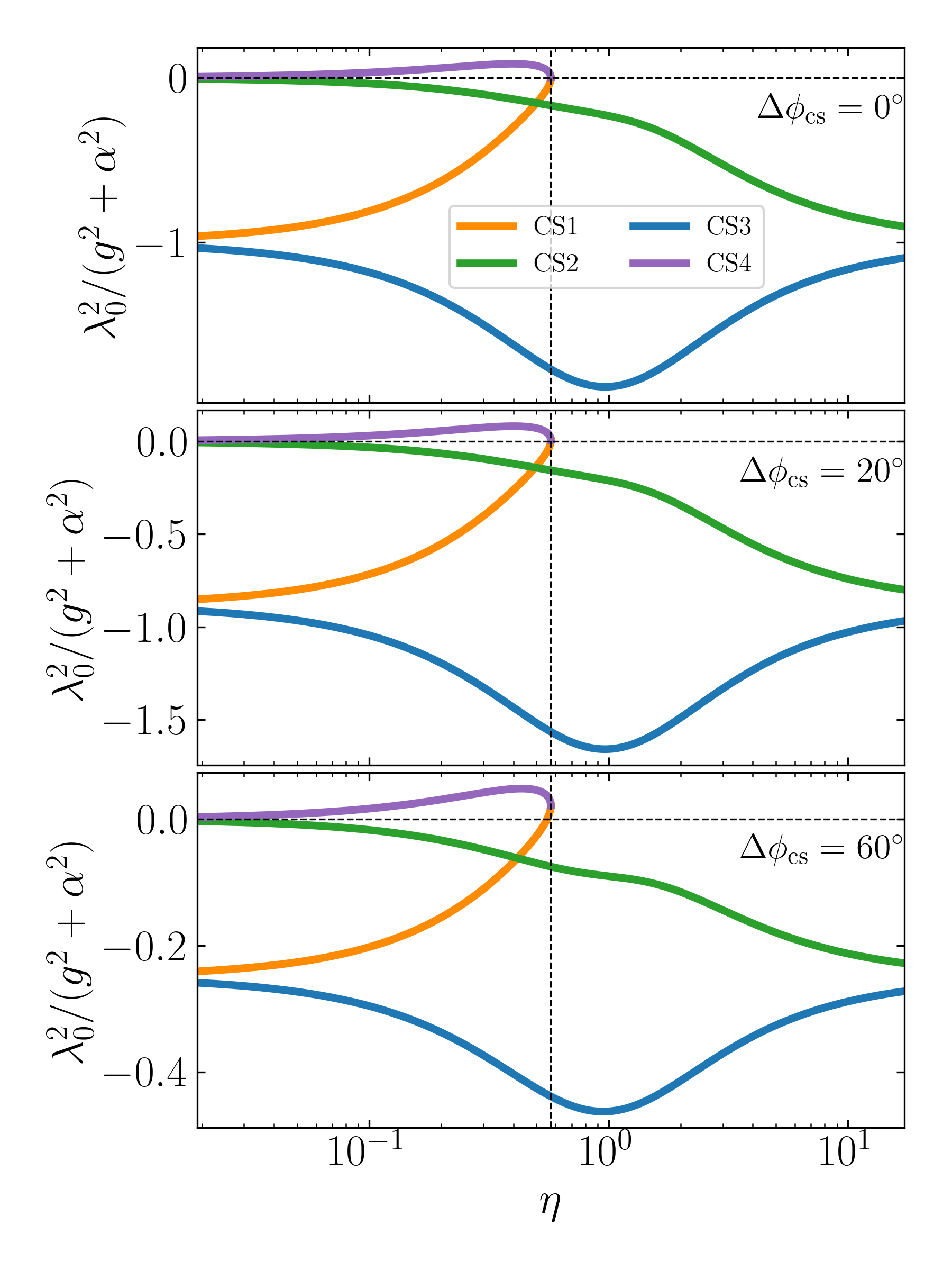}
    \caption{$\lambda_0^2$ (Eq.~\ref{eq:def_l0_sq}) as a function of $\eta$ for
    the four CSs, for three different values of the shift in $\phi_{\rm cs}$
    (e.g.\ for $\Delta \phi_{\rm cs} = 60^\circ$, the phase angles are
    $\phi_{\rm cs} = 120^\circ$ for CS2 and $\phi_{\rm cs} = 60^\circ$ for CSs
    1, 3, and 4). The values of $\Delta \phi_{\rm cs}$ are labeled ($\Delta
    \phi_{\rm cs} = 0$ corresponds to the unmodified CSs).
    }\label{fig:lambda_full}
\end{figure}

The stability of a CS depends on the real part of $\lambda$ in
Eq.~\eqref{eq:lambda_approx}. Equation~\eqref{eq:lambda_approx} $\lambda_0^2$ is
a generalization of Eq.~(A4) in Paper I and generally has the same behavior: it
is negative for CSs 1--3 and positive for CS4, as shown in
Fig.~\ref{fig:lambda_full}. Thus, CS4 is always ``dynamically'' unstable (i.e.\
unstable even in the limit of $t_{\rm al} \to \infty$), as there will always be
at least one positive solution for $\lambda$. On the other hand, CSs 1--3 are
dynamically stable, and their overall stabilities in the presence of the
alignment torque are determined by the sign of $\cos \theta_{\rm cs}$. Using
Fig.~\ref{fig:cs_locs}, we conclude that CS1 and CS2 are stable and attracting
while trajectories near CS3 are driven away by the alignment torque. These
calculations quantify the results long used in the literature
\citep[e.g.][]{ward1975tidal, fabrycky_otides}.

\subsection{Spin Obliquity Evolution Driven by Alignment
Torque}\label{ss:toy_outcomes}

With the above results, we are equipped to ask questions about the dynamics of
Eqs.~(\ref{eq:dqdt_toy}--\ref{eq:dfdt_toy}): what is the long-term evolution of
$\uv{s}$ for a general initial $\uv{s}_{\rm i}$?

For $\eta > \eta_{\rm c}$, the only stable (and attracting) spin state is CS2,
and all initial conditions will evolve asymptotically towards it.

For $\eta < \eta_{\rm c}$, both CS1 and CS2 are stable (assuming the alignment
torque is sufficiently weak that CS2 remains a fixed point; see
Section~\ref{ss:mcs}), and spin evolution may involve separatrix crossing. To
explore the fate of various initial $\uv{s}$ orientations, we numerically
integrate Eqs.~(\ref{eq:dqdt_toy}--\ref{eq:dfdt_toy}) for many random initial
conditions uniformly distributed in $\p{\cos \theta_{\rm i}, \phi_{\rm i}}$ and
determine the nearest CS for each integration after $10t_{\rm al}$. In
Fig.~\ref{fig:toy_phop}, we show the results of this procedure for $\eta = 0.2$,
and $I = 20^\circ$ (we use $t_{\rm al} = 10^3 / \abs{g}$, but the results are
unchanged as long as $t_{\rm al} \gg \abs{g}^{-1}$). It is clear that initial
conditions in zone I evolve into CS1, those in zone II evolve into CS2, while
those in zone III have a probabilistic outcome. These can be understood as
follows:
\begin{figure}
    \centering
    \includegraphics[width=\colummwidth]{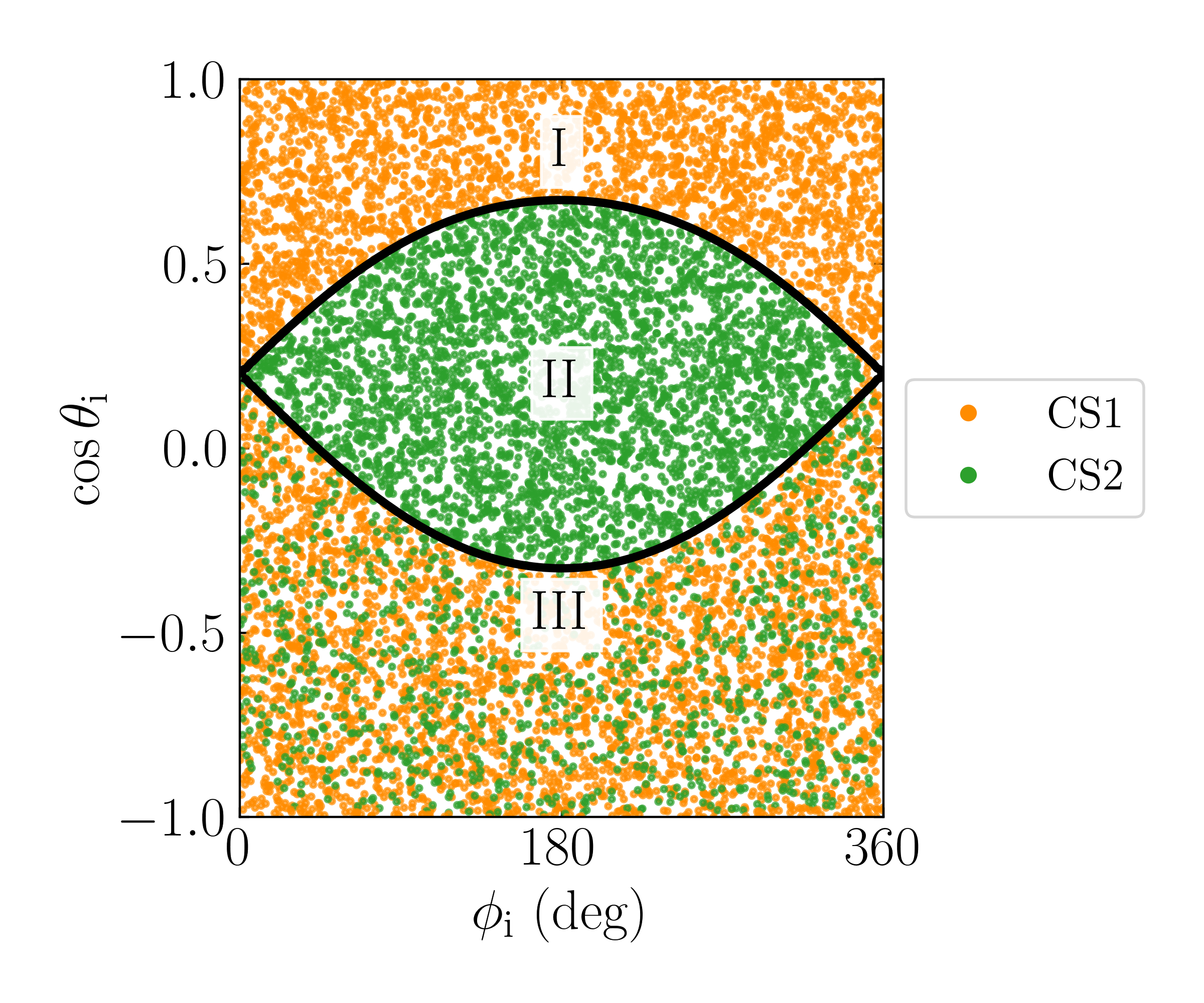}
    \caption{Asymptotic outcomes of spin evolution driven by an alignment torque
    for different initial spin orientations ($\theta_{\rm i}$ and $\phi_{\rm
    i}$) for a system with $\eta = 0.2$ and $I = 20^\circ$. Each dot represents
    an initial spin orientation, and the coloring of the dot indicates which
    stable Cassini State (legend) the system evolves into: initial conditions in
    Zone I evolve into CS1, those in Zone II evolve into CS2, and those in Zone
    III have a probabilistic outcome.}\label{fig:toy_phop}
\end{figure}

For initial conditions in zone I, the spin orientation circulates, and
$\dot{\theta}$ is negative everywhere during the cycle. Thus $\theta$ decreases
until the trajectory has converged to CS1. This is intuitively reasonable, as
CS1 is stable and attracting (see Section~\ref{ss:linear_stab}).

For initial conditions in zone II, our stability analysis in
Section~\ref{ss:linear_stab} shows that when $\uv{s}$ is sufficiently near CS2,
it will converge to CS2 since CS2 is stable and attracting. In fact, this result
can be extended to all initial conditions inside the separatrix, as shown in
Appendix~\ref{app:cs_stab2}.

For initial conditions in zone III, since there are no stable CSs in zone III,
the system must evolve through the separatrix to reach either CS1 or CS2.
The outcome of the separatrix encounter is probabilistic and determines the
final CS\@. Intuitively, this can be understood as probabilistic resonance
capture, as first studied in the seminal work of \citet{henrard1982}: for
$\eta \lesssim \eta_{\rm c} \sim 1$, we have that $\alpha \gtrsim -g$, but
$\alpha \cos \theta$ can become commensurate with $-g$ if $\cos \theta$ becomes
small. This is achieved as $\theta$ evolves from an initially retrograde
obliquity through $90^\circ$ towards $0^\circ$ under the influence of the
alignment torque.

While similar in behavior to previous studies of probabilistic resonance capture
\citep{henrard1982, su2020}, the underlying mechanism is different: In these
previous studies, the phase space structure itself evolves and causes the system
to transition among different phase space zones; here in the problem at hand, a
non-Hamiltonian, dissipative perturbation causes the system to transition among
fixed phase space zones. In the following subsection, we present an analytic
calculation to determine the probability distribution of outcomes upon
separatrix encounter. Readers not interested in the technical details can simply
examine the resulting Fig.~\ref{fig:1hist_toy}.

\subsection{Analytical Calculation of Resonance Capture
Probability}\label{ss:analytic_calculation}

\begin{figure}
    \centering
    \includegraphics[width=\colummwidth]{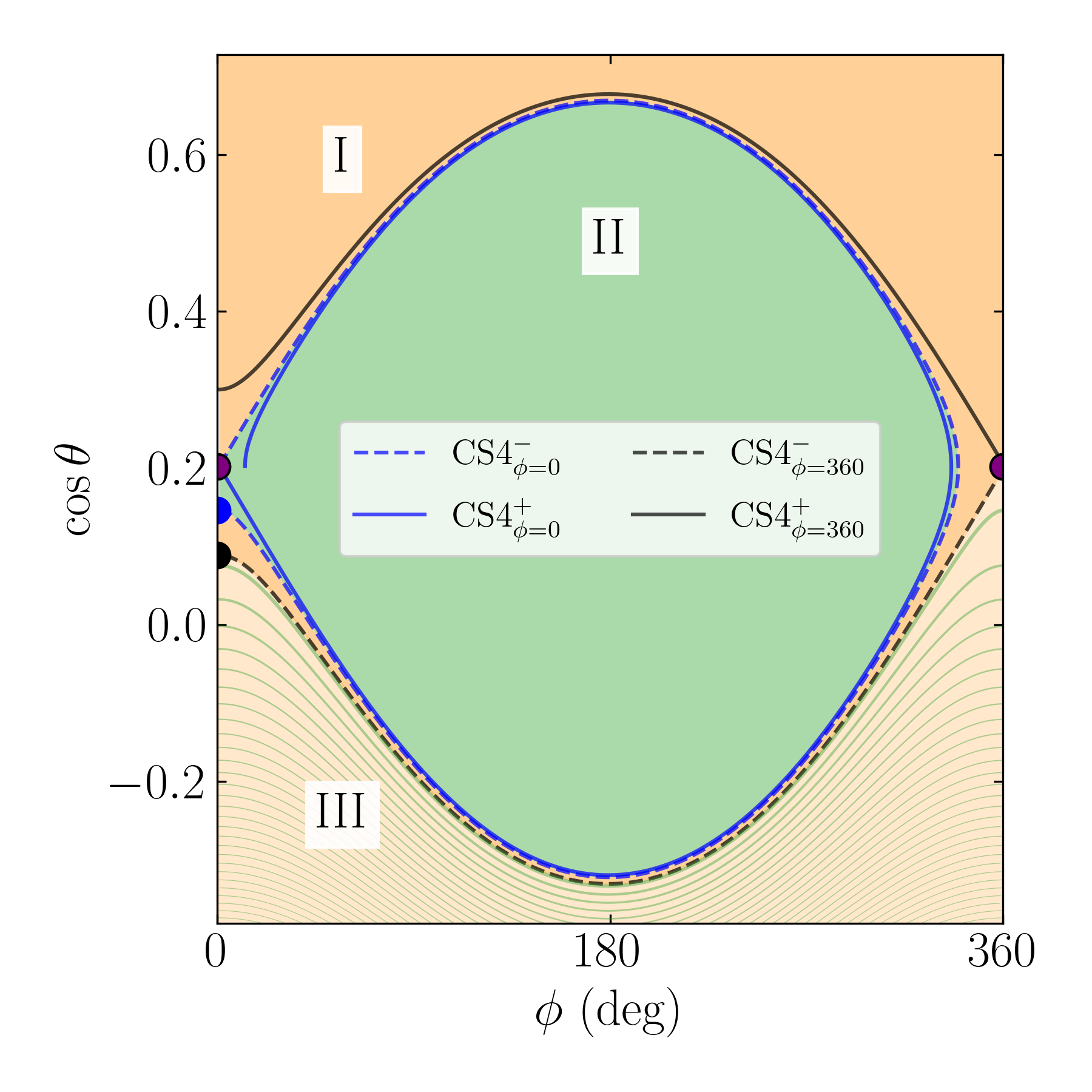}
    \caption{Plot illustrating the probabilistic origin of separatrix
    (resonance) capture for a system with $\eta = 0.2$ and $I = 20^\circ$. The
    orange regions converge to CS1, and the green to CS2. The purple dots denote
    CS4, a saddle point. The boundaries separating the CS1 and CS2-approaching
    regions consist of four critical trajectories (labeled in the legend) that
    are evolved starting from infinitesimal displacements from CS4
    along its stable and unstable eigenvectors going forwards
    and backwards in time: e.g.\ the trajectory labeled CS4$_{\phi=0}^+$ starts
    at $\phi = \epsilon$ (for some small, positive $\epsilon$) and evolves
    forwards in time (with $t_{\rm al} = 10^3 \abs{g}^{-1}$), while the
    trajectory labeled CS4$_{\phi=360}^-$ starts at $\phi = 360^\circ -
    \epsilon$ and evolves backwards in time. The blue and black dots denote the
    intersections of CS4$_{\phi = 0}^-$ and CS4$_{\phi = 360}^-$ critical
    trajectories with the vertical line $\phi = 0$. These critical trajectories
    can be used to understand the probabilistic outcomes that trajectories
    originating in zone III experience upon separatrix encounter, illustrated by
    the tightly spaced orange and green bands in zone III\@; see
    Section~\ref{ss:analytic_calculation} for additional details.
    }\label{fig:toy_hop_manifolds}
\end{figure}

Before discussing our quantitative calculation, we first present a graphical
understanding of the separatrix encounter process.
Figure~\ref{fig:toy_hop_manifolds} shows how the perturbative alignment torque
generates the two outcomes upon separatrix encounter, i.e.\ the zone III to zone
II and the zone III to zone I transition. The critical trajectories in
Fig.~\ref{fig:toy_hop_manifolds} are calculated numerically by integrating from
a point infinitesimally close to CS4 forward and backward in time. In the
absence of the alignment torque, these trajectories would evolve along the
separatrix, but in the presence of the alignment torque, they are perturbed
slightly and cease to overlap. It can be seen in
Fig.~\ref{fig:toy_hop_manifolds} that this splitting opens a path from zone III
into both zones I and II\@: the coloring scheme indicates that the trajectories
within the orange and green regions of phase space stay within their
respective colored regions.

To understand this process more concretely, and to compute the associated
probabilities of the two possible outcomes, we consider the evolution of the
value of the \emph{unperturbed} Hamiltonian (Eq.~\ref{eq:H}) as the spin evolves
due to the alignment torque. A point in zone III evolves such that $H$ is
increasing until $H \approx H_{\rm sep}$, where $H_{\rm sep}$ is the value of
$H$ along the separatrix, given by
\begin{align}
    H_{\rm sep} &\equiv H\p{\cos \theta_{\rm 4}, \phi_{\rm 4}}\nonumber\\
        &\approx g\sin I + \frac{g^2}{2\alpha}\cos^2 I +
            \mathcal{O}\p{\eta^2},\label{eq:def_Hsep}
\end{align}
where
\begin{equation}
    \theta_4 \simeq \pi/2 - \eta \cos I
\end{equation}
(see Section~A.1 of Paper I) and $\phi_4 = 0$ are the coordinates of CS4. As the
system evolves closer to the separatrix, the change in $H$ over each circulation
cycle can be approximated by $\Delta H_-$, the change in $H$ along
$\mathcal{C}_-$ (see Fig.~\ref{fig:1contours}). In general, we define the
quantities $\Delta H_{\pm}$
\begin{equation}
    \Delta H_{\pm} \equiv \oint\limits_{\mathcal{C}_{\pm}}
        \rd{H}{t}\;\mathrm{d}t.\label{eq:def_dHpm}
\end{equation}
Using
\begin{align}
    \rd{H}{t} &=
            \pd{H}{(\cos \theta)}\rd{(\cos \theta)}{t}
            + \pd{H}{\phi}\rd{\phi}{t}\nonumber\\
        &= \p{\rd{(\cos\theta)}{t}}_{\rm tide} \rd{\phi}{t}
\end{align}
and Eq.~\eqref{eq:dqdt_toy}, we find
\begin{align}
    \Delta H_{\pm} &= \mp\frac{1}{t_{\rm al}}
        \int\limits_0^{2\pi} \sin^2\theta\;\mathrm{d}\phi,
        \label{eq:deltaH_pm_toy}
\end{align}
where $\theta = \theta\p{\phi}$ is evolved along $\mathcal{C}_{\pm}$. Thus, if
we evaluate $H$ every time that a trajectory originating in zone III crosses
$\phi = 0$, we see that will initially be $< H_{\rm sep}$ and increase for
each circulation cycle until the system encounters the separatrix. At the
beginning of the separatrix-crossing orbit, the initial value of $H$, denoted
by $H_{\rm i}$, must be greater than $H_{\rm sep} - \Delta H_-$ to encounter the
separatrix on the current orbit. We thus require
\begin{equation}
    H_{\rm i} \in \s{ H_{\rm sep} - \Delta H_-,  H_{\rm sep}}
        \label{eq:Hi_range}.
\end{equation}
The values of $\cos \theta$ corresponding to the lower and upper bounds in this
range are shown as the black and purple dots on the left of
Fig.~\ref{fig:toy_hop_manifolds} respectively.

During the separatrix-crossing orbit, the trajectory first evolves
approximately along $\mathcal{C}_-$ and then along $\mathcal{C}_+$, after which
the final value of $H$, denoted by $H_{\rm f}$, is approximately equal to
\begin{equation}
    H_{\rm f} = H_{\rm i} + \Delta H_+ + \Delta H_-.
\end{equation}
There are two outcomes depending on the value of $H_{\rm f}$:
\begin{itemize}
    \item If $H_{\rm f} < H_{\rm sep}$, then, since $H < H_{\rm
        sep}$ corresponds to the exterior of the separatrix, this implies that
        the trajectory has ended outside of the separatrix. This outcome thus
        corresponds to a zone III to zone I transition. In
        Fig.~\ref{fig:toy_hop_manifolds}, the evolution within the orange shaded
        regions exhibits such an outcome.

    \item If $H_{\rm f} > H_{\rm sep}$, then the trajectory has instead ended
        inside of the separatrix and has executed a zone III to zone II
        transition. This corresponds to evolution within the green shaded
        regions in Fig.~\ref{fig:toy_hop_manifolds}.
\end{itemize}
These two possibilities can be re-expressed in terms of $H_{\rm i}$:
if $H_{\rm i}$ is in the interval $\s{H_{\rm sep} - \Delta H_-, H_{\rm sep} -
\Delta H_{-} - \Delta H_+}$, then the system executes a III $\to$ I transition,
and if it is in the interval $\s{H_{\rm sep} - \Delta H_- - \Delta H_+, H_{\rm
sep}}$, then the system executes a III $\to$ II transition. We see that there is
a critical value of $H_{\rm i}$,
\begin{equation}
    \p{H_{\rm i}}_{\rm crit} = H_{\rm sep} - \Delta H_{-} - \Delta H_+,
        \label{eq:def_Hicrit}
\end{equation}
that separates the two possible outcomes of the separatrix encounter within the
interval given by Eq.~\eqref{eq:Hi_range}. The value of $\cos \theta$ for which
$H$ is equal to $H_{\rm sep} - \Delta H_{-} - \Delta H_+$ is shown as the blue
dot on the left of Fig.~\ref{fig:toy_hop_manifolds}. Finally, if the alignment
torque is weak, then $\abs{\Delta H_{\pm}} \propto t_{\rm al}^{-1}$ is small
compared to any variation in the value of $H$ (e.g.\ when changing the initial
$\phi_{\rm i}$ or $\theta_{\rm i}$ by a small amount), and $H_{\rm i}$ can be
effectively considered as randomly chosen from a uniform distribution over
the range $\s{H_{\rm sep} - \Delta H_-, H_{\rm sep}}$. As a consequence we
obtain the probability of the III $\to$ II transition:
\begin{equation}
    P_{\rm III \to II} = \frac{\Delta H_- + \Delta H_+}{\Delta H_-}.
        \label{eq:def_P32_toy}
\end{equation}

To evaluate Eq.~\eqref{eq:def_P32_toy} analytically, we use the approximate
expression for the separatrix $\eta \ll 1$ (see Eq.~B5 of Paper I)\footnote{
A more exact expression valid for all $\eta \leq \eta_{\rm c}$
can be obtained by using the exact analytical solution to Colombo's Top, see
\citet{ward2004I}. We forgo this approach due to the significant complexity of
the expression involved for a small extension in the regime of validity: our
expression is sufficiently accurate when $\eta \lesssim 0.3$, while $\eta_{\rm c} \lesssim
1$.}:
\begin{equation}
    \p{\cos \theta}_{\mathcal{C}_{\pm}} \approx
        \eta \cos I \pm \sqrt{2\eta\sin I\p{1 - \cos \phi}}.
        \label{eq:sep_theta}
\end{equation}
Using Eq.~\eqref{eq:deltaH_pm_toy}, we find
\begin{align}
    \Delta H_- &\approx \frac{2\pi}{t_{\rm al}}\p{1
        - 2\eta \sin I} + \mathcal{O}(\eta^{3/2}),\\
    \Delta H_+ + \Delta H_- &\approx
        \frac{32 \eta^{3/2}\cos I \sqrt{\sin I}}{t_{\rm al}}
            + \mathcal{O}\p{\eta^{5/2}},
\end{align}
and thus
\begin{align}
    P_{\rm III \to II} &\approx
        \frac{16 \eta^{3/2} \cos I \sqrt{\sin I}}{\pi
            \p{1  - 2\eta \sin I}}.\label{eq:P32_toy}
\end{align}

\begin{figure}
    \centering
    \includegraphics[width=\colummwidth]{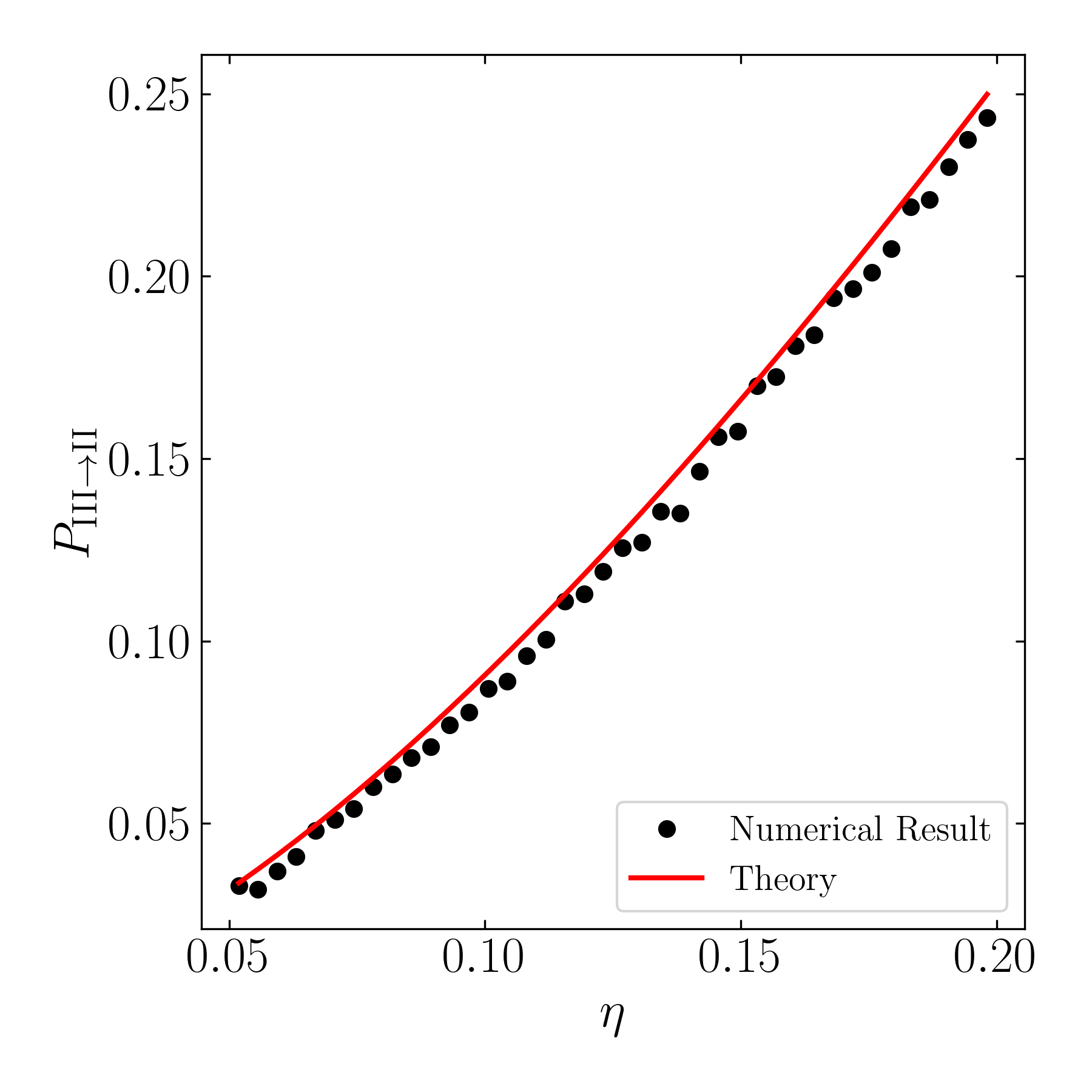}
    \caption{Zone III to Zone II transition probability $P_{\rm III \to II}$
    upon separatrix encounter as a function of $\eta$ driven by an alignment
    torque. For each $\eta$, $1000$ initial random $\p{\theta_{\rm i}, \phi_{\rm
    i}}$ values in zone III are evolved until just after separatrix encounter,
    where the outcome of the encounter is recorded. The red line shows the
    analytical result, Eq.~\eqref{eq:P32_toy}.}\label{fig:1hist_toy}
\end{figure}

To compare Eq.~\eqref{eq:P32_toy} with numerical results, we perform numerical
integrations of Eqs.~(\ref{eq:dqdt_toy}--\ref{eq:dfdt_toy}) while restricting
the initial conditions to those in zone III\@. In Fig.~\ref{fig:1hist_toy}, we
display Eq.~\eqref{eq:P32_toy} alongside the computed $P_{\rm III \to II}$ using
$1000$ initial conditions in zone III for each of $60$ values of $\eta$.
Excellent agreement is observed.

The rigorous connection between the above calculation, focusing
on the evolution of $H$ along the two legs of the separatrix
$\mathcal{C}_{\pm}$, and the graphical picture illustrated in
Fig.~\ref{fig:toy_hop_manifolds} is provided by \emph{Melnikov's Method}
\citep{g_and_h}. Melnikov's Method is a general calculation that gives the
degree of splitting of a ``homoclinic orbit'' (here, the separatrix) of a
Hamiltonian system induced by a small, possibly time-dependent, perturbation.
At a qualitative level, we can state the connection succinctly (see
Fig.~\ref{fig:toy_hop_manifolds}):
\begin{itemize}
    \item The trajectory labeled CS4$_{\phi = 360}^-$ is evolved backwards in
        time from CS4 (where the Hamiltonian has the value $H_{\rm sep}$) along
        $\mathcal{C}_-$, and thus the black dot labels the start of a
        separatrix-crossing orbit with the initial value of the Hamiltonian
        $H_{\rm i} = H_{\rm sep} - H_-$. According to Eq.~\eqref{eq:Hi_range},
        this is exactly the minimum $H_{\rm i}$ such that a trajectory
        experiences a separatrix-crossing orbit. This is consistent with
        Fig.~\ref{fig:toy_hop_manifolds}, where it is clear that any
        trajectories below the black dot at $\phi = 0$ will not experience a
        separatrix encounter on its current circulation cycle.

    \item The trajectory labeled CS4$_{\phi = 0}^-$ is the one evolving
        backwards in time from CS4 along first $\mathcal{C}_+$ then
        $\mathcal{C}_-$, and thus the blue dot labels the start of a
        separatrix-crossing orbit with $H_{\rm i} = H_{\rm sep} - \Delta H_{-} -
        \Delta H_+$. According to Eq.~\eqref{eq:def_Hicrit}, this is exactly the
        critical value of $H_{\rm i}$ that separates trajectories executing a
        III$\to$II transition and a III$\to$I transition. This is also
        consistent with Fig.~\ref{fig:toy_hop_manifolds}, where the region above
        CS4$_{\phi = 0}^-$ is colored green while the region below is colored
        orange.
\end{itemize}

\section{Spin Evolution with Weak Tidal Friction}\label{s:full_tide_prob}

\subsection{Tidal Cassini Equilibria (tCE)}\label{ss:tce}

Having understood the effect of the alignment torque on the spin evolution
(Section~\ref{s:toy_model}), we now implement the full effect of tidal
dissipation, including both tidal alignment and spin synchronization. We use the
weak friction theory of equilibrium tides \citep[e.g.][]{alexander1973weak,
hut1981tidal}. In this model, tides cause both the spin orientation $\uv{s}$ and
frequency $\Omega_{\rm s}$ to evolve on the characteristic tidal timescale
$t_{\rm s}$ following \citep[see][]{lai2012}:
\begin{align}
    \p{\rd{\uv{s}}{t}}_{\rm tide} &= \frac{1}{t_{\rm s}}
                \s{\frac{2n}{\Omega_{\rm s}} - \p{\uv{s} \cdot \uv{l}}}
                    \uv{s} \times \p{\uv{l} \times \uv{s}}\label{eq:dsdt_tide},\\
    \frac{1}{\Omega_{\rm s}}\p{\rd{\Omega_{\rm s}}{t}}_{\rm tide}
        &= \frac{1}{t_{\rm s}} \s{\frac{2n}{\Omega_{\rm s}}\p{\uv{s} \cdot
            \uv{l}} - 1 - \p{\uv{s} \cdot \uv{l}}^2},\label{eq:dWsdt_tide}
\end{align}
where $t_{\rm s}$ is given by
\begin{equation}
    \frac{1}{t_{\rm s}} \equiv \frac{1}{4k}
        \frac{3k_2}{Q}\p{\frac{M_\star}{m}}\p{\frac{R}{a}}^3 n,\label{eq:ts_tide}
\end{equation}
with $k_2$ and $Q$ the tidal Love number\footnote{Note that for rocky planets,
the tidal $k_2$ and the hydrostatic $k_2$ (which is equal to the $3k_{\rm q}$)
need not be equal, e.g.\ for the Earth, $k_2^{\rm tidal} \approx 0.29$
\citep{lainey2016quantification} while the hydrostatic $k_2^{\rm rotational} =
0.94$ \citep{fricke1977joint}. This is due to the Earth's appreciable rigidity.
For higher-mass, more ``fluid'' planets, $k_2^{\rm tidal} \simeq k_2^{\rm
rotational}$.} and tidal quality factor, respectively. We neglect orbital
evolution (thus, $t_{\rm s}$ is a constant) in this section since the time scale
is longer than $t_{\rm s}$ by a factor of $\sim L / S \gg 1$ (we discuss the
effect of orbital evolution in Section~\ref{ss:disc_wasp12b}). We will continue
to consider the case where tidal dissipation is slow, i.e.\ $\abs{g}t_{\rm s}
\gg 1$. The full equations of motion including weak tidal friction can be
written in component form as
\begin{align}
    \rd{\theta}{t} &= g\sin I \sin \phi -
        \frac{1}{t_{\rm s}}\sin \theta\p{\frac{2n}{\Omega_{\rm s}} - \cos \theta}
            ,\label{eq:ds_fullq}\\
    \rd{\phi}{t} &= -\alpha\cos\theta
        - g\p{\cos I + \sin I \cot \theta \cos \phi}\label{eq:ds_fullphi},\\
    \frac{1}{\Omega_{\rm s}}\rd{\Omega_{\rm s}}{t}
        &= \frac{1}{t_{\rm s}} \s{\frac{2n}{\Omega_{\rm s}} \cos \theta
            - \p{1 + \cos^2\theta}}\label{eq:ds_fulls}.
\end{align}

Equation~\eqref{eq:ds_fulls} shows that, at a given obliquity, tides tend to
drive $\Omega_{\rm s}$ towards the pseudo-synchronous equilibrium value, given by
\begin{equation}
     \frac{\Omega_{\rm s}}{n}
        = \frac{2\cos \theta}{1 + \cos^2\theta} \quad
        \p{\dot{\Omega}_{\rm s} = 0}.\label{eq:weaktide_dWszero}
\end{equation}
On the other hand, Eq.~\eqref{eq:ds_fullq} shows that the spin-orbit alignment
timescale $t_{\rm al}$ is related to $t_{\rm s}$ by
\begin{equation}
    t_{\rm al}^{-1} = t_{\rm s}^{-1}\p{\frac{2n}{\Omega_{\rm s}} - \cos \theta}.
        \label{eq:tal_ts_relation}
\end{equation}
Thus, $\dot{\theta}_{\rm tide} < 0$ for $2n / \Omega_{\rm s} > \cos \theta$ and
$\dot{\theta}_{\rm tide} > 0$ for $2n / \Omega_{\rm s} < \cos \theta$.

To understand the long-term evolution of the system, we first consider its
behavior near a CS\@. Specifically, we wish to understand whether initial
conditions near a CS stay near the CS as the evolution of $\Omega_{\rm s}$
causes the CSs (and separatrix) to evolve. We first note that the evolution of
$\Omega_{\rm s}$ alone does not drive $\uv{s}$ towards or away from CSs: As long
as it evolves sufficiently slowly (adiabatically; see Paper I), conservation of
phase space area ensures that trajectories will remain at fixed distances to
stable equilibria of the system. Thus, Eq.~\eqref{eq:dsdt_tide}
or~\eqref{eq:ds_fullq} alone determine whether the system evolves towards or
away from a nearby CS as $\Omega_{\rm s}$ evolves. Then, from
Eq.~\eqref{eq:tal_ts_relation}, we see that CS2 is still always stable (and
attracting), while CS1 is becomes unstable for $\Omega_{\rm s} > 2n \cos
\theta_1 \approx 2n$, where $\theta_1 \approx \eta \sin I$ (Paper I) is the
obliquity of CS1.

\begin{figure*}
    \centering
    \includegraphics[width=\textwidth]{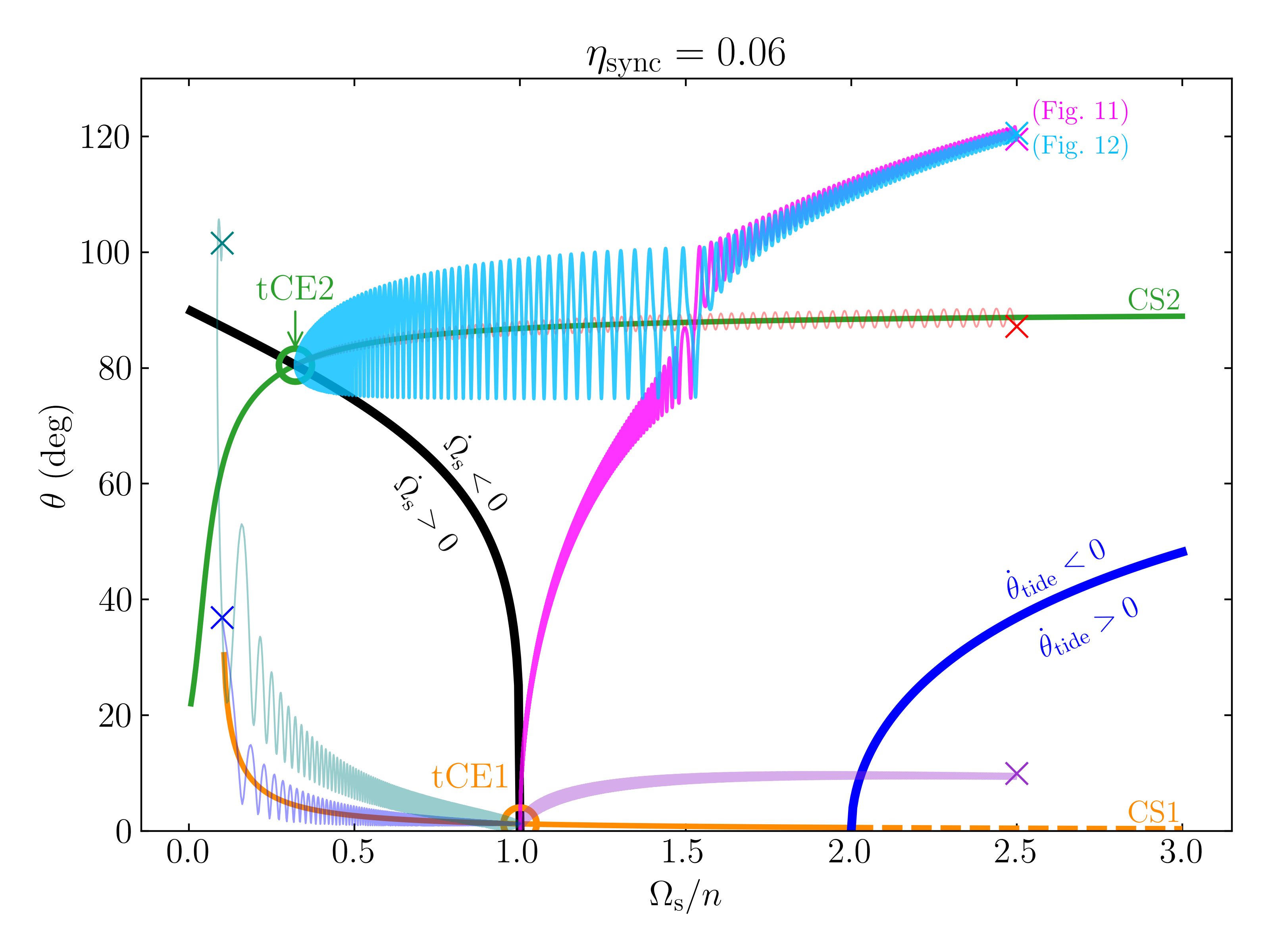}
    \caption{Schematic depiction of the effect of tidal friction on the planet's
    spin evolution in the $\theta$-$\Omega_{\rm s}$ plane for a system with $I =
    20^\circ$ (corresponding to $\eta_{\rm c} = 0.574$; see
    Eq.~\ref{eq:def_etac}) and $\eta_{\rm sync} = 0.06$ (see
    Eq.~\ref{eq:def_etasync}). The black and blue lines denote where the tidal
    $\dot{\Omega}_{\rm s}$ and $\dot{\theta}$ change signs
    (see Eqs.~\ref{eq:ds_fullq} and~\ref{eq:weaktide_dWszero}). The orange and
    green lines give the CS1 and CS2 obliquities respectively (these are the two
    CSs that can be stable in the presence of tidal dissipation). Note that when
    $\dot{\theta}_{\rm tide} > 0$, CS1 becomes unstable, denoted by the dashed
    orange line. The points that lie on CSs and satisfy $\dot{\Omega}_{\rm s}
    = 0$ are called tidal Cassini Equilibria (tCE), which are circled and labeled.
    The various colored crosses and their associated colored lines represent a
    few characteristic examples of the spin evolution under weak
    tidal friction (for illustrative purposes, we have used $\abs{g}t_{\rm s} =
    10^2$ and evolved each example for $5t_{\rm s}$). The phase space evolution
    of the two thicker evolutionary trajectories (cyan and pink; those beginning
    at $\theta_{\rm i} = 120^\circ$) are shown in
    Figs.~\ref{fig:trajs1}--\ref{fig:trajs2}.
    }\label{fig:6equils006}
\end{figure*}

With this consideration, we can identify the long-term equilibria of the
system when tidal torques drive the evolution of both the obliquity and
$\Omega_{\rm s}$ (and thus $\eta$): these equilibria must satisfy
$\dot{\Omega}_{\rm s} = 0$ and be a CS that is stable in the presence of the
tidal torque (i.e.\ satisfying $\rdil{\uv{s}}{t} = 0$); we call such long-term
equilibria \emph{tidal Cassini Equilibria} (tCE). Figure~\ref{fig:6equils006}
depicts the evolution of the system following
Eqs.~(\ref{eq:dsdt_tide}--\ref{eq:dWsdt_tide}) in $(\Omega_{\rm s}, \theta)$
space starting from several representative initial conditions, along with the
locations of CS1 and CS2. The circled points in Fig.~\ref{fig:6equils006} denote
the two tCEs (tCE1 and tCE2, depending on whether it lies on CS1 or CS2).

\begin{figure*}
    \centering
    \includegraphics[width=\textwidth]{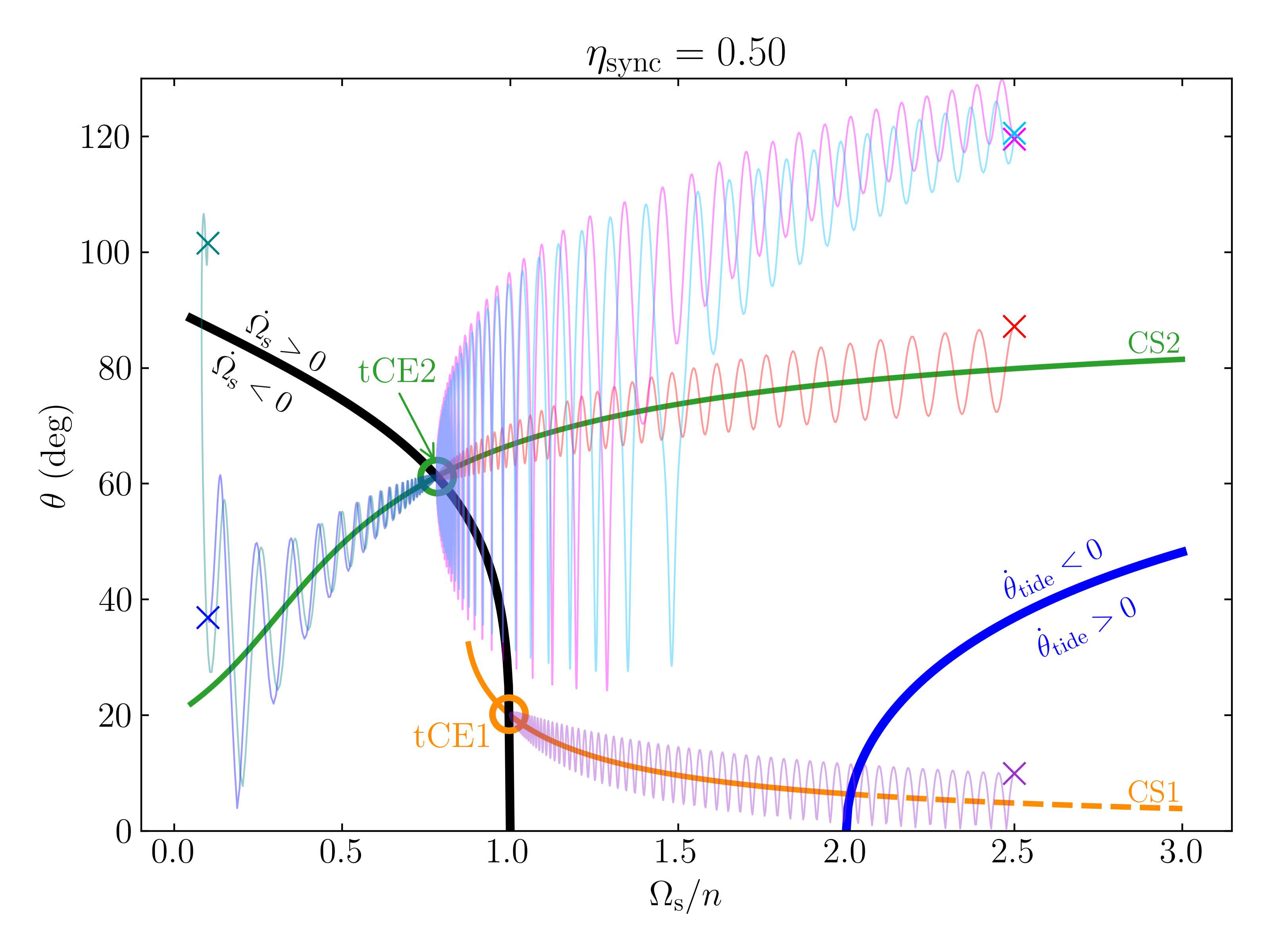}
    \caption{Same as Fig.~\ref{fig:6equils006} but for $\eta_{\rm sync} = 0.5$.
    The crosses and lines correspond to evolutionary trajectories using the same
    initial conditions as those shown in
    Fig.~\ref{fig:6equils006}.}\label{fig:6equils050}
\end{figure*}

\begin{figure*}
    \centering
    \includegraphics[width=\textwidth]{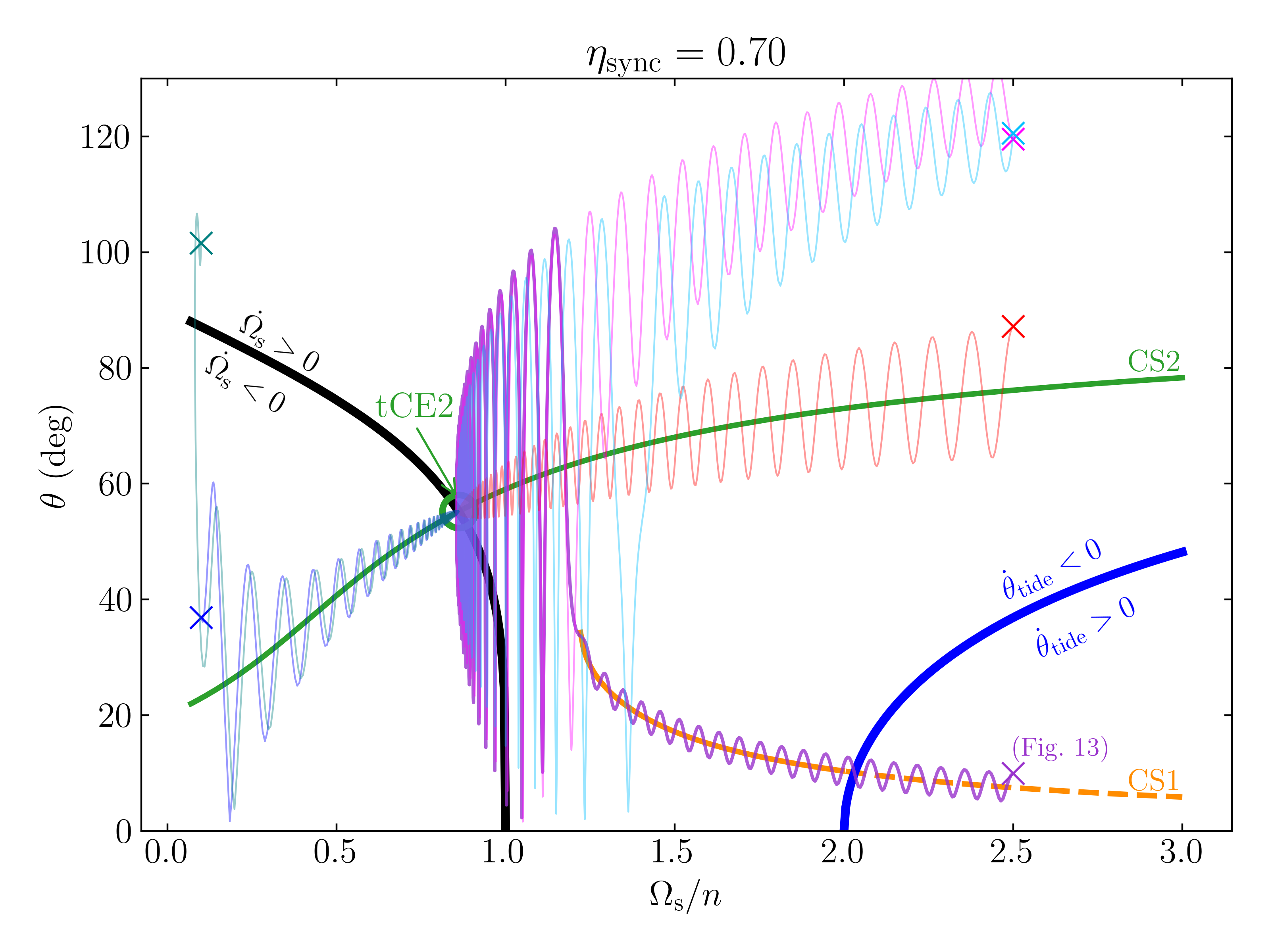}
    \caption{Same as Figs.~\ref{fig:6equils006} but for $\eta_{\rm sync} = 0.7$.
    Note that $\eta_{\rm sync} = 0.7 > \eta_{\rm c} = 0.574$ and tCE1 does
    not exist. The phase space evolution of the thick purple trajectory
    (starting at $\theta_{\rm i} = 10^\circ$) is shown in Fig.~\ref{fig:trajs3}.
    }\label{fig:6equils070}
\end{figure*}

The obliquities of the tCE and the evolutionary track in the
$\theta$-$\Omega_{\rm s}$ plane depend on the parameter
\begin{align}
    \eta_{\rm sync} &\equiv \big(\eta\big)_{\Omega_{\rm s} = n}
        = \eta \frac{\Omega_{\rm s}}{n}\nonumber\\
        &= \frac{k}{2k_{\rm q}}
            \frac{m_{\rm p}m}{M_\star^2}
            \p{\frac{a}{a_{\rm p}}}^3 \p{\frac{a}{R}}^3\cos I.
            \label{eq:def_etasync}
\end{align}
In Fig.~\ref{fig:6equils006}, $\eta_{\rm sync} = 0.06$;
Figs.~\ref{fig:6equils050}--\ref{fig:6equils070} illustrate the cases with
$\eta_{\rm sync} = 0.5$ and $0.7$ respectively.

\begin{figure}
    \centering
    \includegraphics[width=\colummwidth]{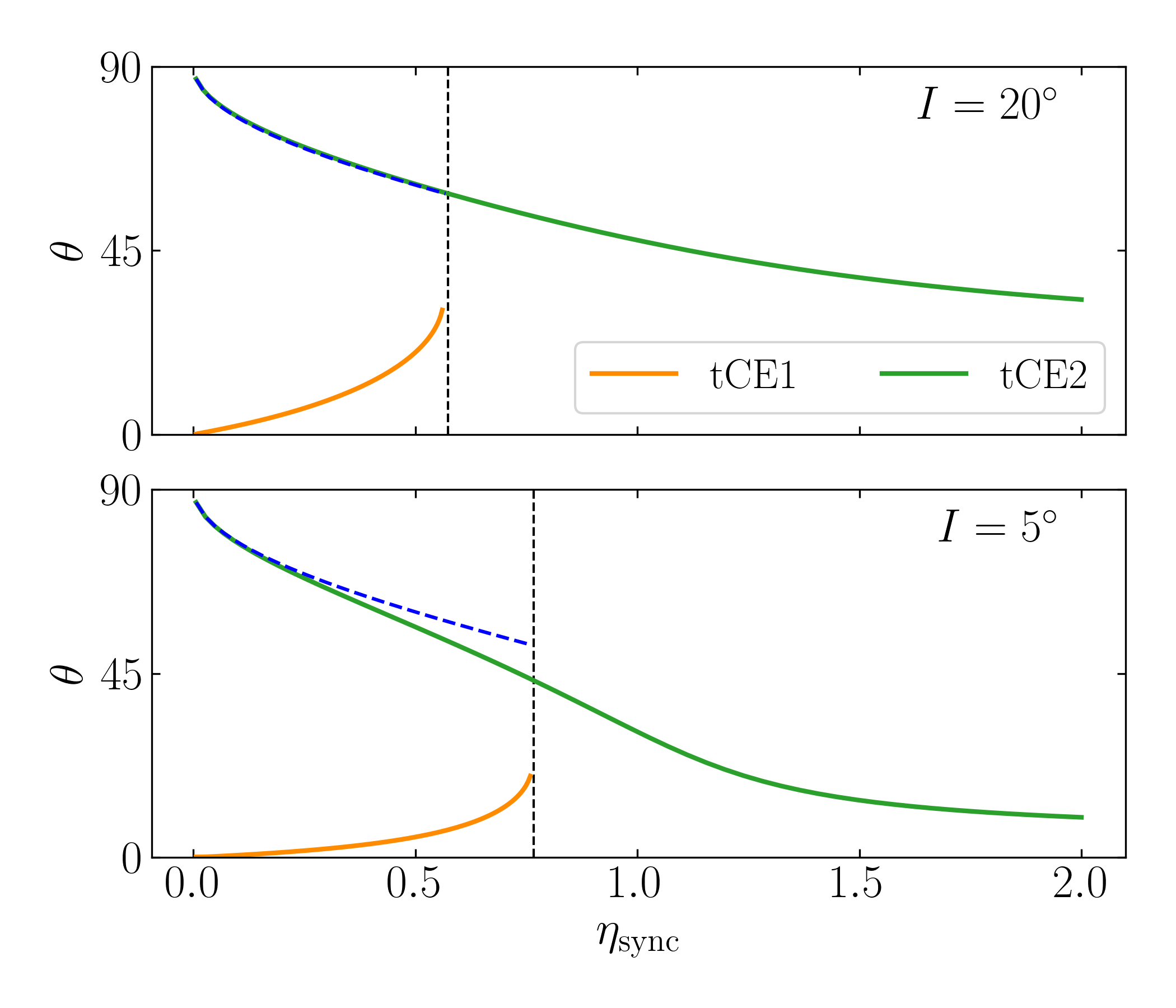}
    \caption{Obliquities of the two tCE as a function of $\eta_{\rm sync}$
    (defined in Eq.~\ref{eq:def_etasync}) for $I = 20^\circ$ (top) and $I =
    5^\circ$ (bottom). The blue dashed lines denote the analytical approximation
    given by Eq.~\eqref{eq:def_tce2_qapprox} and is only valid for $\eta_{\rm
    sync} \ll 1$. The vertical dashed lines denote where $\eta_{\rm sync} =
    \eta_{\rm c}(I)$, above which tCE1 ceases to exist.}\label{fig:tCEs}
\end{figure}

The tCE obliquities as a function of $\eta_{\rm sync}$ are shown in
Fig.~\ref{fig:tCEs} for $I = 20^\circ$ and $I = 5^\circ$. In fact, an analytical
expression for the tCE2 obliquity and rotation rate for $\eta_{\rm sync} \ll 1$
can be obtained using Eqs.~\eqref{eq:weaktide_dWszero}--\eqref{eq:def_etasync}
and $\cos \theta_2 \simeq \eta \cos I$ (valid for $\eta \ll 1$; see Appendix of
Paper I):
\begin{align}
    \cos \theta_{\rm tCE2} &\simeq \sqrt{\frac{\eta_{\rm sync} \cos I}{2}},
        \label{eq:def_tce2_qapprox}\\
    \frac{\Omega_{\rm s, tCE2}}{n} &\simeq
        \sqrt{2 \eta_{\rm sync}\cos I}. \label{eq:def_tce2_approx}
\end{align}
This approximation for $\theta_{\rm tCE2}$ is shown as the blue dashed line in
Fig.~\ref{fig:tCEs}, indicating good agreement with the numerical result
obtained via root finding of Eqs.~\eqref{eq:ds_fullq}--\eqref{eq:ds_fulls} while
assuming $\abs{g}t_{\rm s} \gg 1$.
\begin{figure}
    \centering
    \includegraphics[width=\colummwidth]{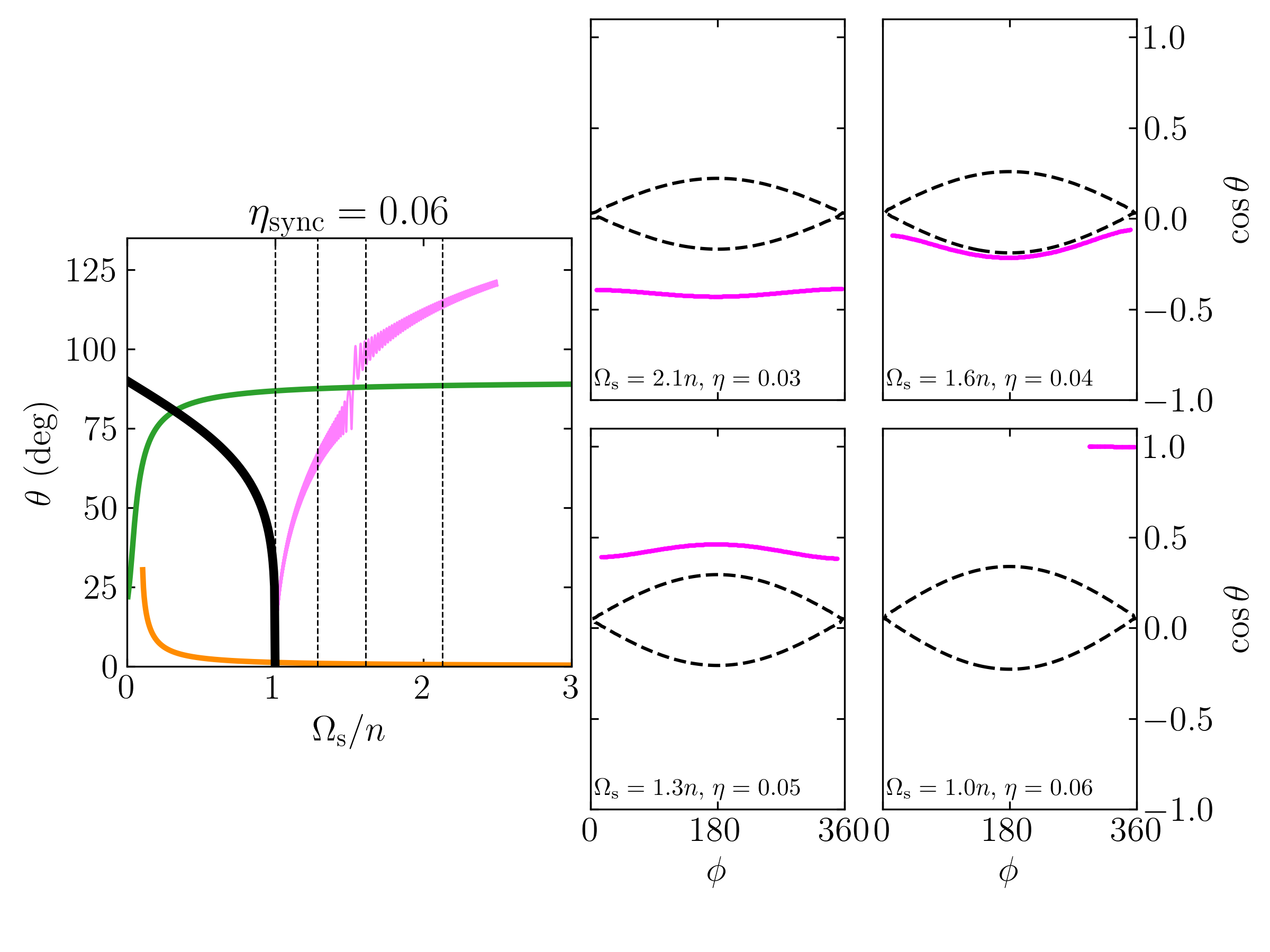}
    \caption{Phase space evolution of the pink trajectory in
    Fig.~\ref{fig:6equils006}, for which $\eta_{\rm sync} = 0.06$ and $I =
    20^\circ$ (corresponding to $\eta_{\rm c} = 0.574$). The initial conditions
    are $\Omega_{\rm s, i} = 2.5n$, $\theta_{\rm i} = 120^\circ$, and $\phi_{\rm
    i} = 0^\circ$, and we have used $\abs{g}t_{\rm s} = 10^2$ and have evolved
    the system for $5t_{\rm s}$. In the left-most panel, the trajectory's
    evolution in the $\theta$-$\Omega_{\rm s}$ plane along with the curves
    indicating CS1, CS2, and $\dot{\Omega}_{\rm s} = 0$ are re-displayed from
    Fig.~\ref{fig:6equils006}. The vertical dashed lines denote the values of
    $\Omega_{\rm s} / n$ for which a few phase space snapshots of the system are
    displayed in the right four panels. In each of these right four panels, the
    trajectory's evolution for a single circulation/libration cycle is displayed
    in the $\cos \theta$-$\phi$ plane for the labeled value of $\Omega_{\rm
    s}$ (and the corresponding value of $\eta$). The system encounters the
    separatrix, undergoes a III $\to$ I transition, and converges to tCE1.
    }\label{fig:trajs1}
\end{figure}
\begin{figure}
    \includegraphics[width=\colummwidth]{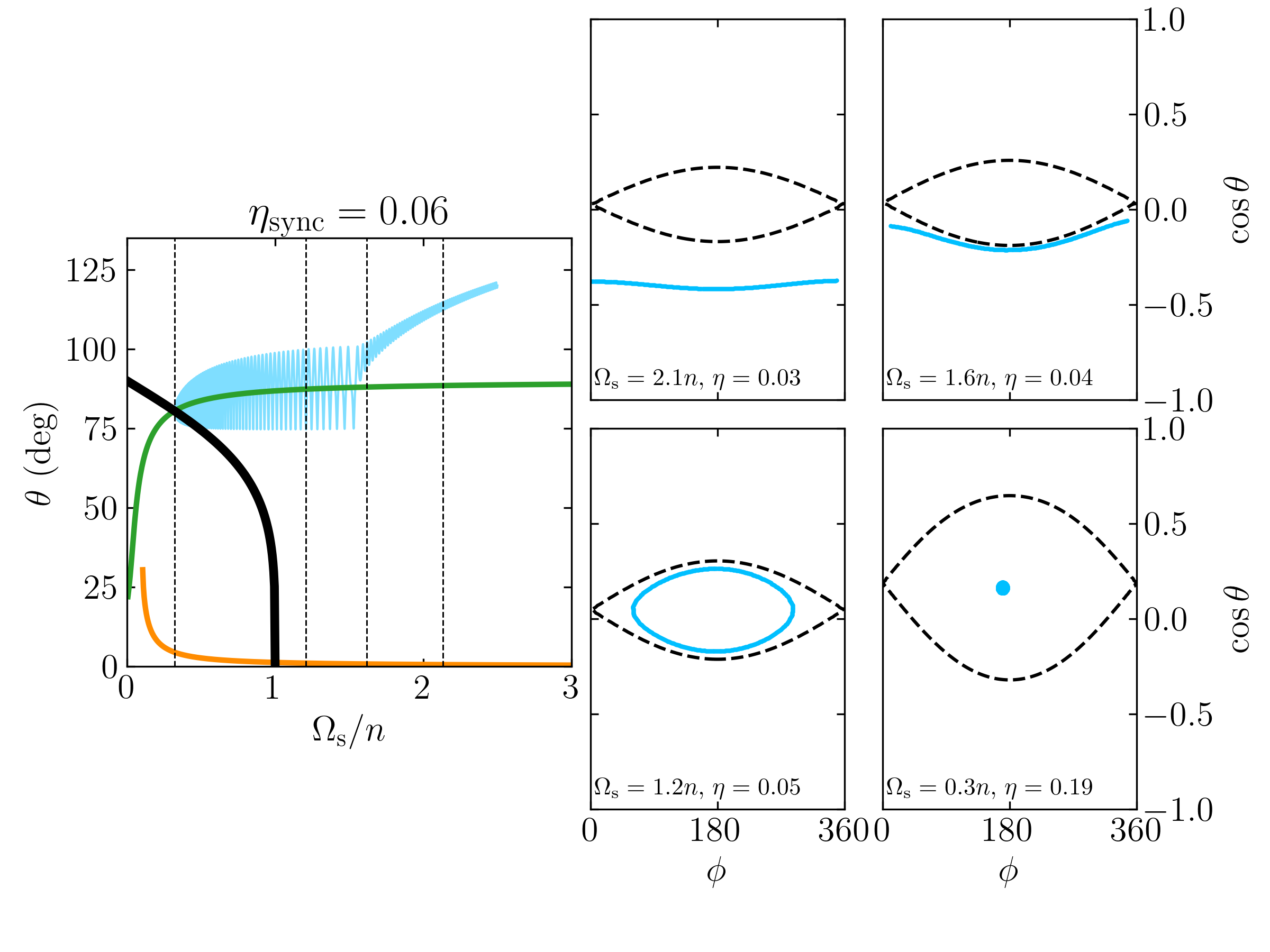}
    \caption{Same as Fig.~\ref{fig:trajs1} but for $\phi_{\rm i} = 286^\circ$,
    corresponding to the cyan trajectory in Fig.~\ref{fig:6equils006}. The
    system encounters the separatrix, undergoes a III $\to$ II transition, and
    converges to tCE2. The small $\phi$ offset of tCE2 from $180^\circ$ arises
    from the alignment torque (see Fig.~\ref{fig:mcs}). Note that the initial
    condition of this trajectory and that displayed in Fig.~\ref{fig:trajs1}
    have the same initial $\theta_{\rm i}$ and $\Omega_{\rm s, i}$ but different
    precessional phases $\phi_{\rm i}$.
    }\label{fig:trajs2}
\end{figure}
\begin{figure}
    \includegraphics[width=\colummwidth]{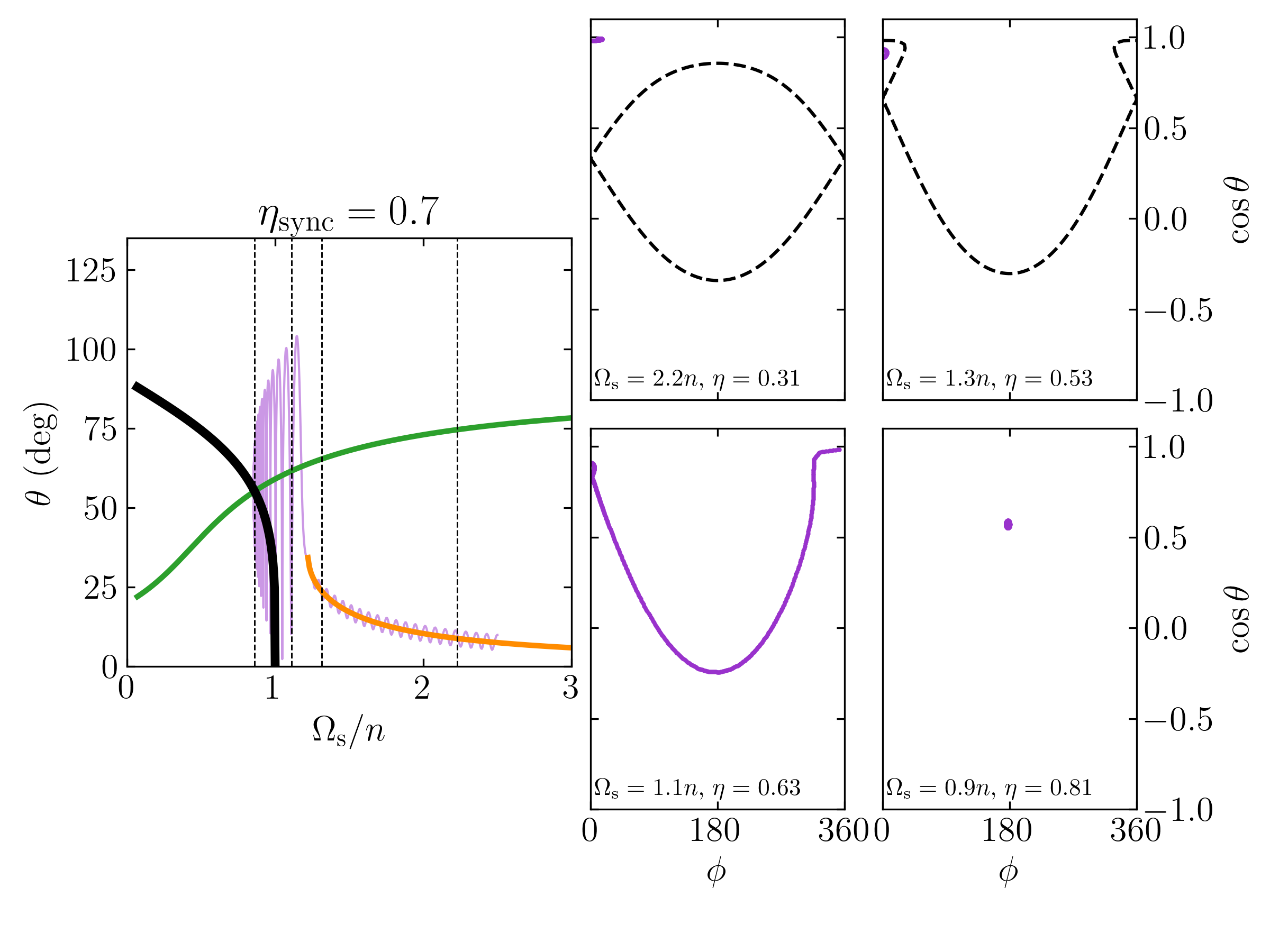}
    \caption{Same as Fig.~\ref{fig:trajs1} but for $\eta_{\rm sync} = 0.7$ and
    $\theta_{\rm i} = 10^\circ$, corresponding to the purple trajectory shown in
    Fig.~\ref{fig:6equils070}. Here, the system evolves along CS1 until the
    separatrix disappears, upon which it experiences large obliquity variations
    that damp due to tidal dissipation. The highly asymmetric shape in the third
    panel arises due to the strong tidal dissipation used in this simulation
    ($\abs{g}t_{\rm s} = 10^2$). The system finally converges to tCE2, the only
    tCE that exists for this value of $\eta_{\rm sync}$.}\label{fig:trajs3}
\end{figure}

There are two important conditions that can influence the existence and stability
of the tCE\@. First, if $\eta_{\rm sync} > \eta_{\rm c}$ (where $\eta_{\rm c}$ is
given by Eq.~\ref{eq:def_etac}), then tCE1 will not
exist (Fig.~\ref{fig:6equils070} gives an example)\footnote{Strictly speaking,
$\eta_{\rm sync}$ can be slightly smaller than $\eta_{\rm c}$, as the planet's
spin is slightly subsynchronous at tCE1 (see Eq.~\ref{eq:weaktide_dWszero}).}.
Second, tCE2 may not be stable if the phase shift due to the alignment torque
(see Section~\ref{ss:mcs}) is too large. Applying the results of
Section~\ref{ss:linear_stab} (see
Eqs.~\ref{eq:mcs_shift}--\ref{eq:mcs_shift_crit}), we find that tCE2 is stable
as long as $t_{\rm s} \geq t_{\rm s, c}$ where
\begin{equation}
    t_{\rm s, c} \equiv \frac{\sin \theta_{\rm tCE2}}{\abs{g} \sin
            I}\p{\frac{2n}{\Omega_{\rm s, tCE2}} - \cos \theta_{\rm tCE2}}.
\end{equation}
When $\eta_{\rm sync} \ll 1$, we can use
Eqs.~(\ref{eq:def_tce2_qapprox}--\ref{eq:def_tce2_approx}) to further simplify
$t_{\rm s, c}$ to\footnote{Note that Eq.~\eqref{eq:def_ts_crit} for the critical
$t_{\rm s}$ agrees with Eq.~(16) of \citet{levrard2007}.}
\begin{equation}
    t_{\rm s, c} \simeq \frac{\tan \theta_{\rm tCE2}}{\abs{g}\sin I} \approx
        \frac{1}{\abs{g}\sin I}\sqrt{\frac{2}{\eta_{\rm sync}
        \cos I}}. \label{eq:def_ts_crit}
\end{equation}

\subsection{Spin and Obliquity Evolution as a Function of Initial Spin
Orientation}\label{ss:weaktide_evolution}

We can now study the final fate of the planet's spin as a function of the
initial condition.
We begin by examining the example trajectories shown in
Figs.~\ref{fig:6equils006}, for which we have integrated
the equations of motion (combining Eqs.~\ref{eq:dsdt_rot}
and~\ref{eq:dsdt_tide} to give $\rdil{\uv{s}}{t}$, and Eq.~\ref{eq:dWsdt_tide})
and set $I = 20^\circ$, $t_{\rm s} = 100 \abs{g}^{-1}$. We discuss each of the
six trajectories in turn:
\begin{itemize}
    \item The trajectory with the initial condition $\Omega_{\rm s, i} = 2.5n$
        and $\theta_{\rm i} = 10^\circ$ (purple) has an initially prograde spin
        (i.e.\ in zone I, see Fig.~\ref{fig:1contours}) and directly evolves to
        tCE1, with the final $\Omega_{\rm s} / n \simeq 1$ and $\theta =
        \theta_{\rm CS1} \simeq \eta_{\rm sync} \sin I$ (for $\eta_{\rm sync}
        \ll 1$; see Appendix~A of paper I).

    \item The trajectory with $\Omega_{\rm s, i} = 2.5n$ and
        $\theta_{\rm i} = 90^\circ$ (red) has an initial condition inside the
        resonance / separatrix (zone II) and evolves to tCE2. Note that the
        obliquity is trapped in a high value due to the stability of CS2 under
        the alignment torque, as shown in Section~\ref{ss:tce}.

    \item We have chosen two trajectories, both with the initial condition
        $\Omega_{\rm s, i} = 2.5n$ and $\theta_{\rm i} = 120^\circ$, but with
        different initial precessional phases $\phi_{\rm i}$. Consider first
        the pink trajectory, for which $\phi_{\rm i} = 0^\circ$. It originates
        in zone III, evolves towards the separatrix as tidal friction damps the
        obliquity, and crosses the resonance (separatrix) without being
        captured, upon which the obliquity continues to damp until the system
        converges to tCE1. The detailed phase space evolution of this trajectory
        is shown in Fig.~\ref{fig:trajs1}, where the outcome of the separatrix
        encounter is very visible.

    \item The light blue trajectory also has $\Omega_{\rm s, i} =
        2.5n$ and $\theta_{\rm i} = 120^\circ$ (like the pink trajectory) but
        with the initial precessional phase $\phi_{\rm i} \approx 286^\circ$.
        It also originates in zone III, encounters the separatrix but is
        captured into the resonance (zone II), upon which tidal friction drives
        the system towards tCE2. The detailed phase space evolution of this
        trajectory is shown in Fig.~\ref{fig:trajs2}, where the resonance
        capture is displayed. Also visible in the final panel of
        Fig.~\ref{fig:trajs2} is the slight phase offset of CS2, i.e.\
        $\phi_{\rm cs} < 180^\circ$, in agreement with the result of
        Section~\ref{ss:mcs} (see Fig.~\ref{fig:mcs}).

    \item For completeness, we also examine some trajectories for initially
        subsynchronous spin rates. The trajectory with $\Omega_{\rm s, i} =
        0.1n$ and $\theta_{\rm i} = 35^\circ$ (blue) has its obliquity rapidly
        damped to zero by tidal friction as it spins up to spin-orbit
        synchronization, eventually converging to tCE1. A subtlety of initially
        subsynchronous spins can be seen here: since the initial $\eta_{\rm i} =
        0.6 > \eta_{\rm c}$ ($= 0.574$), the separatrix and CS1 do not exist
        initially. As such, naively, one expects initial convergence to CS2 and
        subsequent obliquity evolution along CS2 as the spin increases. However,
        due to the strong tidal dissipation adapted in the calculation and the
        proximity of $\eta_{\rm i}$ to $\eta_{\rm c}$, CS1 appears within a
        single circulation cycle, and the obliquity quickly damps to, and
        continues to evolve along CS1.

    \item The trajectory with $\Omega_{\rm s, i} = 0.1n$ and $\theta_{\rm i} =
        100^\circ$ (teal) also has its obliquity damped toward tCE1 as it
        approaches spin-orbit synchronization. We note that if we adopt $t_{\rm
        s} = 10^3\abs{g}^{-1}$, the same initial condition will converge to
        tCE2, agreeing with the intuitive analysis given in the previous
        paragraph.
\end{itemize}
In Figs.~\ref{fig:6equils050}--\ref{fig:6equils070} we show, for each of the six
initial conditions, the evolutionary trajectories for the $\eta_{\rm sync} =
0.5$ and $\eta_{\rm sync} = 0.7$ cases. The qualitative behaviors of these six
examples change in several important ways, so we will discuss a few points of
interest:
\begin{itemize}
    \item For both $\eta_{\rm sync} = 0.5$ and $\eta_{\rm sync} = 0.7$, we see
        that the initial conditions with $\theta_{\rm i} = 120^\circ$
        ($\phi_{\rm i} = 0$, pink; and $\phi_{\rm i} = 286^\circ$, blue)
        converge to tCE2. In fact, for these values of $\eta_{\rm sync}$, all
        initial conditions with $\theta_{\rm i} = 120^\circ$ will converge to
        tCE2 regardless of $\phi_{\rm i}$.

    \item The two subsynchronous initial conditions evolve to tCE2 for both
        $\eta_{\rm sync} = 0.5$ and $\eta_{\rm sync} = 0.7$, as in both cases
        $\eta_{\rm i} \gg \eta_{\rm c}$ and CS2 is the only low-obliquity spin
        equilibrium. The system then continues to evolve along CS2 toward tCE2.

    \item Of particular interest is the trajectory starting from the initial
        condition $\Omega_{\rm s, i} = 2.5n$ and $\theta_{\rm i} = 10^\circ$
        (purple) in the case of $\eta_{\rm sync} = 0.7$. Figure~\ref{fig:trajs3}
        shows the detailed phase space evolution of this trajectory, where it
        can be seen that the system initially evolves along the stable CS1, but
        is ejected when $\Omega_{\rm s}$ becomes sufficiently small that CS1 ceases
        to exist, upon which large obliquity variations eventually lead to
        convergence to tCE2, the only tCE that exists.
\end{itemize}
From the above examples, we see that the spin evolution driven by tides can be
complex and varies greatly depending on the various system parameters and
initial conditions. In the case where the initial spin is subsynchronous, the
detailed outcome depends sensitively on the initial value of $\eta$ and the
tidal dissipation rate. In the following, we restrict our discussion to the more
astrophysically common regime of $\Omega_{\rm s, i} \gg n$, and we adopt the
fiducial value $\Omega_{\rm s, i} = 10n$.

\begin{figure}
    \centering
    \includegraphics[width=\colummwidth]{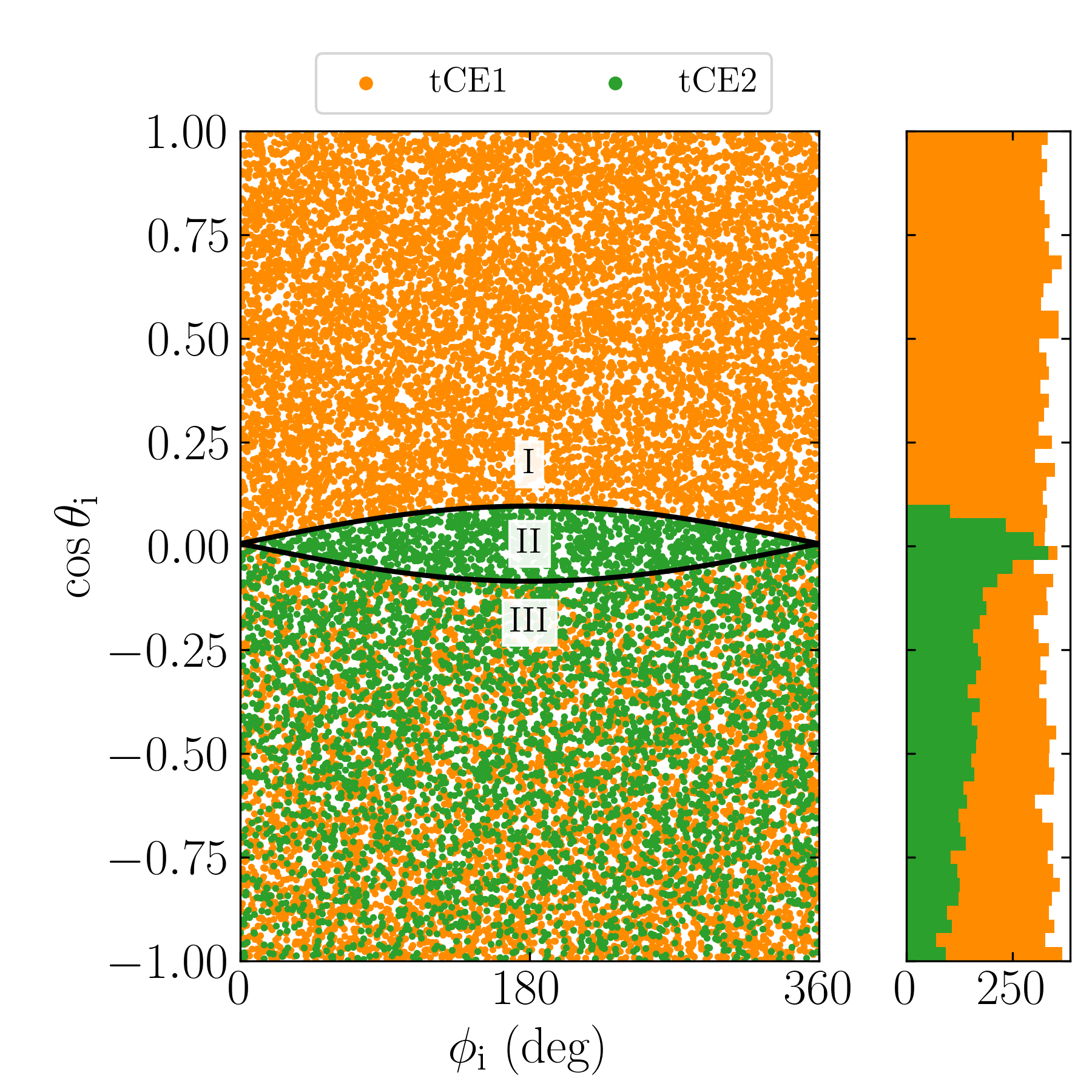}
    \caption{\emph{Left:} Asymptotic outcomes of spin evolution in the presence
    of weak tidal friction for different initial spin orientations ($\theta_{\rm
    i}$ and $\phi_{\rm i}$) for a system with $\eta_{\rm sync} = 0.06$ and $I =
    20^\circ$. Each dot represents an initial spin orientation, and the coloring
    of the dot indicates which tCE (legend) the system evolves into. Similarly
    to Fig.~\ref{fig:toy_phop} initial conditions in Zone I evolve into CS1,
    those in Zone II evolve into CS2, and those in Zone III have a probabilistic
    outcome. \emph{Right:} Histogram of the final tCE that a given initial
    obliquity $\theta_{\rm i}$ evolves into, averaged over $\phi_{\rm i}$.
    }\label{fig:Hhists_0_06}
\end{figure}

\begin{figure}
    \centering
    \includegraphics[width=\colummwidth]{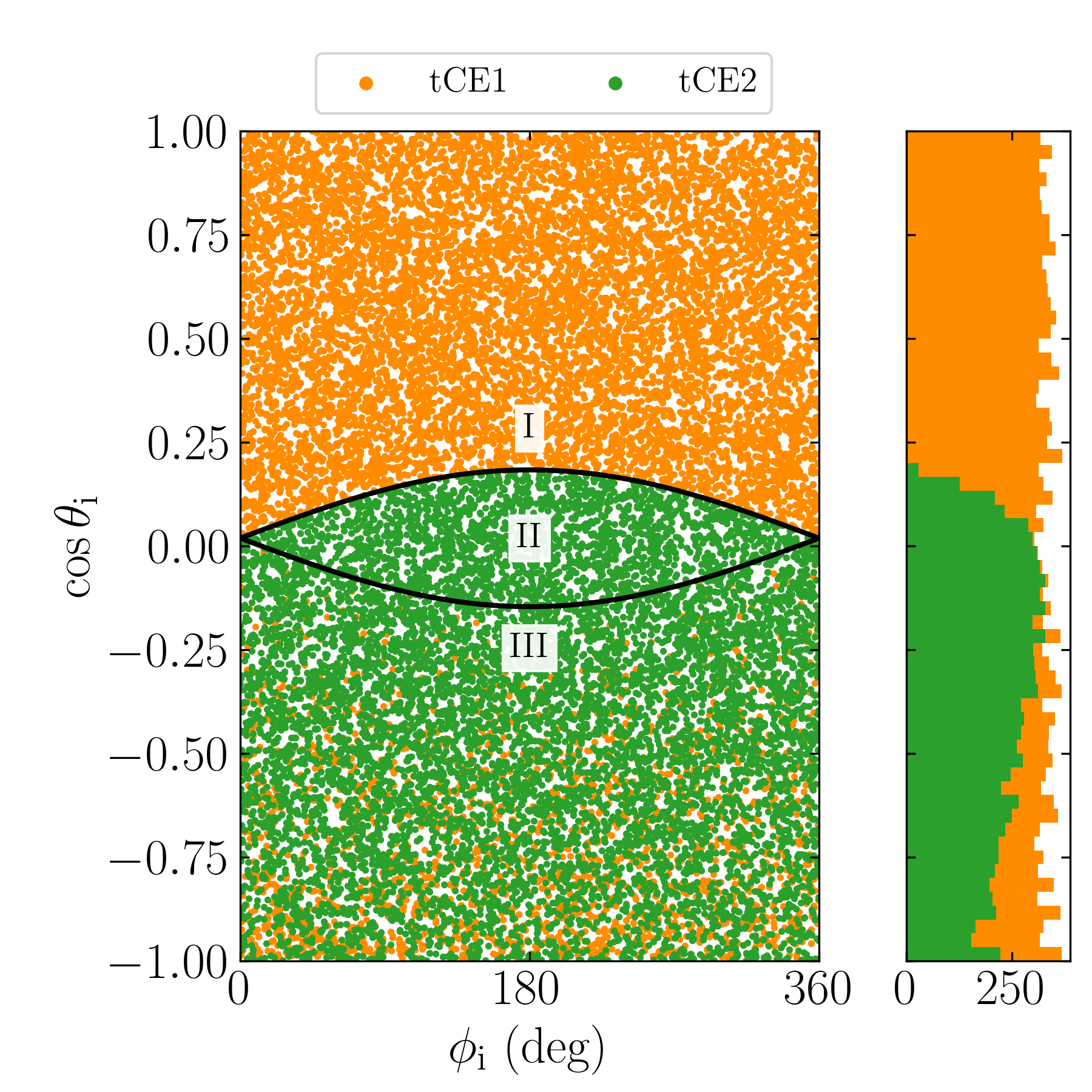}
    \caption{Same as Fig.~\ref{fig:Hhists_0_06} but for $\eta_{\rm sync} =
    0.2$. }\label{fig:Hhists_0_20}
\end{figure}

\begin{figure}
    \centering
    \includegraphics[width=\colummwidth]{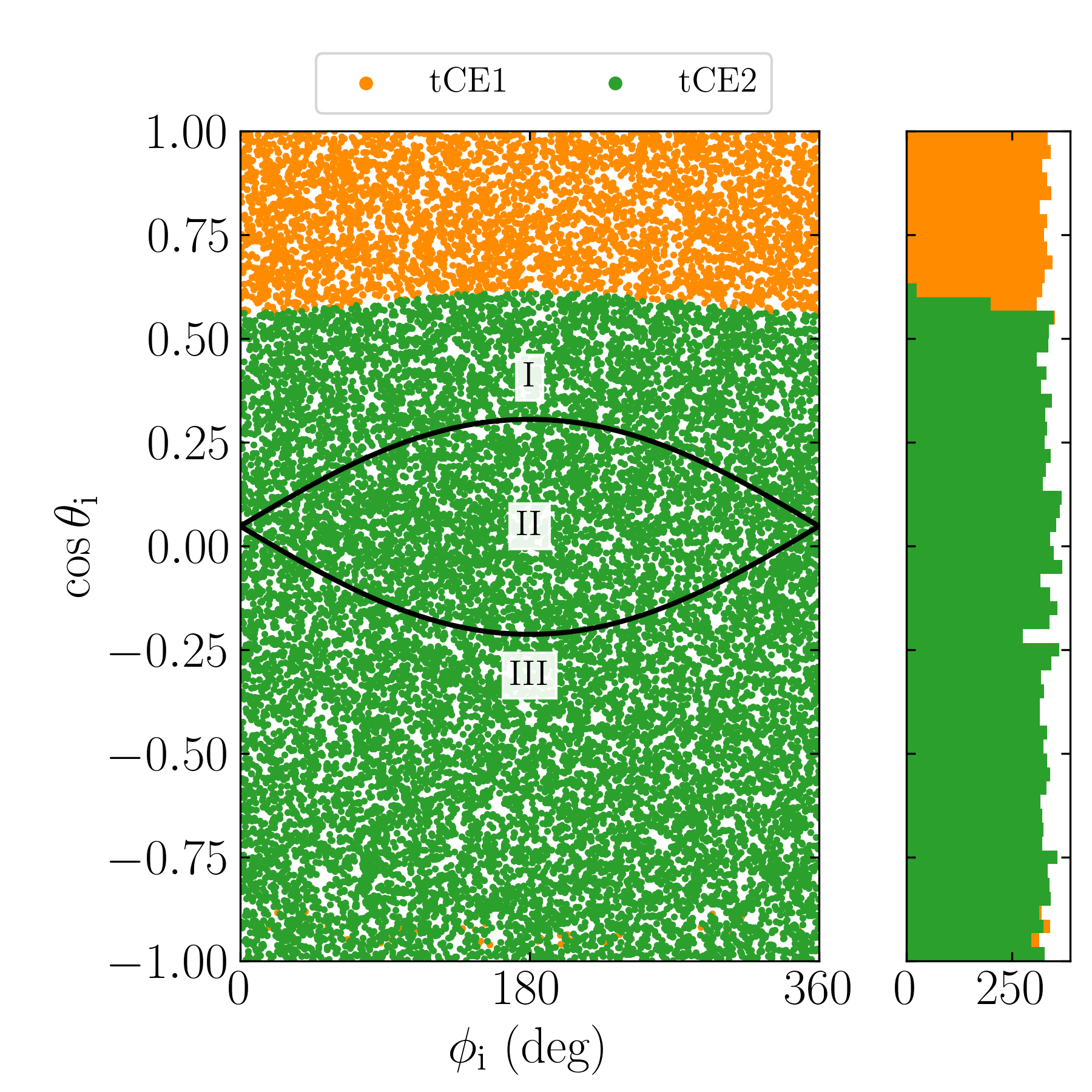}
    \caption{Same as Fig.~\ref{fig:Hhists_0_06} but for $\eta_{\rm sync} =
    0.5$. Note that even points above the separatrix can evolve towards tCE2
    here.}\label{fig:Hhists_0_50}
\end{figure}

Having developed an intuition for a few different possible evolutionary
trajectories, we can attempt to draw general conclusions about the final fate of
the planet's spin as a function of its initial conditions. We do this by again
integrating Eqs.~(\ref{eq:dsdt_rot},~\ref{eq:dsdt_tide}--\ref{eq:dWsdt_tide})
for many initial $\theta_{\rm i}$ and $\phi_{\rm i}$ and examining the final
outcomes. In contrast to the examples shown in
Figs.~\ref{fig:6equils006}--\ref{fig:trajs3}, we use a more gradual tidal
dissipation rate of $\abs{gt_{\rm s}} = 10^3$. In Fig.~\ref{fig:Hhists_0_06}, we
show the final outcome for many randomly chosen $\theta_{\rm i}$ and $\phi_{\rm
i}$ for $\eta_{\rm sync} = 0.06$ and $I = 20^\circ$. We see that the behaviors
seen in the example trajectories of Fig.~\ref{fig:6equils006} are general: tCE1
is generally reached for spins initially in zone I (like the purple trajectory
in Fig.~\ref{fig:6equils006}), tCE2 is generally reached for spins initially in
zone II (like the red trajectory in Fig.~\ref{fig:6equils006}), and a
probabilistic outcome is observed for spins initially in zone III (like the
light blue and pink trajectories in Fig.~\ref{fig:6equils006}).
Figures~\ref{fig:Hhists_0_20} and~\ref{fig:Hhists_0_50} show similar results for
$\eta_{\rm sync} = 0.2$ and $\eta_{\rm sync} = 0.5$. As $\eta_{\rm sync}$ is
increased, more initial conditions reach tCE2. This is both because there are
more systems initially in zone II and because systems initially in zone III have
a higher probability of executing a III $\to$ II transition upon separatrix
encounter. Note also that in Fig.~\ref{fig:Hhists_0_50}, even initial conditions
in zone I are able to reach tCE2; we comment on the origin of this behavior in
the next section.

\subsection{Semi-analytical Calculation of Resonance Capture Probability
}\label{ss:phop_weaktide}

Even when including the evolution of $\Omega_{\rm s}$, and therefore the
parameter $\eta$ (see Eq.~\ref{eq:def_eta}), the probabilities of the III $\to$ I
and III $\to$ II transitions upon separatrix encounter can still be obtained
semi-analytically. The calculation resembles that presented in
Section~\ref{ss:analytic_calculation} but involves several new ingredients. We
describe the calculation below.

In Section~\ref{ss:analytic_calculation}, we found that the evolution of $H$,
the value of the unperturbed Hamiltonian, allowed us to calculate the
probabilities of the various outcomes of separatrix encounter. Specifically, the
outcome upon separatrix encounter is determined by the value of $H$ at the start
of the separatrix-crossing orbit relative to $H_{\rm sep}$, the value of $H$
along the separatrix. However, when the spin $\Omega_{\rm s}$ is also evolving,
$H_{\rm sep}$ also changes during the separatrix-crossing orbit, and the
calculation in Section~\ref{ss:analytic_calculation} must be generalized to
account for this. Instead of focusing on the evolution of $H$ along a
trajectory, we instead follow the evolution of
\begin{equation}
    K \equiv H - H_{\rm sep}.
\end{equation}
Note that $K > 0$ inside the separatrix, and $K < 0$ outside. With this
modification, the outcome of the separatrix-crossing orbit can be determined in
the same way as in Section~\ref{ss:analytic_calculation}. First, we must compute
the change in $K$ along the legs of the separatrix. We define $\Delta K_{\pm}$
by generalizing Eq.~\eqref{eq:def_dHpm} in the natural way:
\begin{align}
    \Delta K_{\pm} &= \oint_{\mathcal{C}_{\pm}} \s{\rd{H}{t}
        - \rd{H_{\rm sep}}{t}}\;\mathrm{d}t.\label{eq:def_dK_weaktide}
\end{align}
Here, however, note that the contours $\mathcal{C}_{\pm}$ depends on the value
of $\Omega_{\rm s}$ at separatrix encounter (or the corresponding value $\eta =
\eta_{\rm cross}$). Since there is no closed form solution for $\Omega_{\rm
s}(t)$, the probabilities of the various outcomes cannot be expressed as a
simple function of the initial conditions.

Continuing the argument presented in Section~\ref{ss:analytic_calculation}, we
consider the outcome of the separatrix-crossing orbit as a function of $K_i$,
the value of $K$ at the start ($\phi = 0$) of the separatrix-crossing orbit. We
find that if $-\Delta K_+ - \Delta K_- < I_{\rm i} < 0$, then the system
undergoes a III $\to$ II transition and eventually evolves towards tCE2, and
if $-\Delta K_- < K_i < -\Delta K_- - \Delta K_+$, then the system undergoes a
III $\to$ I transition and ultimately evolves towards tCE1. Thus, we find that
the probability of a III $\to$ II transition is given by
\begin{equation}
    P_{\rm III \to II} = \frac{\Delta K_+ + \Delta K_-}{\Delta
        K_-}.\label{eq:def_pc_weaktide}
\end{equation}
Again, since $\Delta K_{\pm}$ are evaluated at resonance encounter, and
$\Omega_{\rm s}$ is evolving, there is no way to express $\Delta K_{\pm}$ in a
closed form of the initial conditions. In fact, since many resonance encounters
occur when $\eta = \eta_{\rm cross}$ is $\gtrsim 0.2$, even an approximate
calculation of $\Delta K_{\pm}$ using Eq.~\eqref{eq:sep_theta} (which is valid
only for $\eta \ll 1$) is inaccurate, and we instead calculate $\Delta K_{\pm}$
along the numerically-computed $\mathcal{C}_{\pm}$ for arbitrary $\eta$. Note
that Eqs.~(\ref{eq:def_dK_weaktide},~\ref{eq:def_pc_weaktide}) are equivalent to
the separatrix capture result of \citet{henrard1982} when $\dot{\theta}_{\rm
tide} = 0$ (see also \citealp{henrard1987} and Paper I). In other words, we
argue that this classic calculation can be unified with the calculation given
in Section~\ref{ss:analytic_calculation} to give an accurate prediction of
separatrix encounter outcome probabilities in the presence of both dissipative
perturbation and parametric evolution of the Hamiltonian.

We note that \citet{levrard2007} also presented an analytical expression for the
resonance capture probability (their Eq.~14) following the method of
\citet{goldreich1966spin}. However, their expression is incomplete, as it does
not account for the contribution of the tidal alignment torque to the change of
the integral of motion over a single orbit.

\begin{figure}
    \centering
    \includegraphics[width=\colummwidth]{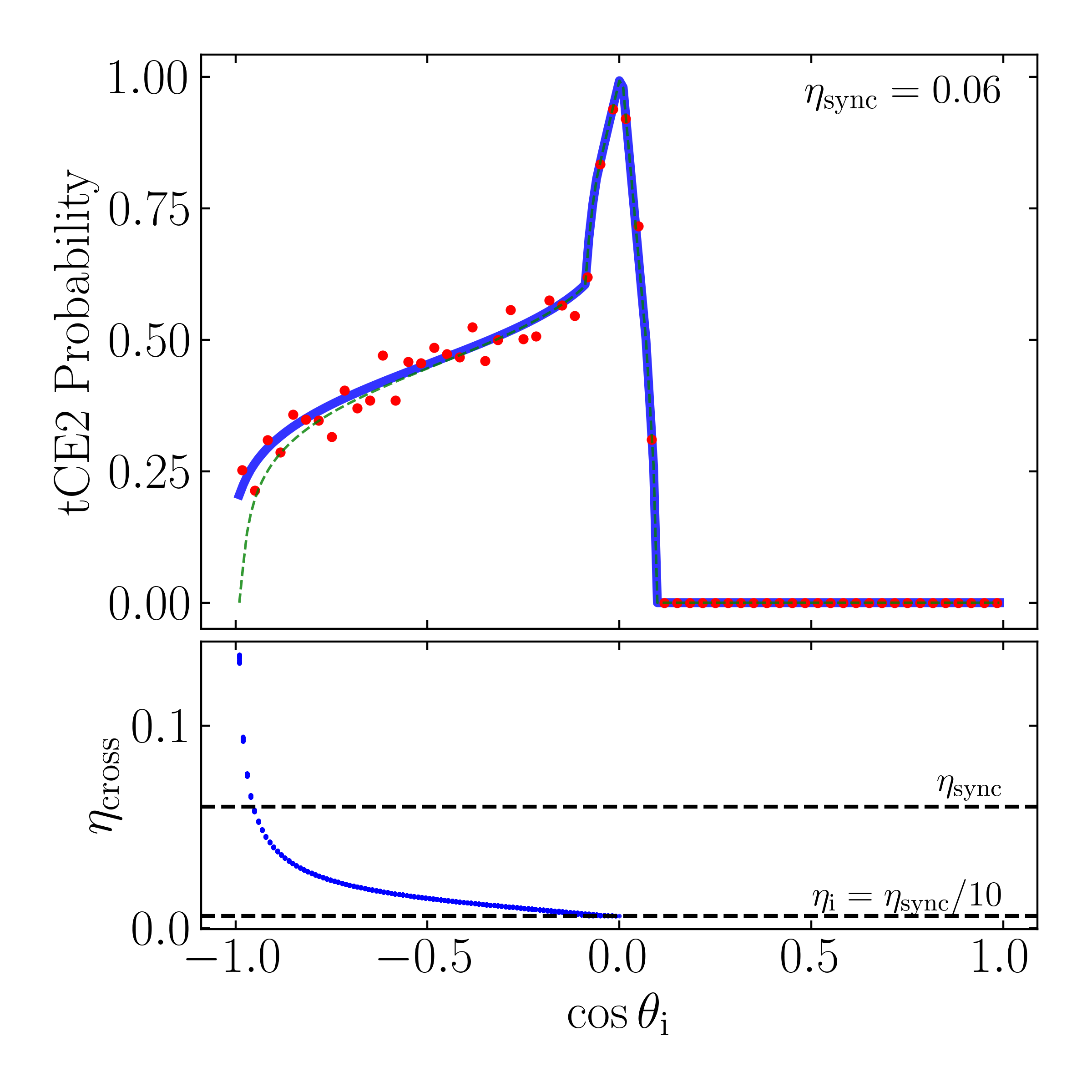}
    \caption{Comparison of the fraction of systems converging to tCE2 obtained
    via numerical simulation (red dots) and obtained via a semi-analytic
    calculation (blue line) for $\eta_{\rm sync} = 0.06$, $I = 20^\circ$, and
    $\Omega_{\rm s, i} = 10n$ (see right panel of Fig.~\ref{fig:Hhists_0_06}).
    The semi-analytic calculation is performed by numerically integrating
    Eqs.~(\ref{eq:dsdt_rot},~\ref{eq:dsdt_tide}--\ref{eq:dWsdt_tide}) on a grid
    of initial conditions uniform in $\cos \theta_{\rm i}$ and $\phi_{\rm i}$
    until the system reaches the separatrix, then calculating the probability of
    reaching tCE2 for each integration using Eq.~\eqref{eq:def_pc_weaktide}. The
    green dashed line in the top panel shows the result of using the analytical
    expression (Eq.~\ref{eq:app_deltaK}) for $\Delta K_{\pm}$, and the bottom
    panel shows the distribution of values of $\eta_{\rm cross}$, the value of
    $\eta$ when a trajectory starting at $\theta_{\rm i}$ encounters the
    separatrix. }\label{fig:pc_fits_0_06}
\end{figure}

\begin{figure}
    \centering
    \includegraphics[width=\colummwidth]{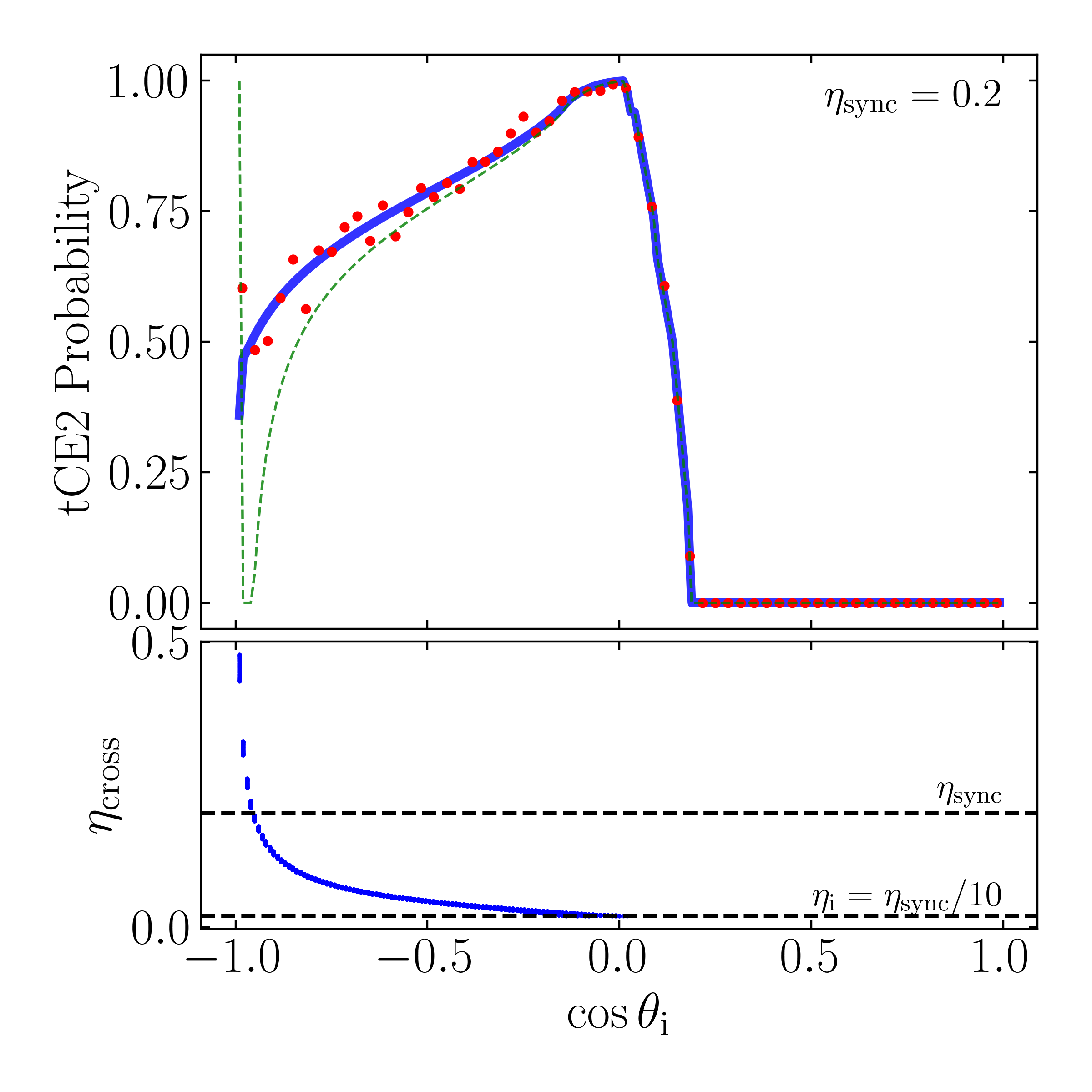}
    \caption{Same as Fig.~\ref{fig:pc_fits_0_06} but for $\eta_{\rm sync} =
    0.2$, corresponding to the right panel of Fig.~\ref{fig:Hhists_0_20}. Note
    that the analytical equation (Eq.~\ref{eq:app_deltaK}; green dashed
    line) exhibits significantly poorer agreement than in
    Fig.~\ref{fig:pc_fits_0_06} when $\eta_{\rm cross} \gtrsim 0.2$.
    }\label{fig:pc_fits_0_20}
\end{figure}

To validate the accuracy of Eq.~\eqref{eq:def_pc_weaktide}, we can compare with
direct numerical integration of
Eqs.~(\ref{eq:dsdt_rot},~\ref{eq:dsdt_tide}--\ref{eq:dWsdt_tide}) for many
initial conditions while evaluating $P_{\rm III \to II}$ (and thus also
obtaining $P_{\rm III \to I} = 1 - P_{\rm III \to II}$) for each simulation at
the moment it encounters the separatrix, if it does so. If the theory is
correct, the total numbers of systems converging to each of tCE1 and tCE2 should
be equal to those predicted by the calculated probabilities. In
Fig.~\ref{fig:pc_fits_0_06}, we show the agreement of this semi-analytic
procedure with the numerical result displayed in the right panel of
Fig.~\ref{fig:Hhists_0_06}. Figure~\ref{fig:pc_fits_0_20} depicts the same for
the parameters of Figs.~\ref{fig:Hhists_0_20}, also showing satisfactory
agreement. Thus, we conclude that the outcomes of separatrix encounter are
accurately predicted by Eq.~\eqref{eq:def_pc_weaktide}.

With the above calculation, we can understand why even some initial conditions
in zone I may converge to tCE2 in certain situations (see
Fig.~\ref{fig:Hhists_0_50}). As long as the initial spin is sufficiently large
($\geq 2n$), Eq.~\eqref{eq:ds_fullq} shows that when $\cos \theta > 2n /
\Omega_{\rm s}$, the obliquity can increase. In particular, when
the critical obliquity $\cos \theta = 2n / \Omega_{\rm s}$ is inside the
separatrix, tidal alignment acts to drive initial conditions in both zones I and
III towards the critical obliquity and into the separatrix, and also towards
larger $H$. As such, when this effect is sufficiently strong,
Eq.~\eqref{eq:def_dK_weaktide} shows that both $\Delta K_\pm > 0$, and both III
$\to$ II and I $\to$ II transitions are guaranteed upon separatrix encounter
(Eq.~\ref{eq:def_pc_weaktide}).

\subsection{Spin Obliquity Evolution for Isotropic Initial Spin Orientations
}\label{ss:tce2_etasync}

\begin{figure}
    \centering
    \includegraphics[width=\colummwidth]{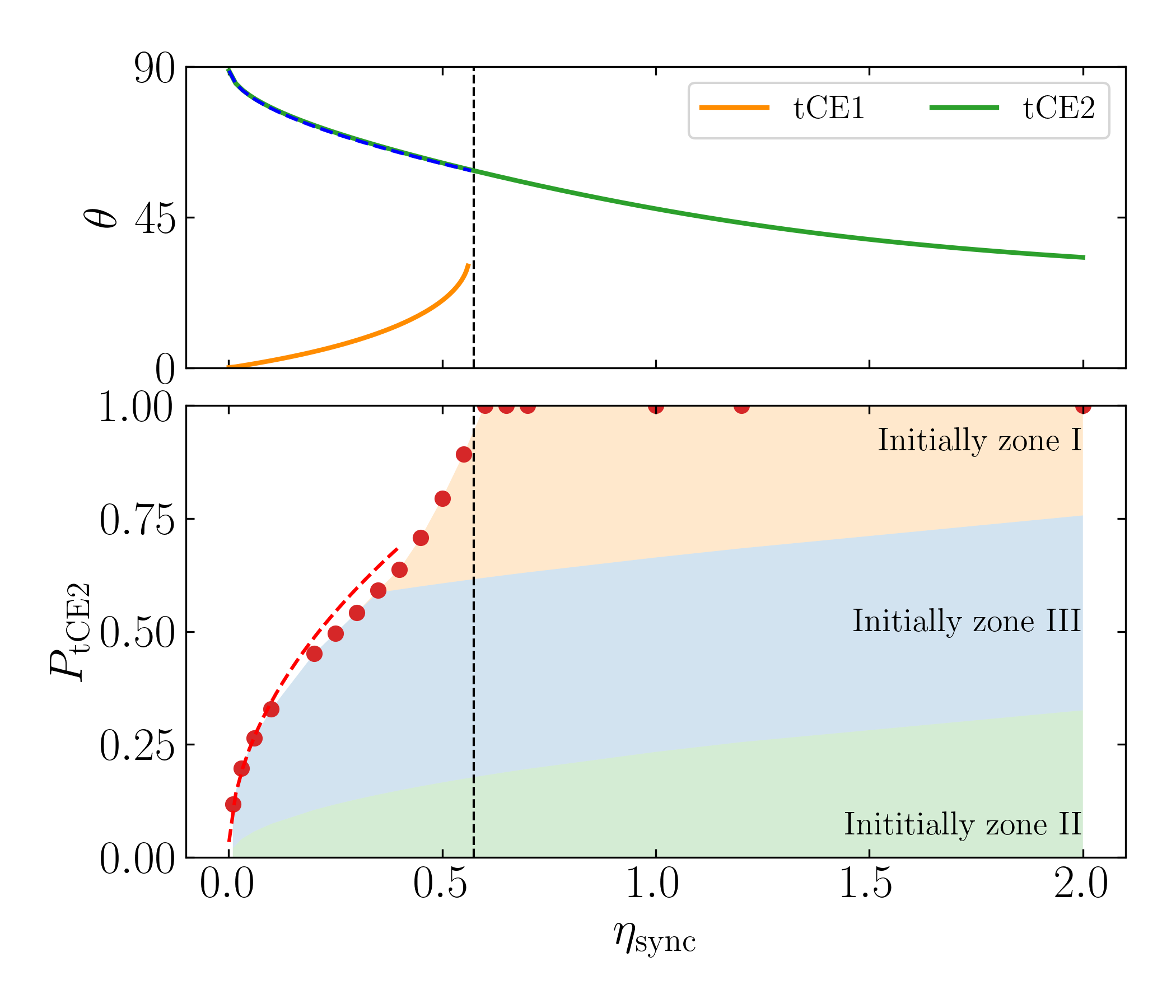}
    \caption{\emph{Top:} Same as top panel of Fig.~\ref{fig:tCEs}.
    \emph{Bottom:} Total probability of the system ending up in tCE2 ($P_{\rm
    tCE2}$; red dots) as a function of $\eta_{\rm sync}$
    (Eq.~\ref{eq:def_etasync}), averaged over an isotropic initial spin
    orientation and taking $\Omega_{\rm s, i} = 10n$. The red dashed line shows
    the analytical prediction Eq.~\eqref{eq:app_tce2_p_tot}. The three shaded
    regions denote the contributions of initial conditions in zones I/II/III
    (labeled) to the total tCE2 probability. For example, among systems that
    converge to tCE2 for $\eta_{\rm sync} = 0.06$, more originate in zone III
    than zone II, and none originate in zone I.}\label{fig:probs20}
\end{figure}

In Sections~\ref{ss:weaktide_evolution}--\ref{ss:phop_weaktide}, we considered
the outcome of the spin evolution driven by tidal torque as a function of the
initial spin orientation, specified by $\theta_{\rm i}$ and $\phi_{\rm i}$.
Here, we calculate the probability of evolution into tCE2 when averaging over a
distribution of initial spin orientations, which we denote by $P_{\rm tCE2}$.
For simplicity, we assume $\uv{s}$ to be isotropically distributed (see
Section~\ref{s:summary} for discussions concerning impact of more physically
realistic distributions of $\uv{s}$). The bottom panel of Fig.~\ref{fig:probs20}
shows $P_{\rm tCE2}$ for $I = 20^\circ$ as a function of $\eta_{\rm sync}$. We
see that, e.g., tCE2 with a large obliquity ($\sim 70^\circ$) can be reached
with substantial probability ($\gtrsim 50\%$).
\begin{figure}
    \centering
    \includegraphics[width=\colummwidth]{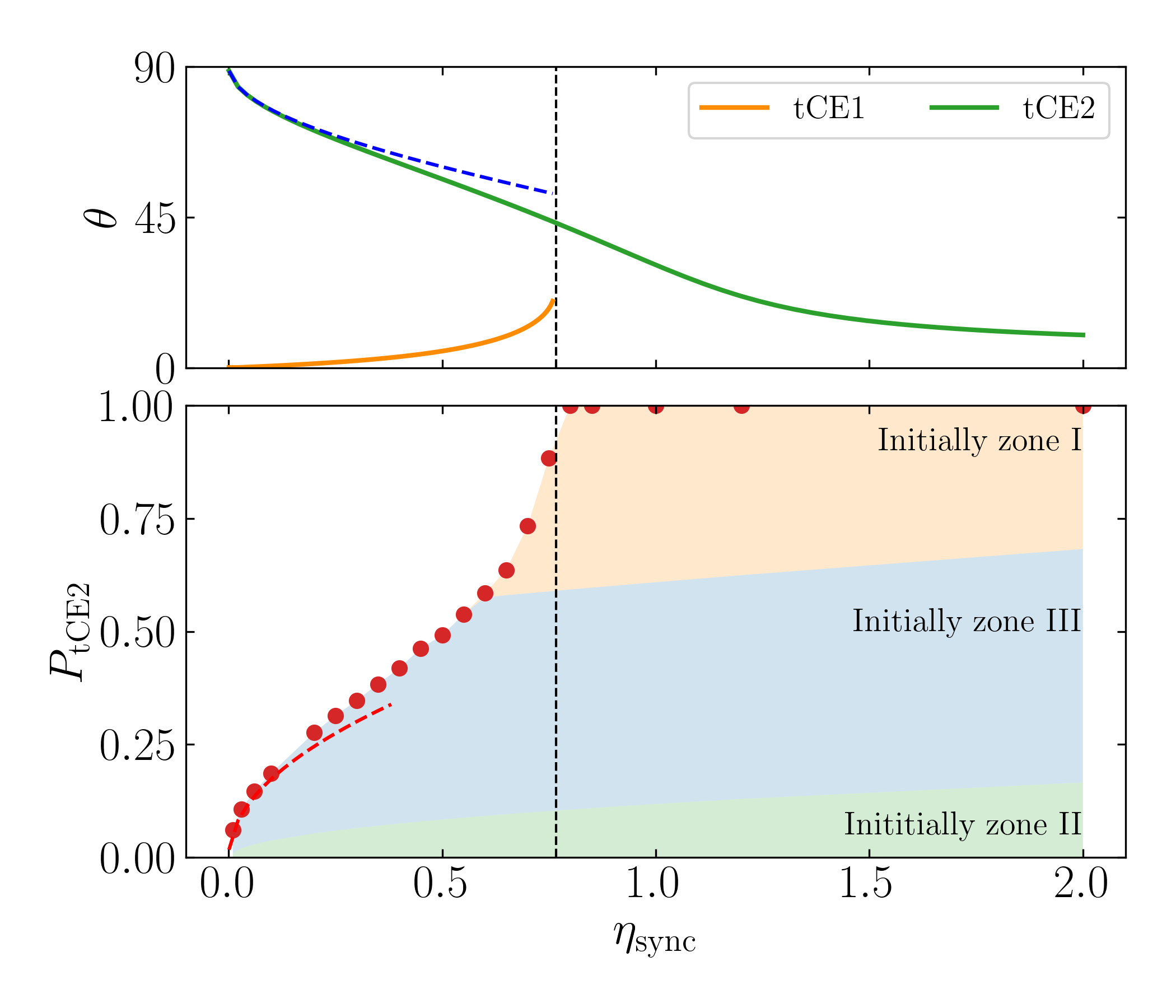}
    \caption{Same as Fig.~\ref{fig:probs20} but for $I =
    5^\circ$.}\label{fig:probs5}
\end{figure}

\begin{figure}
    \centering
    \includegraphics[width=\colummwidth]{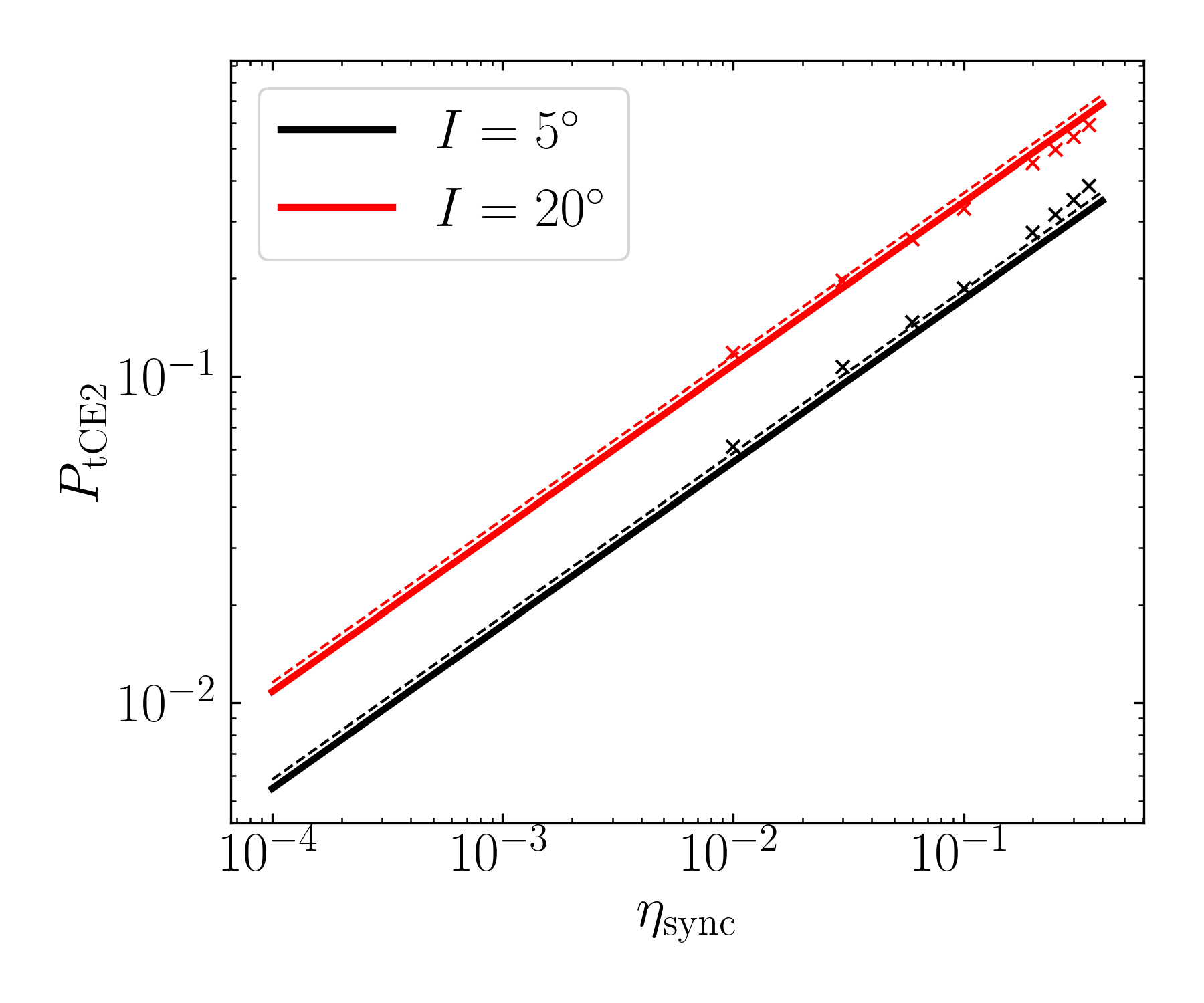}
    \caption{$P_{\rm tCE2}$ as a function of $\eta_{\rm sync}$ for $I = 5^\circ$
    and $I = 20^\circ$ shown on a log-log plot, to emphasize the scaling at
    small $\eta_{\rm sync}$. The crosses are the results of numerical
    integrations as shown in Figs.~\ref{fig:probs20}--\ref{fig:probs5}, the
    solid lines are Eq.~\eqref{eq:app_tce2_p_tot} for $\Omega_{\rm s, i} = 10n$
    and the dashed lines are for $\Omega_{\rm s, i} = 3n$.}\label{fig:anal_ptce}
\end{figure}

When $\eta_{\rm sync} \ll 1$ and $\Omega_{\rm s, i} \gtrsim n$, an approximate
analytical formula for $P_{\rm tCE2}$ can be obtained (see
Appendix~\ref{app:ptce2}):
\begin{align}
    P_{\rm tCE2} &\simeq
            \frac{4\sqrt{\eta _{\rm sync} \sin I}}{\pi}\s{
                \sqrt{n / \Omega_{\rm s, i}}
                + \frac{3}{2\p{1
                + \sqrt{n / \Omega_{\rm s, i}}}}}.\label{eq:app_tce2_p_tot}
\end{align}
Eq.~\eqref{eq:app_tce2_p_tot} is shown in
Figs.~\ref{fig:probs20}--\ref{fig:probs5} as the red dashed lines; it
agrees well with the numerical results (red dots) for $\eta_{\rm sync} \lesssim
0.4$. To illustrate the predicted values of $P_{\rm tCE2}$ for small $\eta_{\rm
sync}$, we display $P_{\rm tCE2}$ for $\eta_{\rm sync} \in \s{10^{-4}, 0.4}$ for
both $I = 20^\circ$ and $I = 5^\circ$ in Fig.~\ref{fig:anal_ptce}. Note that
for $\eta_{\rm sync} \leq 10^{-2}$, numerical results for $P_{\rm tCE2}$ are
difficult to obtain, as the integration of
Eqs.~\eqref{eq:ds_fullq}--\eqref{eq:ds_fulls} slows down dramatically due to the
rapid precession of $\uv{s}$ about $\uv{l}$.

\section{Applications}\label{s:applications}

\subsection{Obliquities of Super-Earths with Exterior Companions
}\label{ss:disc_sehj}

Consider a system consisting of an inner Super-Earth (SE) with semi-major axis $a
\lesssim 0.5\;\mathrm{AU}$ and an exterior companion. For concreteness, we
assume the companion (with mass $m_{\rm p}$) to be a cold
Jupiter (CJ) with $a_{\rm p} \gtrsim 1\;\mathrm{AU}$. Such systems are quite
abundant \citep{zhu2018super, bryan2019excess}. A phase of giant impacts may
occur in the formation of such SEs \citep{inamdar2015formation,
izidoro2017breaking}, leading to a wide range of initial obliquities for the
SEs. We are interested in the ``final'' obliquities of the SEs driven by tidal
dissipation.

\begin{figure}
    \centering
    \includegraphics[width=\colummwidth]{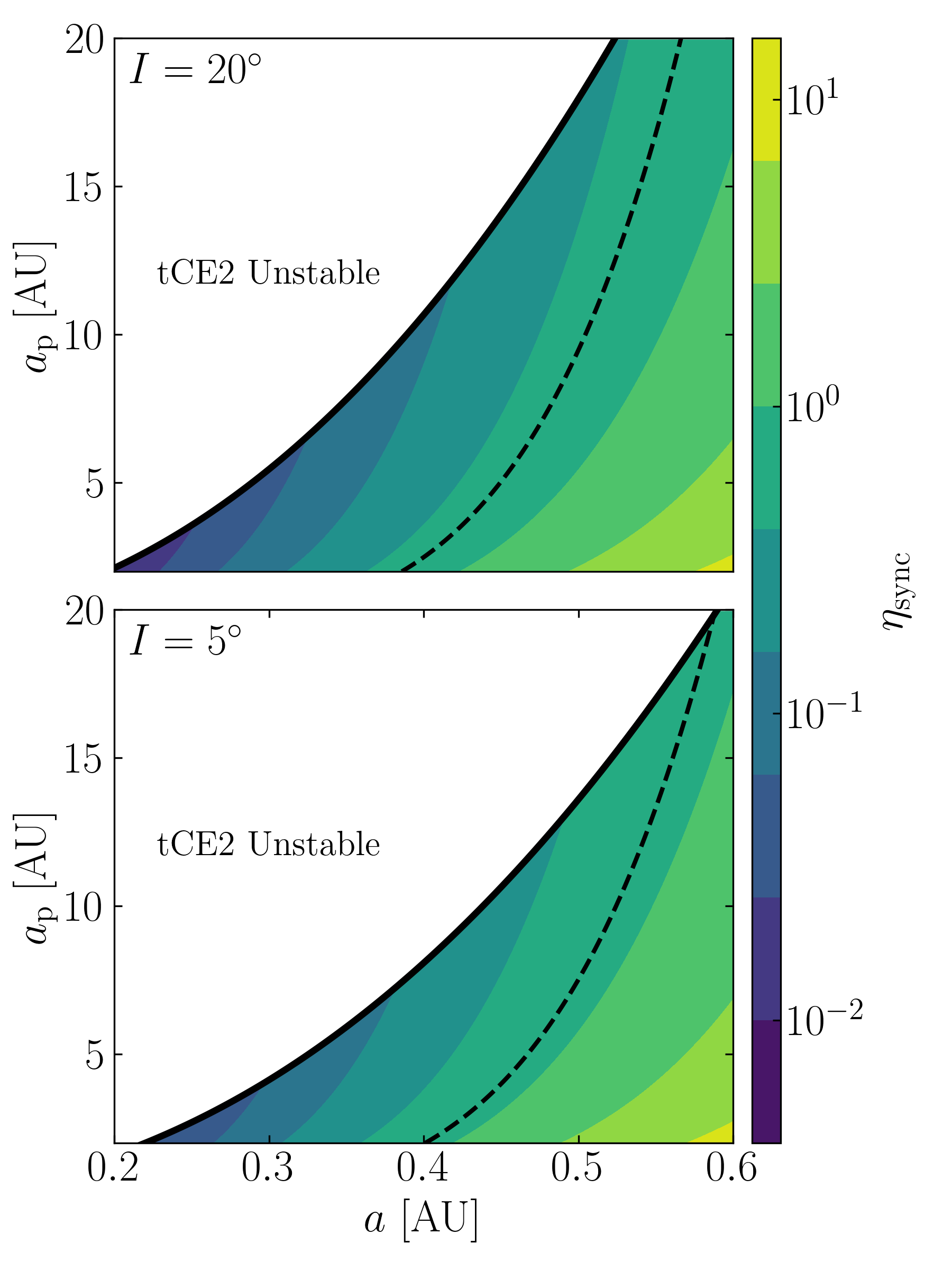}
    \caption{Depiction of the values of $\eta_{\rm sync}$ for the Super Earth +
    Cold Jupiter systems as a function of $a$ and $a_{\rm p}$ for $I = 20^\circ$
    (top) and $I = 5^\circ$ (bottom). The SE is taken to have $m = 4M_{\oplus}$
    and $R = 2R_{\oplus}$ while the CJ is taken to have $m_{\rm p} = M_{\rm J}$,
    and we have taken $k \approx k_{\rm q}$ for the SE\@. We only show the
    regions satisfying $t_{\rm s} \geq t_{\rm s, c}$ (the stability condition
    for tCE2; Eqs.~\ref{eq:sehj_ts}--\ref{eq:sehj_tsc}). The line satisfying
    $\eta_{\rm sync} = \eta_{\rm c}$ (Eq.~\ref{eq:def_etac}) is shown as the
    black dashed line. Systems with $\eta_{\rm sync} \gtrsim 0.1$ have
    appreciable probabilities of being captured in permanent tCE2 with
    significant obliquities (see Figs.~\ref{fig:probs20}--\ref{fig:anal_ptce}).
    }\label{fig:sehj_region}
\end{figure}

For typical SE parameters, the spin evolution timescale due to tidal dissipation
is given by
\begin{align}
    \frac{1}{t_{\rm s}} \simeq{}& \frac{1}{3 \times 10^7\;\mathrm{yr}}
            \p{\frac{1}{4k}}
            \p{\frac{2k_2/Q}{10^{-3}}}
            \p{\frac{M_\star}{M_{\odot}}}^{3/2}
            \p{\frac{m}{4M_{\oplus}}}^{-1}\nonumber\\
        &\times \p{\frac{R}{2R_{\oplus}}}^3
            \p{\frac{a}{0.4\;\mathrm{AU}}}^{-9/2}.
            \label{eq:sehj_ts}
\end{align}
This occurs well within the age of SE-CJ systems. On the other hand,
the orbital evolution of the SE occurs on the timescale \citep[e.g.][]{lai2012}
\begin{align}
    -\frac{\dot{a}}{a} ={}& \frac{3k_2}{Q}\frac{M_\star}{m}
            \p{\frac{R}{a}}^5n \p{1 - \frac{\Omega_{\rm s}}{n}\cos \theta}\nonumber\\
        \simeq{}& \frac{1}{7 \times 10^{14}\;\mathrm{yr}}
            \p{1 - \frac{\Omega_{\rm s}}{n}\cos \theta}
            \p{\frac{2k_2/Q}{10^{-3}}}
            \p{\frac{M_\star}{M_{\odot}}}^{3/2}\nonumber\\
        &\times \p{\frac{m}{4M_{\oplus}}}^{-1}
            \p{\frac{R}{2R_{\oplus}}}^5
            \p{\frac{a}{0.4\;\mathrm{AU}}}^{-13/2}.\label{eq:sehj_adot}
\end{align}
Thus, $a$ does not evolve within the age of the SE-CJ system (for $a \gtrsim
0.06\;\mathrm{AU}$), and we shall treat $a$ as a constant in this subsection
(but see Sections~\ref{ss:disc_usp}--\ref{ss:disc_wasp12b}). With typical
SE-CJ parameters, Eq.~\eqref{eq:def_etasync} can be evaluated:
\begin{align}
    \eta_{\rm sync} ={}& 0.303 \cos I
            \p{\frac{k}{k_{\rm q}}}
            \p{\frac{m_{\rm p}}{M_{\rm J}}}
            \p{\frac{m}{4M_{\oplus}}}
            \p{\frac{M_\star}{M_{\odot}}}^{-2}
            \p{\frac{a}{0.4\;\mathrm{AU}}}^{6}\nonumber\\
        &\times \p{\frac{a_{\rm p}}{5\;\mathrm{AU}}}^{-3}
            \p{\frac{R}{2R_{\oplus}}}^{-3}.\label{eq:sehj_etasync}
\end{align}
We see from Figs.~\ref{fig:probs20}
and~\ref{fig:probs5} that this value of $\eta_{\rm sync}$ can lead to a
high-obliquity tCE2 with significant probability, assuming the SE has a wide
range of initial obliquities. In addition, Eq.~\eqref{eq:def_ts_crit} shows that
tCE2 is stable if $t_{\rm s} \gtrsim t_{\rm s, c}$, where
\begin{align}
    \frac{1}{t_{\rm s, c}} ={}& \frac{\sin I \cos^2 I}{3 \times
        10^5\;\mathrm{yr}}
            \p{\frac{k}{k_{\rm q}}}
            \p{\frac{m_{\rm p}}{M_{\rm J}}}^{3/2}
            \p{\frac{m}{4 M_{\oplus}}}^{1/2}\nonumber\\
        &\times \p{\frac{M_\star}{M_{\odot}}}^{-3/2}
            \p{\frac{a}{0.4\;\mathrm{AU}}}^6
            \p{\frac{a_{\rm p}}{5\;\mathrm{AU}}}^{-9/2}
            \p{\frac{R}{2 R_{\oplus}}}^{-3/2}.
            \label{eq:sehj_tsc}
\end{align}
In Fig.~\ref{fig:sehj_region}, we show the value of $\eta_{\rm sync}$ in the
regions of $\p{a, a_{\rm p}}$ parameter space that satisfy the stability
condition for tCE2. We see that a generous portion of parameter space is
able to generate and sustain SEs in stable tCE2 with significant obliquities. In
summary, we predict that a large fraction of SEs with exterior CJ
companions can have long-lived, significant obliquities ($\gtrsim 60^\circ$) due
to being trapped in tCE2.

\subsection{Formation of Ultra-short-period Planet Formation via Obliquity Tides
}\label{ss:disc_usp}

Ultra-short period planets (USPs), Earth-sized planets with sub-day periods,
constitute a statistically distinct subsample of Kepler planets
\citep[e.g.][]{winn2018kepler, dai2018larger}. It is generally thought that USPs
evolved from close-in SEs through orbital decay, driven by tidal dissipation in
their host stars \citep{lee2017magnetospheric} or in the planets
\citep{petrovich2019ultra, pu2019low}. In particular, \citet{pu2019low} showed
that a ``low-eccentricity migration'' mechanism can successfully produce USPs
with the observed properties. In this scenario, USPs evolve from a subset of SE
systems: a low-mass planet with an initial period of a few days maintains a
small but finite eccentricity due to secular forcings from exterior companion
planets (SEs or sub-Neptunes) and evolve to become a USP due to orbital decay
driven by tidal dissipation.

\citet{millholland2020formation} proposed an alternative formation mechanism of
USPs based on obliquity tides (instead of eccentricity tides as in
\citealp{pu2019low}). This mechanism consists of three stages:
\begin{itemize}
    \item A proto-USP (with two external companions) is assumed to be rapidly
        captured into CS2 with appreciable obliquity and a pseudo-synchronous
        spin rate.

    \item The inner planet undergoes runaway tidal migration as a result of the
        decreasing semi-major axis and increasing obliquity while following CS2.

    \item The inward migration stalls when the tidal torque becomes
        sufficiently strong to destroy CS2.
\end{itemize}

Here, we evaluate the viability of the obliquity-driven migration scenario for
USPs using the general results presented earlier in this paper.

First, the spin evolution timescale for typical proto-USP parameters is
\begin{align}
    \frac{1}{t_{\rm s}} ={}& \frac{1}{1200\;\mathrm{yr}}
            \p{\frac{1}{4k}}
            \p{\frac{2k_2/Q}{10^{-3}}}
            \p{\frac{M_\star}{M_{\odot}}}^{3/2}
            \p{\frac{\rho}{\rho_{\oplus}}}^{-1}
            \p{\frac{a}{0.035\;\mathrm{AU}}}^{-9/2}.
\end{align}
where $\rho$ is the density of the proto-USP, $\rho_{\oplus}$ is the density
of the Earth, and we have adopted the (approximately) largest possible value for
$a$ (the semi-major axis of the proto-SE) to ensure that orbital decay can
happen within the age of the system (see Eq.~\ref{eq:adot_usp}). This is much
shorter than the age of the system, and so the proto-USP can quickly evolve into
one of the stable tCE (either tCE1 or tCE2).

Next, to determine which tCE the planet evolves into, we need to evaluate
$\eta_{\rm sync}$ (see Eq.~\ref{eq:def_etasync}). For simplicity, we consider
the case where the proto-USP is surrounded by a single external planetary
companion with $a_{\rm p} \gtrsim a$ (typical of Kepler multi-planet systems) but
with $L_{\rm p} \gg L$ (this condition can easily be relaxed; see
Section~\ref{ss:disc_wasp12b}). To account for such a close-by companion,
Eq.~\eqref{eq:wlp} must be modified to \citep[see e.g.][]{lai_2017}:
\begin{equation}
    \omega_{\rm lp} = \frac{3m_{\rm p}}{4M_{\star}}
        \p{\frac{a}{a_{\rm p}}}^3 n f(\alpha),
        \label{eq:def_g_usp}
\end{equation}
where $\alpha = a / a_{\rm p}$ and
\begin{equation}
    f(\alpha) \equiv \frac{b_{\rm 3/2}^{(1)}(\alpha)}{3\alpha}
        \approx 1 + \frac{15}{8}\alpha^2 + \frac{175}{64}\alpha^4\dots
\end{equation}
with $b_{\rm 3/2}^{(1)}$ the Laplace coefficient. With this modification,
$\eta_{\rm sync}$ (Eq.~\ref{eq:def_etasync}) is given by
\begin{align}
    \eta_{\rm sync} ={}& 0.011 f(\alpha)
            \p{\frac{k}{k_{\rm q}}}
            \p{\frac{\rho}{\rho_{\oplus}}}
            \p{\frac{a}{0.035 \;\mathrm{AU}}}^{3}\cos I \nonumber\\
        &\times
            \p{\frac{m_{\rm p}}{10 M_{\oplus}}}
            \p{\frac{1.3 a}{a_{\rm p}}}^{3}
            \p{\frac{M_\star}{M_{\odot}}}^{-2}
            ,\label{eq:etasync_usp}
\end{align}
where we have normalized $a_{\rm p}/ a$ to $1.3$ (corresponding to a period
ratio $P_{\rm p} / P = 1.5$), for which $f(\alpha) \approx 5.5$. As $k / k_{\rm
q} \sim 1$ (see footnote 1) for the close-in proto-USP, we have $\eta_{\rm sync}
\lesssim 0.06$, much less than $\eta_{\rm c} \sim 1$ under most
conditions\footnote{One can make $\eta_{\rm sync}$ larger by choosing a larger
initial value for $a$, e.g.\ $a = 0.05\;\mathrm{AU}$. However, the planet would
not be able to experience orbital decay for such a large value, see
Eq.~\eqref{eq:adot_usp}. Also note that Kepler systems of SEs have adjacent
period ratios in the range of $1.3$--$4$ \citep{fabrycky2014architecture},
corresponding to semi-major axis ratios of $1.2$--$2.5$.}. As such, if the
initial planetary obliquity is prograde, the planet is guaranteed to evolve into
tCE1, and not tCE2 (see
Figs.~\ref{fig:Hhists_0_06},~\ref{fig:pc_fits_0_06}--\ref{fig:pc_fits_0_20}). If
we assume instead a randomly oriented initial planetary spin,
Figs.~\ref{fig:anal_ptce}--\ref{fig:probs20} suggest that the probability of
capture into tCE2 is small ($\lesssim 20\%$). A more sophisticated calculation
including the effect of a third planet does not greatly modify these results.

The second stage of the proposed mechanism, runaway inward migration after
attaining tCE2, requires that the initial orbital decay timescale be
sufficiently fast. Evaluating Eq.~\eqref{eq:sehj_adot} for the relevant physical
parameters, we find
\begin{align}
    -\frac{\dot{a}}{a} ={}& \frac{1}{8 \times 10^8\;\mathrm{yr}}
            \p{1 - \frac{\Omega_{\rm s}}{n}\cos \theta}
            \p{\frac{2k_2/Q}{10^{-3}}}
            \p{\frac{M_\star}{M_{\odot}}}^{3/2}\nonumber\\
        &\times \p{\frac{m}{M_{\oplus}}}^{-1}
            \p{\frac{R}{R_{\oplus}}}^5
            \p{\frac{a}{0.035\;\mathrm{AU}}}^{-13/2}.
            \label{eq:adot_usp}
\end{align}
For $\eta_{\rm sync} \ll \eta_{\rm c}$, Eqs.~\eqref{eq:def_tce2_approx} imply
that $\Omega_{\rm s}\cos \theta / n \ll 1$ in tCE2, so indeed the orbit of the
proto-USP is able to decay within the lifetime of the system. On the other hand,
in tCE1, $\omega_{\rm s} \approx n$ and $\cos \theta \simeq 1 - \eta_{\rm
sync}^2 \sin^2 I / 2$, so $\dot{a} / a$ is suppressed by a factor of $\sim
\eta_{\rm sync}^2 \sin^2 I$. This shows that a proto-USP in tCE1 is unable to
initiate runaway orbital decay within the age of the system. Note that this
constraint also implies $\eta_{\rm sync}$ (Eq.~\ref{eq:etasync_usp}) cannot be
increased by considering proto-USPs with larger values of $a$, as the initial
orbital decay will become too slow.

Finally, we compute the orbital separation at which tCE2 becomes unstable when
the tidal alignment torque is too strong. Evaluating Eq.~\eqref{eq:def_ts_crit},
we find that tCE2 breaks ($t_{\rm s} \lesssim t_{\rm s, c}$) when the semi-major
axis is smaller than $a_{\rm break}$, where
\begin{align}
    a_{\rm break} \simeq{}&
        \s{\frac{k_{\rm q}}{k^3f^3(\alpha)}}^{1/18}
        \p{\frac{2k_2}{Q}}^{1/9}
        \p{\sin I \cos^2 I}^{-1/9}\nonumber\\
    &\times \p{\frac{M_\star^2}{m_{\rm p}m}}^{1/6}
        \p{Ra_{\rm p}}^{1/2}\nonumber\\
    \simeq{}& 0.028 \;\mathrm{AU}
        \p{\frac{2k_2/Q}{10^{-3}}}^{1/9}
        \p{\sin I \cos^2 I}^{-1/9}
        \p{\frac{M_\star}{M_{\odot}}}^{1/3}\nonumber\\
        &\times \p{\frac{m_{\rm p}}{10M_{\oplus}}}^{-1/6}
        \p{\frac{\rho}{\rho_{\oplus}}}^{-1/6}
        \p{\frac{a_{\rm p}}{0.05\;\mathrm{AU}}}^{1/2},\label{eq:abreak_usp}
\end{align}
where we have used $k \sim k_{\rm q} \sim 0.4$ and $\alpha = 0.028 / 0.05$. Once
the system exits tCE2, it rapidly evolves to tCE1, in which orbital decay is
severely suppressed (Eq.~\ref{eq:adot_usp}). This final orbital separation does
not qualify as a USP ($P \lesssim \mathrm{day}$). To reduce $a_{\rm break}$ to
$0.0195\;\mathrm{AU}$ (corresponding to a 1 day orbital period for $M_{\star} =
1M_{\odot}$) would require the value of $a_{\rm p} / m_{\rm p}^{1/3}$ to be
$\sim 2$ times smaller than that adopted in Eq.~\eqref{eq:abreak_usp} (e.g.\ for
$m_{\rm p}$ to be larger by a factor of $8$ for the same $a_{\rm p}$). Note that
observed USPs almost always have $a_{\rm p} / a \gtrsim 3$
\citep{steffen2013lack, winn2018kepler}.

In summary, our results suggest that only proto-USPs with large primordial
obliquities have a nonzero probability of evolving into tCE2
initially\footnote{The probability is small even for isotropic primordial
obliquities. This low probability may not be an issue, as the occurence rate of
USPs is only $\sim 0.5\%$ around solar type stars \citep{sanchis2014study,
winn2018kepler}.}. More importantly, proto-USPs that successfully initiate
runaway tidal migration after reaching tCE2 will likely cease their inward
migration before becoming a USP\@.

\subsection{Orbital decay of WASP-12b Driven by Obliquity Tides}\label{ss:disc_wasp12b}

WASP-12b is a hot Jupiter (HJ) with mass $m = 1.41M_{\rm J}$ and radius $R =
1.89R_{\rm J}$ orbiting a host star (with mass $M_\star
= 1.36M_{\odot}$ and radius $R_\star = 1.63R_{\odot}$) on a $P =
1.09\;\mathrm{day}$ ($a = 0.023 \;\mathrm{AU}$) orbit \citep{hebb2009wasp,
maciejewski2013multi}. Long-term observations have revealed
that its orbit is undergoing decay with $P / \dot{P} = -3.2\;\mathrm{Myr}$
\citep{maciejewski2016departure, patra2017apparently, Patra12b, turner12b}. Such
a rapid orbital decay puts useful constraints on the physics of tidal
dissipation in the host star \citep[e.g.][]{weinberg2017tidal, barker2020tidal}.

\citet{millholland2019obliquity} considered the possibility that the measured
orbital decay of WASP-12b is caused by tidal dissipation in the HJ trapped in a
high-obliquity CS due to an undetected planetary companion. We now evaluate the
plausibility of this scenario. We begin with the planetary spin evolution
timescale, which is given by (see Eq.~\eqref{eq:dsdt_tide}):
\begin{align}
    \frac{1}{t_{\rm s}} ={}& \frac{1}{6000\;\mathrm{yr}}
            \p{\frac{1}{4k}}
            \p{\frac{2k_2/Q}{10^{-6}}}
            \p{\frac{M_\star}{1.36 M_{\odot}}}^{3/2}\nonumber\\
        &\times \p{\frac{m}{1.41 M_{\rm J}}}^{-1}
            \p{\frac{R}{1.89 R_{\rm J}}}^{3}
            \p{\frac{a}{0.023\;\mathrm{AU}}}^{-9/2}.
\end{align}
Thus, the spin of WASP-12b has plenty of time to find a tCE\@. We also wish to
calculate $\eta_{\rm sync}$, but there are two uncertainties: (i) the
properties of the hypothetical planet companion (mass $m_{\rm p}$) to WASP-12b
are unknown, and it is likely that $L_{\rm p}$ is smaller than $L$; and (ii) we
should evaluate $\eta_{\rm sync}$ using the ``primordial'' / initial value of
$a$ for WASP-12b at the start of its orbital migration, not necessarily its
present day value. Concerning (i), we express the precession of $\uv{l}$ about
$\bm{J} = J\uv{\jmath} \equiv \bm{L} + \bm{L}_{\rm p}$, the total angular
momentum axis, as
\begin{equation}
    \rd{\uv{l}}{t} = \omega_{\rm lp} \frac{J}{L_{\rm p}}
        \p{\uv{l} \times \uv{\jmath}}\cos I,
        \label{eq:wasp12b_g}
\end{equation}
where $\omega_{\rm lp}$ is given by Eq.~\eqref{eq:def_g_usp}. Thus, we see
that the precession frequency $g$ in
Sections~\ref{s:theory}--\ref{s:full_tide_prob} is changed to
(cf.\ Eq.~\ref{eq:wlp})
\begin{equation}
    g = -\omega_{\rm lp}\frac{J}{L_{\rm p}} \cos I
        = -\frac{3m_{\rm p}}{4M_{\star}}
        \p{\frac{a}{a_{\rm p}}}^3 n f(\alpha) \frac{J}{L_{\rm p}}
        \cos I\label{eq:hj_g}.
\end{equation}
Concerning (ii), we use the fiducial values for the initial semi-major
axis $a_{\rm i} = 0.038\;\mathrm{AU}$ and initial semi-major axis ratio $a_{\rm
p} / a_{\rm i} = 1.29$, to be justified \emph{a posteriori}. Assuming $J /
L_{\rm p} \simeq L / L_{\rm p}$ (i.e.\ $L \gg L_{\rm p}$), we have
\begin{align}
    \eta_{\rm sync, i}
        \simeq{}&
            \frac{k}{2k_{\rm q}}\p{\frac{m}{M_\star}}^2
            \p{\frac{a_{\rm i}}{a_{\rm p}}}^{7/2}
            \p{\frac{a_{\rm i}}{R}}^3
            f\p{\alpha_{\rm i}}
            \cos I,\nonumber\\
        ={}& 0.015  f(\alpha_{\rm i})
            \p{\frac{m}{1.41M_{\rm J}}}^2
            \p{\frac{M_\star}{1.36M_{\odot}}}^{-2}
            \nonumber\\
        &\times
            \p{\frac{a_{\rm i}}{0.038\;\mathrm{AU}}}^{3}
            \p{\frac{a_{\rm p}}{1.29 a_{\rm i}}}^{-7/2}
            \p{\frac{R}{1.89R_{\rm J}}}^{-3}\cos I,\label{eq:hj_etasync}
\end{align}
where we have used $k / k_{\rm q} \simeq 1$. For the adopted fiducial of $a_{\rm
i}$ and $a_{\rm p}$, $\alpha_{\rm i} = a_{\rm i} / a_{\rm p}$ and $f(\alpha_{\rm
i}) \simeq 5$.

\begin{figure}
    \centering
    \includegraphics[width=\colummwidth]{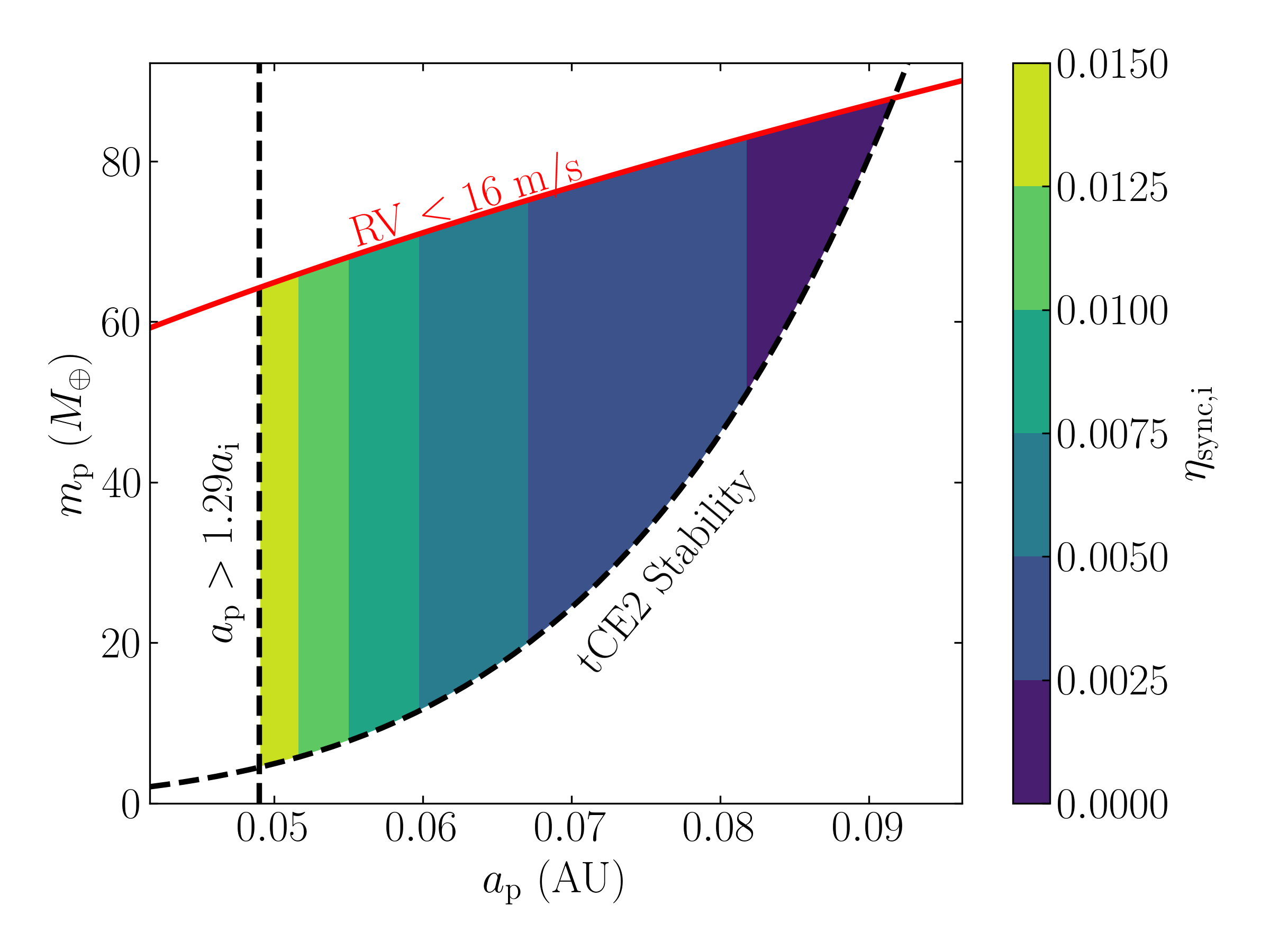}
    \caption{Constraints on the companion of WASP-12b and the values of
    $\eta_{\rm sync, i}$ (Eq.~\ref{eq:hj_etasync}) in the obliquity tidal decay
    scenario. The right dashed line is from Eq.~\eqref{eq:hj_tce2_stab},
    required for the current system to be locked in a stable tCE2; the left
    vertical line is from Eq.~\eqref{eq:hj_orbstab}, required for the dynamical
    stability of the ``primordial'' system; the red line is from
    Eq.~\eqref{eq:hj_rv}. The probability of capture into tCE2 (starting from a
    isotropic distribution of spin orientation) is proportional to $\eta_{\rm
    sync,i}^{1/2}$ (see Eq.~\ref{eq:app_tce2_p_tot} with $I$ replaced by $I_{\rm
    J}$, and note that $I_{\rm J} \ll 1$ for the WASP-12b system).  The plot
    adopts the largest possible value of $a_{\rm i}$ ($= 0.038\;\mathrm{AU} >
    0.023\;\mathrm{AU} = a$); using a smaller $a_{\rm i}$ would significantly
    reduce $\eta_{\rm sync, i}$, making capture into tCE2 even less likely.
    }\label{fig:wasp12b_constr}
\end{figure}

We next work towards justifying these choices of fiducial parameters. There are
four physical and observational constraints on the ``WASP-12b + companion''
system (see Fig.~\ref{fig:wasp12b_constr}):

(i) The HJ must have had a sufficiently small initial semi-major axis such that
its orbital decay timescale is less than the age of the system. The orbital
decay rate is given by
\begin{align}
    -\p{\frac{\dot{a}}{a}}_{\rm i}
        =& \frac{1}{\mathrm{Gyr}}
            \p{\frac{2k_2/Q}{10^{-6}}}
            \p{\frac{M_\star}{1.36 M_{\odot}}}^{3/2}
            \p{\frac{m}{1.41M_{\rm J}}}^{-1}\nonumber\\
        &\times \p{\frac{R}{1.89R_{\rm J}}}^5
            \p{\frac{a_{\rm i}}{0.038\;\mathrm{AU}}}^{-13/2}
            \p{1 - \frac{\Omega_{\rm s}}{n}\cos \theta}.
            \label{eq:hj_adot}
\end{align}
Thus, the initial semi-major axis for the HJ cannot exceed $0.038\;\mathrm{AU}$
even when $\p{1 - \Omega_{\rm s}\cos \theta/n} \approx 1$.

(ii) The exterior planet must be sufficiently massive to keep the HJ in the
high-obliquity tCE2 today, i.e.\ the tCE2 must be stable under the influence of
the exterior planet. With the amended precession frequency $\abs{g}$ given by
Eq.~\eqref{eq:hj_g}, the stability of tCE2 requires (see
Eq.~\ref{eq:def_ts_crit})
\begin{equation}
    \frac{1}{t_{\rm s}} \lesssim \omega_{\rm lp}\cos I \frac{J}{L_{\rm p}}
        \sin I_{\rm J} \sqrt{\frac{\eta_{\rm sync}\cos I_{\rm J}}{2}},
\end{equation}
where $\cos I_{\rm J} \equiv \uv{l} \cdot \uv{\jmath}$. Using $\sin I_{\rm J} =
\p{L_{\rm p} / J} \sin I \ll 1$ for $L \gg L_{\rm p}$, this yields
\begin{align}
    \frac{a_{\rm p}}{a}
        \lesssim{}& \s{
            \frac{\p{kf(\alpha)}^{3/2}}{k_{\rm q}^{1/2}}
            \sin I \cos^{3/2} I
            \frac{Q}{2k_2}
            \frac{m_{\rm p}m^2}{M_\star^3}
            \p{\frac{a}{R}}^{9/2}}^{4/19}\nonumber\\
        \simeq{}&
        3.5
            \s{\frac{k^3f^3(\alpha)}{k_{\rm q}}}^{2/19}
            \p{\sin I \cos^{3/2} I}^{4/19}
            \p{\frac{m}{1.41M_{\rm J}}}^{8/19}\nonumber\\
        &\times \p{\frac{m_{\rm p}}{80M_{\oplus}}}^{4/19}
            \p{\frac{M_\star}{M_{\odot}}}^{-12/19}
            \p{\frac{2k_2/Q}{10^{-6}}}^{-4/19}\nonumber\\
        &\times \p{\frac{a}{0.023\;\mathrm{AU}}}^{18/19}
            \p{\frac{R}{1.89 R_{\rm J}}}^{-18/19},\label{eq:hj_tce2_stab}
\end{align}
where in the second equality, we have used the currently observed values for
$a$, $m$, $R$, and $M_\star$, and have set $\p{k^3f^3/k_{\rm q}}^{2/19} \simeq
1$.

(iii) The RV signal of the exterior planet must be smaller than the residuals of
the published RVs, $\sim 16\;\mathrm{m/s}$ \citep{hebb2009wasp,
husnoo2011orbital, knutson2014friends, bonomo2017gaps}. This requires
\begin{equation}
    \p{\frac{a_{\rm p}}{0.076\;\mathrm{AU}}}^{-1/2}
    \p{\frac{m_{\rm p}}{80M_{\oplus}}}
    \p{\frac{M_\star}{1.36M_{\odot}}}^{-1/2}
    \p{\frac{\sin i_{\rm p}}{1 / \sqrt{2}}} \lesssim 1,
    \label{eq:hj_rv}
\end{equation}
where $i_{\rm p}$ is the line-of-sight inclination angle of $m_{\rm p}$. Here,
we have taken $a_{\rm p}$ to be the maximum value ($3.3 \times
0.023\;\mathrm{AU}$) permitted by Eq.~\eqref{eq:hj_tce2_stab}. For these extreme
values of $a_{\rm p}$ and $m_{\rm p}$, we still have $L_{\rm p} / L \simeq 0.4$,
and so $L \gg L_{\rm p}$ is satisfied for the permitted parameter space.

(iv) Finally, we require that the initial orbital configuration of the two
planets be dynamically stable. We use the Hill stability criterion
\citep[e.g.][]{gladman1993dynamics, petit2020path},
\begin{equation}
    a_{\rm p} - a_{\rm i} > 2\sqrt{3} \p{\frac{a_{\rm p} + a_{\rm
        i}}{2}} \p{\frac{m + m_{\rm p}}{3M_{\star}}}^{1/3}.
\end{equation}
Assuming $m_{\rm p} \ll m$, this yields
\begin{equation}
    \frac{a_{\rm p}}{a_{\rm i}} > 1.29.\label{eq:hj_orbstab}
\end{equation}
The combination of the two constraints in
Eqs.~(\ref{eq:hj_adot},~\ref{eq:hj_orbstab}) justify the fiducial parameters
used in Eq.~\eqref{eq:hj_tce2_stab}.

We next address the implications of the rather small ``initial'' $\eta_{\rm
sync}$ value found in Eq.~\eqref{eq:hj_etasync}. When evaluating $\eta_{\rm
sync, i}$, it is possible that $R$ is larger today than its ``primordial'' value
(at semi-major axis $a_{\rm i} > a$) due to inflation induced by increased
stellar irradiation. However, if a smaller value of $R$ is used in
Eq.~\eqref{eq:hj_etasync}, the value of $a_{\rm i}$ must also be decreased such
that $R^5 / a_{\rm i}^{13/2}$ is constant in order to maintain the same
$\p{\dot{a} / a}_{\rm i}$ (see Eq.~\ref{eq:hj_adot}), which further decreases
$\eta_{\rm sync, i}$.

For $\eta_{\rm sync, i} \sim 0.075$ (corresponding to the fiducial parameters
used in Eq.~\ref{eq:hj_etasync}), we can infer that prograde primordial
obliquities will evolve towards tCE1 (see
Figs.~\ref{fig:Hhists_0_06},~\ref{fig:probs20}--\ref{fig:probs5}). On the other
hand, if the primordial obliquity of the HJ is assumed to be isotropically
distributed, then Fig.~\ref{fig:probs20} suggests that the probability of entry
into tCE2 is $\lesssim 25\%$ even if the perturbing planet is misaligned by
$I_{\rm J} \sim I \sim 20^\circ$. In reality, $\sin I_{\rm J} = \p{L_{\rm p} /
J}\sin I \ll \sin I$ for $L_{\rm p} \ll L$, so the probability is likely much
smaller (see Eq.~\ref{eq:app_tce2_p_tot} with $I$ replaced by $I_{\rm J}$).

Figure~\ref{fig:wasp12b_constr} illustrates the joint constraints on the
possible companion to WASP-12b and the resulting range of $\eta_{\rm sync, i}$
values. These small $\eta_{\rm sync, i}$ values suggest that capture of WASP-12b
into the high-obliquity tCE2 is unlikely from either an isotropic or
prograde-favoring initial obliquity distribution, and the observed orbital decay
of WASP-12b is unlikely to be driven by obliquity tides in the planet. For
obliquity tides to be operating today, we would have to imagine a scenario where
dynamical effects when the WASP-12b system was young may have preferentially
generated tCE2-producing systems, i.e.\ systems with $\theta_{\rm i} \simeq
90^\circ$. While the scenario considered by
\citet{millholland_disk} and \citet{su2020} with an exterior, dissipating
protoplanetary disk does not directly apply here due to the slow disk dispersal
time scale, a similar effect (decreasing $\eta$) can be accomplished by
simultaneous disk-driven migration of an inner HJ and exterior companion. The
exploration of such a scenario in the context of HJ formation is beyond the
scope of this paper.

\section{Summary and Discussion}\label{s:summary}

We have presented a comprehensive study on the evolution of a planet's spin
(both magnitude and direction) due to the combined effects of tidal dissipation
and gravitational interaction with an exterior companion/perturber.  This paper
extends our previous study \citep{su2020} of Colombo's Top (``spin + companion''
system) to include dissipative tidal effects, for which we have adopted the weak
friction theory of the equilibrium tide. Our paper contains several new general
theoretical results that can be adapted to various situations, as well as three
applications to exoplanetary systems of current interest.

We summarize our general theoretical results and provide a guide to the key
equations and figures as follows:
\begin{enumerate}
    \item In the presence of a spin-orbit alignment torque (such as that
        arising from tidal dissipation), our linear analysis
        (Section~\ref{ss:linear_stab}) shows explicitly that only two of the
        equilibrium spin orientations (called ``Cassini States'', CSs) are
        stable and attracting (see Fig.~\ref{fig:cs_locs}): the ``simple'' CS1
        (which typically has a low obliquity) and the ``resonant'' CS2 (which
        can have a large obliquity). The latter arises from the spin-orbit
        resonance, which occurs when the spin precession frequency of the planet
        is comparable to the orbital precession frequency driven by the
        companion. However, when the alignment torque is too strong (or the
        alignment timescale $t_{\rm al}$ too short), the CSs themselves can be
        significantly modified. In particular, when $t_{\rm al}$ is shorter than
        a critical value (of order the planet's orbital precession period; see
        Eq.~\ref{eq:mcs_shift_crit}), CS2 becomes destabilized and ceases to
        exist.

    \item We compute the long-term evolution of the planetary spin
        obliquity driven by the alignment torque for an arbitrary initial
        spin orientation (Section~\ref{ss:toy_outcomes}). When neglecting the
        evolution of the planet's spin magnitude, which implies that the spin
        and orbital precession frequencies $\alpha$, $g$ (see
        Eqs.~\ref{eq:wsl}--\ref{eq:wlp}) and the ratio $\eta = -g/\alpha$ are
        held constant, the asymptotic outcomes of the obliquity evolution (CS1 or
        CS2) can be analytically determined from the initial spin orientation
        (see Fig.~\ref{fig:toy_phop}), and we have obtained a new analytical
        expression for the probability of resonance capture into CS2
        (Eq.~\ref{eq:P32_toy} and Fig.~\ref{fig:1hist_toy}).

    \item In general, tidal torques act on both the obliquity and magnitude of
        the planetary spin, thus the ratio $\eta = -g/\alpha$ (which determines
        the phase-space structure of the system) evolves in time. Still, there
        are at most two equilibrium configurations (spin magnitude and
        obliquity) that are stable under the effect of tidal dissipation. We
        call these \emph{tidal Cassini Equilibria} (tCE\@; see
        Fig.~\ref{fig:6equils006}). The locations of these equilibria are
        determined by the system architecture and are parameterized by
        $\eta_{\rm sync}$ (Eq.~\ref{eq:def_etasync}), the ratio $\eta$ evaluated
        for $\Omega_s = n$ (fully synchronized spin rate).

    \item We show that if tCE1 exists (which requires $\eta_{\rm sync} <
        \eta_{\rm c}$, where $\eta_{\rm c}$ is given by Eqs.~\ref{eq:def_etac};
        Section~\ref{ss:tce}), which tCE a given initial planetary spin
        configuration asymptotically evolves towards depends on which of the
        phase space zones (see Fig.~\ref{fig:1contours}) the initial spin
        orientation belongs to (see
        Figs.~\ref{fig:Hhists_0_06}--\ref{fig:Hhists_0_50}): (i) If the spin
        originates in zone I, then it generally evolves towards tCE1 (unless
        $\eta_{\rm sync}$ very near $\eta_{\rm c}$, e.g.\ see
        Fig.~\ref{fig:Hhists_0_50}); (ii) if the spin originates in zone II,
        then it evolves towards tCE2 (which has a nontrivial obliquity); and
        (iii) if the spin originates in zone III, the outcome is generally
        probabilistic.

    \item For initial conditions in zone III, the probability of approaching
        either tCE can be determined by careful study of the dynamics upon
        separatrix encounter (Sections~\ref{ss:analytic_calculation}
        and~\ref{ss:phop_weaktide}); Figs.~\ref{fig:pc_fits_0_06}
        and~\ref{fig:pc_fits_0_20} give two example results. Assuming that the
        initial spin orientation is isotropically distributed, we have computed
        the overall probability of the system evolving into tCE2 as a function
        of $\eta_{\rm sync}$: Figs.~\ref{fig:probs20} and~\ref{fig:probs5} give
        the results for two different planet mutual inclinations, and
        Eq.~\eqref{eq:app_tce2_p_tot} gives an approximate analytical expression
        valid for $\eta_{\rm sync} \ll 1$.
\end{enumerate}

Applying our general theoretical results to three types of exoplanetary systems,
our key findings are (see Section~\ref{s:applications}):
\begin{enumerate}
    \item We show that over a wide range of parameter space, a super-Earth (SE)
        with an exterior cold Jupiter companion (or other types of companions
        with a similar $m_{\rm p}/a_{\rm p}^3$) has a substantial probability of
        being trapped in a permanent tCE2 with a significant obliquity, assuming
        that SEs have a wide range of primordial obliquities (e.g.\ due to giant
        impacts or collisions).

    \item We show that, in general, the formation of ultra-short-period planets
        (USPs) via runaway orbital decay driven by obliquity tides is difficult
        due to the low probability of capture into the high-obliquity tCE2. More
        importantly, proto-USPs that happen to be captured into tCE2 and
        initiate runaway tidal migration will likely break away from tCE2 and
        cease their inward migration before becoming a USP\@.

    \item The hot Jupiter WASP-12b is unlikely to be undergoing enhanced orbital
        decay due to obliquity tides, as the capture into tCE2 has a low
        probability or requires rather special initial conditions.
\end{enumerate}

Finally, we mention some possible caveats of our study.  We have adopted
dissipative tidal torques according to the (parameterized) weak friction theory
of the equilibrium tide. Other mechanisms of tidal dissipation may be dominant,
depending on the internal property of the planet and the nature of tidal forcing
\citep[e.g.][]{papaloizou_ivanov_inertial, Ogilvie2014, Storch2014}. We expect
that, with proper parameterization and rescaling, our theoretical results
presented in Sections~\ref{s:toy_model}--\ref{s:full_tide_prob} are largely
unaffected by the details of the tidal model. In any case, a different tidal
model is amenable to the same analysis as presented in this paper: The tCEs can
still be found by an analysis similar to that shown in
Fig.~\ref{fig:6equils006}, and the probabilistic outcome of a separatrix
encounter can still be solved using the techniques developed in
Sections~\ref{ss:analytic_calculation} and~\ref{ss:phop_weaktide}.

Some of results presented in Section~\ref{s:full_tide_prob}, such as
Figs.~\ref{fig:probs20}--\ref{fig:anal_ptce}, pertain to the probabilistic
outcomes of an initially isotropic distribution of spin orientations, assuming
that giant impacts or planet collisions effectively randomize a planet's
primordial spin. More physically accurate distributions can be used in the case
of planetary mergers \citep{li2020planetary} or many smaller impacts
\citep{dones1993does}. Figures~\ref{fig:probs20}--\ref{fig:probs5} can be
updated accordingly by convolving any initial obliquity distribution with the
tCE2 capture probability distributions, such as those shown in the right panels
of Figs.~\ref{fig:Hhists_0_06}--\ref{fig:Hhists_0_50} or the upper panels of
Figs.~\ref{fig:pc_fits_0_06}--\ref{fig:pc_fits_0_20}. The qualitative results
are unlikely to change, though the detailed probabilities for tCE2 capture can
increase (decrease) if the initial obliquity distribution favors (disfavors)
$\theta_{\rm i} \approx 90^\circ$ compared to the isotropic distribution.

\section{Acknowledgements}

We thank the anonymous referee for their useful comments. We thank Alexandre
Correia, Sarah Millholland, and Phil Nicholson for useful discussions and
comments. This work has been supported in part by NSF grant AST-2107796 and NASA
grant 80NSSC19K0444. YS is supported by the NASA FINESST grant 19-ASTRO19-0041.

\section{Data Availability}

The data referenced in this article will be shared upon reasonable request to
the corresponding author.

\bibliography{Su_weak_tides}

\begin{thebibliography}{}
\expandafter\ifx\csname natexlab\endcsname\relax\def\natexlab#1{#1}\fi
\providecommand{\url}[1]{\href{#1}{#1}}
\providecommand{\dodoi}[1]{doi:~\href{http://doi.org/#1}{\nolinkurl{#1}}}
\providecommand{\doeprint}[1]{\href{http://ascl.net/#1}{\nolinkurl{http://ascl.net/#1}}}
\providecommand{\doarXiv}[1]{\href{https://arxiv.org/abs/#1}{\nolinkurl{https://arxiv.org/abs/#1}}}

\bibitem[{Adams {et~al.}(2019)Adams, Millholland, \&
  Laughlin}]{millholland_signatures}
Adams, A.~D., Millholland, S., \& Laughlin, G.~P. 2019, arXiv preprint
  arXiv:1906.07615

\bibitem[{Alexander(1973)}]{alexander1973weak}
Alexander, M. 1973, Astrophysics and Space Science, 23, 459

\bibitem[{Anderson \& Lai(2018)}]{anderson2018teeter}
Anderson, K.~R., \& Lai, D. 2018, Monthly Notices of the Royal Astronomical
  Society, 480, 1402

\bibitem[{Barker(2020)}]{barker2020tidal}
Barker, A.~J. 2020, Monthly Notices of the Royal Astronomical Society, 498,
  2270

\bibitem[{Benz {et~al.}(1989)Benz, Slattery, \& Cameron}]{benz1989tilting}
Benz, W., Slattery, W., \& Cameron, A. 1989, Meteoritics, 24, 251

\bibitem[{Bonomo {et~al.}(2017)Bonomo, Desidera, Benatti, Borsa, Crespi,
  Damasso, Lanza, Sozzetti, Lodato, Marzari, {et~al.}}]{bonomo2017gaps}
Bonomo, A.~S., Desidera, S., Benatti, S., {et~al.} 2017, Astronomy \&
  Astrophysics, 602, A107

\bibitem[{Bryan {et~al.}(2018)Bryan, Benneke, Knutson, Batygin, \&
  Bowler}]{bryan2018constraints}
Bryan, M.~L., Benneke, B., Knutson, H.~A., Batygin, K., \& Bowler, B.~P. 2018,
  Nature Astronomy, 2, 138

\bibitem[{Bryan {et~al.}(2019)Bryan, Knutson, Lee, Fulton, Batygin, Ngo, \&
  Meshkat}]{bryan2019excess}
Bryan, M.~L., Knutson, H.~A., Lee, E.~J., {et~al.} 2019, The Astronomical
  Journal, 157, 52

\bibitem[{Bryan {et~al.}(2020)Bryan, Chiang, Bowler, Morley, Millholland,
  Blunt, Ashok, Nielsen, Ngo, Mawet, {et~al.}}]{bryan2020obliquity}
Bryan, M.~L., Chiang, E., Bowler, B.~P., {et~al.} 2020, The Astronomical
  Journal, 159, 181

\bibitem[{Colombo(1966)}]{colombo1966}
Colombo, G. 1966, The Astronomical Journal, 71, 891

\bibitem[{Dai {et~al.}(2018)Dai, Masuda, \& Winn}]{dai2018larger}
Dai, F., Masuda, K., \& Winn, J.~N. 2018, The Astrophysical Journal Letters,
  864, L38

\bibitem[{Dones \& Tremaine(1993)}]{dones1993does}
Dones, L., \& Tremaine, S. 1993, Science, 259, 350

\bibitem[{Fabrycky {et~al.}(2007)Fabrycky, Johnson, \&
  Goodman}]{fabrycky_otides}
Fabrycky, D.~C., Johnson, E.~T., \& Goodman, J. 2007, The Astrophysical
  Journal, 665, 754

\bibitem[{Fabrycky {et~al.}(2014)Fabrycky, Lissauer, Ragozzine, Rowe, Steffen,
  Agol, Barclay, Batalha, Borucki, Ciardi, {et~al.}}]{fabrycky2014architecture}
Fabrycky, D.~C., Lissauer, J.~J., Ragozzine, D., {et~al.} 2014, The
  Astrophysical Journal, 790, 146

\bibitem[{Fricke(1977)}]{fricke1977joint}
Fricke, W. 1977, Transactions of the International Astronomical Union, Series
  B, 16, 56

\bibitem[{Gladman(1993)}]{gladman1993dynamics}
Gladman, B. 1993, Icarus, 106, 247

\bibitem[{Goldreich \& Peale(1966)}]{goldreich1966spin}
Goldreich, P., \& Peale, S. 1966, The Astronomical Journal, 71, 425

\bibitem[{Groten(2004)}]{groten2004fundamental}
Groten, E. 2004, Journal of Geodesy, 77, 724, \dodoi{10.1007/s00190-003-0373-y}

\bibitem[{Guckenheimer \& Holmes(1983)}]{g_and_h}
Guckenheimer, J., \& Holmes, P.~J. 1983, Nonlinear oscillations, dynamical
  systems, and bifurcations of vector fields (New York: Springer-Verlag)

\bibitem[{Hamilton \& Ward(2004)}]{ward2004II}
Hamilton, D.~P., \& Ward, W.~R. 2004, The Astronomical Journal, 128, 2510

\bibitem[{Hebb {et~al.}(2009)Hebb, Collier-Cameron, Loeillet, Pollacco,
  H{\'e}brard, Street, Bouchy, Stempels, Moutou, Simpson,
  {et~al.}}]{hebb2009wasp}
Hebb, L., Collier-Cameron, A., Loeillet, B., {et~al.} 2009, The Astrophysical
  Journal, 693, 1920

\bibitem[{Henrard(1982)}]{henrard1982}
Henrard, J. 1982, Celestial Mechanics and Dynamical Astronomy, 27, 3

\bibitem[{Henrard \& Murigande(1987)}]{henrard1987}
Henrard, J., \& Murigande, C. 1987, Celestial Mechanics, 40, 345

\bibitem[{Husnoo {et~al.}(2011)Husnoo, Pont, H{\'e}brard, Simpson, Mazeh,
  Bouchy, Moutou, Arnold, Boisse, D{\'\i}az, {et~al.}}]{husnoo2011orbital}
Husnoo, N., Pont, F., H{\'e}brard, G., {et~al.} 2011, Monthly Notices of the
  Royal Astronomical Society, 413, 2500

\bibitem[{Hut(1981)}]{hut1981tidal}
Hut, P. 1981, Astronomy and Astrophysics, 99, 126

\bibitem[{Inamdar \& Schlichting(2015)}]{inamdar2015formation}
Inamdar, N.~K., \& Schlichting, H.~E. 2015, Monthly Notices of the Royal
  Astronomical Society, 448, 1751

\bibitem[{Izidoro {et~al.}(2017)Izidoro, Ogihara, Raymond, Morbidelli, Pierens,
  Bitsch, Cossou, \& Hersant}]{izidoro2017breaking}
Izidoro, A., Ogihara, M., Raymond, S.~N., {et~al.} 2017, Monthly Notices of the
  Royal Astronomical Society, 470, 1750

\bibitem[{Knutson {et~al.}(2014)Knutson, Fulton, Montet, Kao, Ngo, Howard,
  Crepp, Hinkley, Bakos, Batygin, {et~al.}}]{knutson2014friends}
Knutson, H.~A., Fulton, B.~J., Montet, B.~T., {et~al.} 2014, The Astrophysical
  Journal, 785, 126

\bibitem[{Korycansky {et~al.}(1990)Korycansky, Bodenheimer, Cassen, \&
  Pollack}]{korycansky1990one}
Korycansky, D., Bodenheimer, P., Cassen, P., \& Pollack, J. 1990, Icarus, 84,
  528

\bibitem[{Lai(2012)}]{lai2012}
Lai, D. 2012, Monthly Notices of the Royal Astronomical Society, 423, 486

\bibitem[{Lai \& Pu(2017)}]{lai_2017}
Lai, D., \& Pu, B. 2017, The Astronomical Journal, 153, 42,
  \dodoi{10.3847/1538-3881/153/1/42}

\bibitem[{Lainey(2016)}]{lainey2016quantification}
Lainey, V. 2016, Celestial Mechanics and Dynamical Astronomy, 126, 145

\bibitem[{Lee \& Chiang(2017)}]{lee2017magnetospheric}
Lee, E.~J., \& Chiang, E. 2017, The Astrophysical Journal, 842, 40

\bibitem[{Levrard {et~al.}(2007)Levrard, Correia, Chabrier, Baraffe, Selsis, \&
  Laskar}]{levrard2007}
Levrard, B., Correia, A., Chabrier, G., {et~al.} 2007, Astronomy \&
  Astrophysics, 462, L5

\bibitem[{Li \& Lai(2020)}]{li2020planetary}
Li, J., \& Lai, D. 2020, The Astrophysical Journal Letters, 898, L20

\bibitem[{Li {et~al.}(2021)Li, Lai, Anderson, \& Pu}]{li2021giant}
Li, J., Lai, D., Anderson, K.~R., \& Pu, B. 2021, Monthly Notices of the Royal
  Astronomical Society, 501, 1621

\bibitem[{Maciejewski {et~al.}(2013)Maciejewski, Dimitrov, Seeliger, Kitze,
  Errmann, Nowak, Niedzielski, Popov, Marka, Go{\'z}dziewski,
  {et~al.}}]{maciejewski2013multi}
Maciejewski, G., Dimitrov, D., Seeliger, M., {et~al.} 2013, Astronomy \&
  Astrophysics, 551, A108

\bibitem[{Maciejewski {et~al.}(2016)Maciejewski, Dimitrov, Fern{\'a}ndez, Sota,
  Nowak, Ohlert, Nikolov, Hinse, Pall{\'e}, Tingley,
  {et~al.}}]{maciejewski2016departure}
Maciejewski, G., Dimitrov, D., Fern{\'a}ndez, M., {et~al.} 2016, Astronomy \&
  Astrophysics, 588, L6

\bibitem[{Millholland \& Batygin(2019)}]{millholland_disk}
Millholland, S., \& Batygin, K. 2019, The Astrophysical Journal, 876, 119

\bibitem[{Millholland \& Laughlin(2018)}]{millholland_wasp12b}
Millholland, S., \& Laughlin, G. 2018, The Astrophysical Journal Letters, 869,
  L15

\bibitem[{Millholland \& Laughlin(2019)}]{millholland2019obliquity}
---. 2019, Nature Astronomy, 3, 424

\bibitem[{Millholland \& Spalding(2020)}]{millholland2020formation}
Millholland, S.~C., \& Spalding, C. 2020, The Astrophysical Journal, 905, 71

\bibitem[{Morbidelli {et~al.}(2012)Morbidelli, Tsiganis, Batygin, Crida, \&
  Gomes}]{morbidelli_gi}
Morbidelli, A., Tsiganis, K., Batygin, K., Crida, A., \& Gomes, R. 2012,
  Icarus, 219, 737

\bibitem[{{Ogilvie}(2014)}]{Ogilvie2014}
{Ogilvie}, G.~I. 2014, \araa, 52, 171,
  \dodoi{10.1146/annurev-astro-081913-035941}

\bibitem[{Ohno \& Zhang(2019)}]{ohno_infer_obl}
Ohno, K., \& Zhang, X. 2019, The Astrophysical Journal, 874, 2

\bibitem[{Papaloizou \& Ivanov(2010)}]{papaloizou_ivanov_inertial}
Papaloizou, J. C.~B., \& Ivanov, P.~B. 2010, Monthly Notices of the Royal
  Astronomical Society, 407, 1631, \dodoi{10.1111/j.1365-2966.2010.17011.x}

\bibitem[{Patra {et~al.}(2017)Patra, Winn, Holman, Yu, Deming, \&
  Dai}]{patra2017apparently}
Patra, K.~C., Winn, J.~N., Holman, M.~J., {et~al.} 2017, The Astronomical
  Journal, 154, 4

\bibitem[{{Patra} {et~al.}(2020){Patra}, {Winn}, {Holman}, {Gillon},
  {Burdanov}, {Jehin}, {Delrez}, {Pozuelos}, {Barkaoui}, {Benkhaldoun},
  {Narita}, {Fukui}, {Kusakabe}, {Kawauchi}, {Terada}, {Bouma}, {Weinberg}, \&
  {Broome}}]{Patra12b}
{Patra}, K.~C., {Winn}, J.~N., {Holman}, M.~J., {et~al.} 2020, \aj, 159, 150,
  \dodoi{10.3847/1538-3881/ab7374}

\bibitem[{Peale(2008)}]{peale2008obliquity}
Peale, S. 2008, in Extreme Solar Systems, Vol. 398, 281

\bibitem[{Peale(1969)}]{peale1969}
Peale, S.~J. 1969, The Astronomical Journal, 74, 483

\bibitem[{Peale(1974)}]{peale1974possible}
---. 1974, The Astronomical Journal, 79, 722

\bibitem[{Petit {et~al.}(2020)Petit, Pichierri, Davies, \&
  Johansen}]{petit2020path}
Petit, A.~C., Pichierri, G., Davies, M.~B., \& Johansen, A. 2020, Astronomy \&
  Astrophysics, 641, A176

\bibitem[{Petrovich {et~al.}(2019)Petrovich, Deibert, \&
  Wu}]{petrovich2019ultra}
Petrovich, C., Deibert, E., \& Wu, Y. 2019, The Astronomical Journal, 157, 180

\bibitem[{Pu \& Lai(2019)}]{pu2019low}
Pu, B., \& Lai, D. 2019, Monthly Notices of the Royal Astronomical Society,
  488, 3568

\bibitem[{Rogoszinski \& Hamilton(2019)}]{hamilton_tilting_ice}
Rogoszinski, Z., \& Hamilton, D.~P. 2019, arXiv preprint arXiv:1908.10969

\bibitem[{Safronov \& Zvjagina(1969)}]{original_gi}
Safronov, V., \& Zvjagina, E. 1969, Icarus, 10, 109

\bibitem[{Saillenfest {et~al.}(2021)Saillenfest, Lari, \&
  Bou{\'e}}]{saillenfest2021large}
Saillenfest, M., Lari, G., \& Bou{\'e}, G. 2021, Nature Astronomy, 5, 345

\bibitem[{Saillenfest {et~al.}(2020)Saillenfest, Lari, \&
  Courtot}]{saillenfest2020future}
Saillenfest, M., Lari, G., \& Courtot, A. 2020, Astronomy \& Astrophysics, 640,
  A11

\bibitem[{Sanchis-Ojeda {et~al.}(2014)Sanchis-Ojeda, Rappaport, Winn, Kotson,
  Levine, \& El~Mellah}]{sanchis2014study}
Sanchis-Ojeda, R., Rappaport, S., Winn, J.~N., {et~al.} 2014, The Astrophysical
  Journal, 787, 47

\bibitem[{Seager \& Hui(2002)}]{seager2002constraining}
Seager, S., \& Hui, L. 2002, The Astrophysical Journal, 574, 1004

\bibitem[{Snellen {et~al.}(2014)Snellen, Brandl, de~Kok, Brogi, Birkby, \&
  Schwarz}]{snellen2014fast}
Snellen, I.~A., Brandl, B.~R., de~Kok, R.~J., {et~al.} 2014, Nature, 509, 63

\bibitem[{Steffen \& Farr(2013)}]{steffen2013lack}
Steffen, J.~H., \& Farr, W.~M. 2013, The Astrophysical Journal Letters, 774,
  L12

\bibitem[{{Storch} \& {Lai}(2014)}]{Storch2014}
{Storch}, N.~I., \& {Lai}, D. 2014, \mnras, 438, 1526,
  \dodoi{10.1093/mnras/stt2292}

\bibitem[{Su \& Lai(2020)}]{su2020}
Su, Y., \& Lai, D. 2020, The Astrophysical Journal, 903, 7

\bibitem[{{Turner} {et~al.}(2021){Turner}, {Ridden-Harper}, \&
  {Jayawardhana}}]{turner12b}
{Turner}, J.~D., {Ridden-Harper}, A., \& {Jayawardhana}, R. 2021, \aj, 161, 72,
  \dodoi{10.3847/1538-3881/abd178}

\bibitem[{Vokrouhlick{\`y} \& Nesvorn{\`y}(2015)}]{vokrouhlicky2015tilting}
Vokrouhlick{\`y}, D., \& Nesvorn{\`y}, D. 2015, The Astrophysical Journal, 806,
  143

\bibitem[{Ward(1975)}]{ward1975tidal}
Ward, W.~R. 1975, The Astronomical Journal, 80, 64

\bibitem[{Ward \& Canup(2006)}]{ward_jupiter}
Ward, W.~R., \& Canup, R.~M. 2006, The Astrophysical Journal Letters, 640, L91

\bibitem[{Ward \& Hamilton(2004)}]{ward2004I}
Ward, W.~R., \& Hamilton, D.~P. 2004, The Astronomical Journal, 128, 2501

\bibitem[{Weinberg {et~al.}(2017)Weinberg, Sun, Arras, \&
  Essick}]{weinberg2017tidal}
Weinberg, N.~N., Sun, M., Arras, P., \& Essick, R. 2017, The Astrophysical
  Journal Letters, 849, L11

\bibitem[{Winn {et~al.}(2018)Winn, Sanchis-Ojeda, \&
  Rappaport}]{winn2018kepler}
Winn, J.~N., Sanchis-Ojeda, R., \& Rappaport, S. 2018, New Astronomy Reviews,
  83, 37

\bibitem[{Zhu \& Wu(2018)}]{zhu2018super}
Zhu, W., \& Wu, Y. 2018, The Astronomical Journal, 156, 92

\end{thebibliography}
\bibliographystyle{aasjournal}

\appendix

\onecolumn

\section{Convergence of Initial Conditions Inside the Separatrix to CS2
}\label{app:cs_stab2}

In Section~\ref{ss:linear_stab}, we studied the stability of the CSs under of
tidal alignment torque given by Eq.~\eqref{eq:dsdt_tide_toy}, finding that CS2
is locally stable. Later, in Section~\ref{ss:toy_outcomes}, we found that all
initial conditions within the separatrix converge to CS2, which is not
guaranteed by local stability of CS2. In this section, we give an analytic
demonstration that all points inside the separatrix indeed converge to CS2,
focusing on the case where $\eta \ll 1$.

Similarly to the analytic calculation in Section~\ref{ss:analytic_calculation}, we seek
to compute the change in the unperturbed Hamiltonian over a single libration
cycle. To calculate the evolution of $H$, we first parameterize the unperturbed
trajectory (similarly to Eq.~\ref{eq:sep_theta}). For initial conditions inside
the separatrix, the value of $H$ can be written $H = H_{\rm sep} + \Delta H$
where $\Delta H > 0$, and the two legs of the libration trajectory can be
written:
\begin{align}
    \cos \theta_{\pm} &\approx
        \eta \cos I \pm \sqrt{2\eta\s{\sin I\p{1 - \cos \phi} - \Delta H}}.
        \label{eq:lib_cycle_toy}
\end{align}
We have taken $\sin \theta \approx 1$, a good approximation in zone II when
$\eta \ll 1$. Note that there are some values of $\phi$ for which no solutions
of $\theta$ exist, reflecting the fact that the libration cycle does not extend
over the full interval $\phi \in [0, 2\pi]$. During a libration cycle,
$\theta_-$ [$\theta_+$] is traversed while $\phi' > 0$ [$\phi' < 0$], i.e.\ the
trajectory librates counterclockwise in $(\cos \theta, \phi)$ phase space (see
Fig.~\ref{fig:1contours}).

The leading order change to $H$ over a single libration cycle can then computed by
integrating $\rdil{H}{t}$ along this trajectory, yielding:
\begin{align}
    \oint \rd{H}{t}\;\mathrm{d}t
        &= \oint \p{\rd{(\cos \theta)}{t}}_{\rm tide}
            \;\mathrm{d}\phi,\nonumber\\
        &= \int\limits_{\phi_{\min}}^{\phi_{\max}}
                \frac{1}{t_{\rm s}}
                \p{\sin^2\theta_- - \sin^2\theta_+} \;\mathrm{d}\phi\nonumber\\
        &\approx \frac{1}{t_{\rm s}}
            \int\limits_{\phi_{\min}}^{\phi_{\max}}
                4\eta \cos I \sqrt{2\eta\s{\sin I\p{1 - \cos \phi} - \Delta H}}
                \;\mathrm{d}\phi > 0.
\end{align}
Here, $\phi_{\min} > 0$ and $\phi_{\max} < 2\pi$ are defined such that the
trajectory librates over $\phi \in \s{\phi_{\min}, \phi_{\max}}$. Thus, $H$ is
strictly increasing for all initial conditions inside the separatrix, and they
all converge to CS2.

\section{Approximate TCE2 Probability for Small $\eta_{\rm sync}$
}\label{app:ptce2}

In this appendix, we seek a tentative analytic understanding for the
probability of convergence to tCE2 when $\eta_{\rm sync}$ is small, i.e.\ the
left extremes of Figs.~\ref{fig:probs20} and~\ref{fig:probs5}. In this regime,
following the discussions in Sections~\ref{ss:analytic_calculation}
and~\ref{ss:phop_weaktide}, we understand that initial conditions (ICs) in zone I
always converge to tCE1, ICs in zone II always converge to tCE2,
and ICs in zone III experience separatrix encounter and
probabilistically converge to either one of the tCE\@. To further proceed, we
will assume an isotropic distribution of initial spin orientations; different
distributions again will only change the quantitative but not qualitative
character of the discussion. Then the tCE2 probability, which we denote by
$P_{\rm tCE2}$, can be expressed as the sum of: (i) the probability that an IC
is in zone II, and (ii) the probability that an IC is both in zone III and
undergoes a III $\to$ II transition. To simplify the discussion, we will
approximate that $P_{\rm tCE2}$ can be calculated as
\begin{equation}
    P_{\rm tCE2} \sim \frac{A_{\rm II}}{4\pi}
            + \frac{A_{\rm III}}{4\pi}\ev{P_{\rm III \to II}},
            \label{eq:app_ptce2_schematic}
\end{equation}
where $A_{\rm II}$ and $A_{\rm III}$ are the phase space areas of zones II and
III respectively, and $\ev{P_{\rm III \to II}}$ is the \emph{average} III $\to$
II transition probability for a random IC in zone III\@. Next, we evaluate each
of the expressions in Eq.~\eqref{eq:app_ptce2_schematic}.

We first consider $A_{\rm II}$ and $A_{\rm III}$. Exact analytic forms for both
$A_{\rm II}$ and $A_{\rm III}$ is known (\citealp{ward2004I}, Paper I), but an
accurate approximation can be obtained using Eq.~\eqref{eq:sep_theta} since
$\eta_{\rm i} \ll 1$. We obtain that:
\begin{align}
    \frac{A_{\rm II}}{4\pi} &= \frac{4}{\pi}\sqrt{\eta_{\rm i} \sin I},\\
    \frac{A_{\rm III}}{4\pi}
        &= \frac{1 + \eta_{\rm i} \cos I}{2} -
            \frac{2}{\pi}\sqrt{\eta_{\rm i} \sin I}.
\end{align}

Next, we need to evaluate $\ev{P_{\rm III \to II}}$, for which we must
understand the outcomes of the separatrix encounters that ICs in zone III
experience. We proceed by analytically calculating $\Delta K_{\pm}$
(Eq.~\ref{eq:def_dK_weaktide}) for use in Eq.~\eqref{eq:def_pc_weaktide} to
obtain the probabilities of the outcomes of separatrix encounter. We first
rewrite Eq.~\eqref{eq:def_dK_weaktide} as:
\begin{align}
    \Delta K_{\pm} &= \oint_{\mathcal{C}_{\pm}} \rd{H}{t}
        - \rd{H_{\rm sep}}{t}\;\mathrm{d}t
        \nonumber\\
        &= \oint_{\mathcal{C}_{\pm}}
           \p{\rd{(\cos\theta)}{t}}_{\rm tide}
            + \frac{\dot{\Omega}_{\rm s}}{\dot{\phi}}
            \p{\pd{H}{\Omega_{\rm s}} - \pd{H_{\rm sep}}{\Omega_{\rm s}}}\;\mathrm{d}\phi
                \label{eq:app_dhpm}.
\end{align}
Then, using the full equations of motion for the planet's spin including weak
tidal friction in component form, given by
Eqs.~(\ref{eq:ds_fullq}--\ref{eq:ds_fulls}), we can evaluate $\Delta K_{\pm}$ by
integrating along the two legs of the separatrix $\mathcal{C}_{\pm}$ (see
Fig.~\ref{fig:1contours}). Note that we must use the value of $\eta$ at the
moment of separatrix encounter, which we denote $\eta_{\rm cross}$, as the
evolution of $\Omega_{\rm s}$ changes the spin-orbit precession frequency
$\alpha$ and thus $\eta$ itself:
\begin{align}
    t_{\rm s} \Delta K_{\pm} \approx{}&
        \frac{\eta_{\rm cross}^2}{\eta_{\rm sync}}\s{
            -2\cos I\p{\pm 2\pi \eta_{\rm cross} \cos I
                + 8\sqrt{\eta_{\rm cross} \sin I}}
            \mp 4\pi \sin I
            - 8 \cos I \sqrt{\eta_{\rm cross}\sin I}
                + \frac{4\eta_{\rm sync}}{\eta_{\rm cross}}
                    \sqrt{\sin I/\eta_{\rm cross}}}\nonumber\\
        &+ \frac{2\eta_{\rm cross}}{\eta_{\rm sync}}
            \p{\mp 2\pi\p{1 - 2\eta_{\rm cross} \sin I}
            + 16\cos I \eta_{\rm cross}^{3/2}\sqrt{\sin I}}
            + 8\sqrt{\eta_{\rm cross} \sin I}
            \pm 2 \pi \eta_{\rm cross} \cos I
            - \frac{64}{3} \p{\eta_{\rm cross} \sin I}^{3/2}.
                \label{eq:app_deltaK}
\end{align}
The resulting $P_{\rm III \to II}$ obtained using this analytic $\Delta K_{\pm}$
in Eq.~\eqref{eq:def_pc_weaktide} is shown as the green dashed line in the top
panel of Fig.~\ref{fig:pc_fits_0_06}, where it can be seen that agreement is
reasonable for $\eta_{\rm cross} \lesssim 0.05$. For the purposes of this
section, we drop all but the leading order terms in both the numerator and
denominator of Eq.~\eqref{eq:def_pc_weaktide} and obtain:
\begin{equation}
    P_{\rm III \to II} \simeq
        \frac{6\eta_{\rm sync}}{\pi} \sqrt{\frac{\sin I}{\eta_{\rm cross}}}.
\end{equation}

However, $\eta_{\rm cross}$ cannot be expressed in closed form as a function of
the ICs. Based on the bottom panel of Fig.~\ref{fig:pc_fits_0_06}, we make the
crude approximation that $\eta_{\rm cross}$ is uniformly distributed between
$\eta_{\rm i}$ and $\eta_{\rm sync}$. Note that if $\Omega_{\rm s} \simeq n$,
then this approximation is invalid: since nearly anti-aligned spins ($\theta_{\rm
i} \approx 180^\circ$) will undergo significant spin-down before tidal friction
can realign the spin orientation, $\Omega_{\rm s, i}$ being too close to $n$
results in $\eta_{\rm cross} \ll \eta_{\rm sync}$. We thus obtain:
\begin{align}
    \ev{P_{\rm III \to II}}
        &\sim \frac{1}{\eta_{\rm sync} - \eta_{\rm i}}
            \int\limits_{\eta_{\rm i}}^{\eta_{\rm sync}}
            P_{\rm III \to II}\;\mathrm{d}\eta_{\rm cross}\nonumber\\
        &= \frac{12 \sqrt{\eta_{\rm sync}\sin I}}{\pi
            \p{1 + \sqrt{n / \Omega_{\rm s, i}}}}.
\end{align}

With this result, we can finally express Eq.~\eqref{eq:app_ptce2_schematic} as:
\begin{align}
    P_{\rm tCE2} &\simeq
            \frac{4\sqrt{\eta _{\rm sync} \sin I}}{\pi}\s{
                \sqrt{n / \Omega_{\rm s, i}}
                + \frac{3}{2\p{1
                + \sqrt{n / \Omega_{\rm s, i}}}}}
            + \mathcal{O}\p{\eta_{\rm sync}}.
\end{align}
This is exactly Eq.~\eqref{eq:app_tce2_p_tot}. We remark again that this is
valid in the regime where $\eta_{\rm sync} \ll 1$ and $\Omega_{\rm s} \gtrsim
n$.

\label{lastpage} 
\end{document}